\documentclass[apj]{emulateapj}
\usepackage{graphics}
\usepackage{amsmath}
\usepackage{relsize}

\shortauthors{Nevin et al.}
\shorttitle{Feedback from Biconical AGN Outflows}
\begin{document}
\title{
The Origin of Double-Peaked Narrow Lines in Active Galactic Nuclei III: Feedback from Biconical AGN Outflows
}

\author{R. Nevin}
\affil{Department of Astrophysical and Planetary Sciences, University of Colorado, Boulder, CO 80309, USA}

\author{J. M. Comerford}
\affil{Department of Astrophysical and Planetary Sciences, University of Colorado, Boulder, CO 80309, USA}

\author{F. M{\"u}ller-S{\'a}nchez}
\affil{Department of Astrophysical and Planetary Sciences, University of Colorado, Boulder, CO 80309, USA}

\author{R. Barrows}
\affil{Department of Astrophysical and Planetary Sciences, University of Colorado, Boulder, CO 80309, USA}

\and

\author{M. Cooper}
\affil{Center for Cosmology, Department of Physics \& Astronomy, 4129 Reines Hall, University of California, Irvine, CA 92697, USA}

\begin{abstract}

We apply an analytic Markov Chain Monte Carlo model to a sample of 18 AGN-driven biconical outflows that we identified from a sample of active galaxies with double-peaked narrow emission lines at $z < 0.1$ in the Sloan Digital Sky Survey. We find that 8/18 are best described as asymmetric bicones, 8/18 are nested bicones, and 2/18 are symmetric bicones. From the geometry and kinematics of the models, we find that these moderate-luminosity AGN outflows are large and energetic. The biconical outflows axes are randomly oriented with respect to the photometric major axis of the galaxy, implying a randomly oriented and clumpier torus to collimate the outflow, but the torus also allows some radiation to escape equatorially. We find that 16/18 (89\%) outflows are energetic enough to drive a two-staged feedback process in their host galaxies. All of these outflows geometrically intersect the photometric major axis of the galaxy, and 23\% of outflow host galaxies are significantly redder or have significantly lower specific star formation rates when compared to a matched sample of active galaxies.

\end{abstract}


\section{Introduction}

The tight observational correlations between stellar bulge properties such as mass and velocity dispersion and supermassive black hole (SMBH) mass indicate that SMBHs can be powerful drivers of galaxy evolution (e.g., \citealt{Merritt2000,McConnell2013}). Since the sphere of influence of the SMBH's gravity is miniscule, a physical coupling between the host galaxy and the energy of active galactic nuclei (AGNs), which are the active phase of SMBHs, must explain these scaling relations. This coupling is known as AGN feedback.

Both theoretical models and observations have investigated the role of AGN feedback in galaxy evolution. Observationally, the bimodal color distribution of galaxies in the nearby universe and the lack of massive galaxies in the galaxy mass function require quenching of star formation in galaxies via a feedback mechanism (e.g., \citealt{Bell2004,Brown2007,Faber2007,Silk2011}). In models, AGN-driven feedback provides a mechanism to evacuate gas from a galaxy and quench star formation and the growth of the SMBH (e.g., \citealt{diMatteo2005,Hopkins2005,Springel2005,Croton2006}). Despite the utility of AGN feedback in regulating galaxy and SMBH growth, there is limited direct evidence for feedback operating on host galaxies. Additionally, despite many proposed mechanisms to deliver energy from the AGN to the interstellar medium (ISM) of the host galaxy, little is known about the energy, geometry, and efficiency of these mechanisms.

Recent work has focused on a handful of very energetic AGN-driven outflows and winds, such as UV and X-ray Broad Absorption Line (BAL) QSO outflows (e.g., \citealt{Crenshaw2012,Arav2013,Crenshaw2015}) as well as narrow line region (NLR) outflows (e.g., \citealt{Muller-Sanchez2011,Fischer2013,Crenshaw2015,Muller-Sanchez2016,Fischer2017}). This work has carefully measured the outflow velocity, and therefore, kinetic energy, of these objects. Theoretical studies predict a 0.5\% threshold as the ratio of kinetic luminosity, or $\mathrm{L}_{\mathrm{KE}}$, associated with an outflow to the AGN bolometric luminosity, or $\mathrm{L}_{\mathrm{bol}}$, of the AGN outflows necessary to drive a powerful two-staged feedback process (\citealt{Hopkins2010a}). Some high energy outflows exceed this energy threshold (e.g., \citealt{Crenshaw2012,Arav2013,Borguet2013,Liu2016}).

While some energetic outflows have the potential to disrupt the molecular gas in the disk of the host galaxy, these are extreme cases, and it remains difficult to find direct evidence for feedback in most of these galaxies. To fully address how the overall population of AGNs drive feedback in their host galaxies, it is necessary to characterize the amount of energy entrained in outflows, determine the efficiency of energy delivery using the geometry of the outflow, and find direct evidence for the effects of feedback on the host galaxy.

It is important to determine both the geometry of the outflow and how the outflow is oriented with respect to the star forming disk of the galaxy to establish how and where the energy is delivered to the ISM. Some authors find that a spherical geometry with a 180$^{\circ}$ opening angle describes NLR outflows while others measure narrower opening angles associated with a biconical outflow (e.g., \citealt{Muller-Sanchez2011,Liu2013b}). A bicone model for the NLR of an outflow is expected from the unified model of AGNs; a thick torus provides the collimation necessary to produce a biconical outflow (\citealt{Antonucci1985}). \citet{Fischer2013} find a homogeneous distribution of orientations for their sample of Seyfert galaxies with ionized outflows, suggesting that AGN outflows may have a random orientation with respect to the star forming disk of the galaxy. The AGNIFS group also finds that ionized gas outflows are oriented at random angles to the galactic plane (\citealt{Storchi-Bergmann2010,Riffel2011a,Riffel2011b,Riffel2013,Riffel2015,Schonell2014}). Other observational studies of ionized outflows suggest outflows that are aligned with the photometric major axis of the galaxy (e.g., \citealt{Elitzur2012}).

Our sample of AGN outflows from \citet{Nevin2016} are moderate-luminosity (42 $<$ log L$_{\mathrm{bol}}$ (erg s$^{-1}$) $<$ 46) AGN outflows in the local universe. They were originally selected from the Sloan Digital Sky Survey (SDSS) as $z < 0.1$ Type 2 AGNs with double-peaked narrow emission lines. Moderate-luminosity AGNs such as these account for 10\% of the total AGN population at low redshifts ($z<0.1$) (e.g., \citealt{Silverman2006,Ueda2006}); they are more ubiquitous than high-luminosity AGNs (1\% of the total AGN population), which includes the BAL QSO population. In addition to representing a larger fraction of the AGN population, moderate-luminosity outflows also operate on kpc-scales, coincident with circumnuclear star formation. This enables us to directly assess the effects of outflows on the ISM of the host galaxies (\citealt{Crenshaw2015}). If moderate-luminosity AGNs are capable of driving feedback, they are so common that they could contribute significantly to the explanation for observed galaxy-SMBH scaling relations. 
 
The remainder of this paper is organized as follows. We describe the sample selection, biconical models, and Markov Chain Monte Carlo analytic modeling technique in Section \ref{methods}. In Section \ref{results} we calculate sample statistics for orientation of the outflows and energy diagnostics of the outflows. We discuss the implications of the best fit biconical outflow models, energy diagnostics, and geometry in Section \ref{discuss}. We present our conclusions in Section \ref{conclude}. A cosmology with $\Omega_m = 0.3$, $\Omega_{\Lambda}=0.7$, and $h=0.7$ is assumed throughout.

\section{Methods}
\label{methods}

The 71 Type 2 AGNs with double-peaked narrow emission lines at $z < 0.1$ in the SDSS were introduced and classified in \citet{Nevin2016}. We observed each galaxy at two position angles with optical longslit spectroscopy (\citealt{Comerford2012,Nevin2016}). For each galaxy, we reduced the 2d spectra and extracted the [OIII]$\lambda$5007 profiles. We calculated the velocity offset of the [OIII]$\lambda$5007 emission line relative to the systemic velocity derived in the SDSS DR7 value-added catalogues (OSSY) from absorption lines (\citealt{Oh2011}). The 71 galaxies were classified as Outflow, Outflow Composite, Rotation-Dominated + Obscuration, Rotation-Dominated + Disturbance, or Ambiguous. Here we focus only on the 61 galaxies that were classified under the outflow-dominated classifications of Outflow and Outflow Composite.

\subsection{Selection Criteria for Analytic Modeling}
\label{selection}

We initially model all of the 61 outflow-dominated AGNs as biconical outflows. We are motivated to use a biconical outflow model because each of the double-peaked emission lines are kinematically described as outflows on all spatial scales (\citealt{Nevin2016}). We apply the kinematic classification method from \citet{Nevin2016} to the velocity dispersions and velocity offsets of both components on all spatial scales. We find that unlike in some nearby Seyfert galaxies, where the NLR kinematics are best described by a small-scale outflows and large-scale illuminated disk rotation (\citealt{Fischer2017}), both components can only be described by outflow kinematics for our sample. We analyze the lack of rotation-dominated structure further in Appendix \ref{APa}. Therefore, we rule out illuminated rotating structure as the origin for the double peaks at all positions observed along the slit and ensure that we are physically motivated to model the two components as the walls of a biconical outflow.

Then, we establish selection criteria to determine which galaxies are well-modeled, and identify 18 galaxies for further analysis in this paper. We present these 18 galaxies with their PAs, spatial apertures, spatial resolution, and spectral resolution in Table \ref{obsproperties}. We select these 18 galaxies based upon three requirements for the spatially resolved spectra as described in the following paragraphs. 

\begin{deluxetable*}{cccccccccc}

\tabletypesize{\scriptsize}
\tablewidth{0pt}
\tablecolumns{5}
\tablecaption{Observational Properties}
\tablehead{
\colhead{SDSS ID} & 
\colhead{PA 1} & 
\colhead{PA 2} &
\colhead{Instrument} &
\colhead{Slitwidth ($^{\prime \prime}$)} &
\colhead{Seeing ($^{\prime \prime}$)$^a$} &
\colhead{Dispersion (\AA \ pix$^{-1}$)}}

\startdata

J0009$-$0036 &23 &67 & MMT/ Blue Channel & 1.0 & 0.8 & 0.50 \\
J0803+3926& 50& 140& Palomar/DBSP & 1.5 & 2.5 &0.55 \\
J0821+5021& 43&133 & Palomar/DBSP & 1.5 & 2.5  &0.55\\
J0854+5026& 16&106 & Palomar/DBSP & 1.5 & 2.5 &0.55 \\
J0930+3430& 21&111 & Palomar/DBSP & 1.5 &2.5 &0.55\\
J0959+2619& 28&118 & MMT/ Blue Channel & 1.0 &1.0  &0.50 \\
J1027+1049& 75&165 & APO/DIS  & 1.5 & 1.8  & 0.62\\

J1109+0201& 31&121 & APO/DIS & 1.5 & 1.2 & 0.62\\

J1152+1903& 17&107 & Palomar/DBSP & 1.5 &1.9  &0.55\\

J1315+2134&74 &164 & APO/DIS & 1.5 & 1.5 & 0.62\\
J1328+2752& 39&129 & APO/DIS & 1.5 & 1.7  &0.62 \\
J1352+0525& 162& 252& Keck/DEIMOS & 0.75 &0.9  &0.33 \\
J1420+4959& 79&169 & APO/DIS & 1.5 & 1.5 & 0.62\\
J1524+2743& 12&102 & APO/DIS & 1.5 &1.5 &0.62\\
J1526+4140 & 38 &128 & APO/DIS & 1.5 &2.0  &0.62\\

J1606+3427& 17 &107 & Keck/DEIMOS & 0.75 & 0.9 &0.33\\
J1630+1649& 30 &120 & Keck/DEIMOS & 0.75 & 1.0 &0.33 \\
J1720+3106& 152&242 & Keck/DEIMOS  & 0.75 &0.6 & 0.33

\enddata

\tablecomments{The observed PAs and associated instrument used for the observations for each galaxy. Column 1: galaxy name, also SDSS ID. Column 2: the first observed PA, also often the PA of the photometric major axis of the galaxy in SDSS. Column 3: the second observed PA. Column 4: observatory and instrument. Column 5: the slitwidth of the longslit used. Column 6: the approximate seeing of each observation derived from the FWHM of the stellar PSF. Column 7:  the dispersion of the instrument. } 
\tablenotetext{a}{Seeing for Keck/DEIMOS is derived from the FWHM of the PSF of acquisition stars.}
\label{obsproperties} 
\end{deluxetable*}

First, we include only rows in the longslit spectra in which the Akaike statistic from \citet{Nevin2016} demonstrates that a two Gaussian fit is significantly better than a one Gaussian fit. This ensures that our goal of producing an analytic model of a cone for a two Gaussian profile is met. Second, we require that the kinematic model directly model the two Gaussian centroids as the two walls of a cone. We assign a joint velocity and dispersion tracking method to associate each single Gaussian component of the two Gaussian fit with a physical wall of the cone. 

Second, in previous work, groups have used a dispersion association method to track components that belong to a given wall of a bicone. In \citet{Westmoquette2011}, the component with the larger dispersion always corresponds to the same wall of the bicone. For example, one wall of the bicone has a larger velocity dispersion and is the redder emission component at one spatial position. If at a different spatial position, the emission components switch relative velocity dispersions, where the bluer component now has a larger dispersion, suddenly the bluer component is now associated with the wall with the larger velocity dispersion. 

We choose to use a stricter tracking method than \citet{Westmoquette2011}. In addition to the dispersion tracking method, we  restrict our modeling to galaxies where the components with similar dispersion across the slit are also related in velocity. This combination of velocity and dispersion tracking restricts our physical model to a model of bulk motion of the biconical outflow. For instance, the dispersion tracking method in \citet{Westmoquette2011} allows material from a given wall of the bicone to suddenly move at a new velocity that may be $\sim$1000 km s$^{-1}$ different from the bulk motion of that wall. In contrast, we require that both of the walls of the cone move at the velocity set by the velocity law (Section \ref{analyticModels}), and we choose to model only galaxies whose spectra match this physical explanation. 

We eliminate rows of data that violate the velocity and dispersion association requirements. For example, if a narrower emission component is the blueshifted component in one row but then in a subsequent row becomes the redshifted component, we eliminate the exterior rows by truncating the data at the last row of dispersion association. The justification is that the flux in these exterior rows is low enough to confuse emission components, and we may not be associating components with their proper physical wall of the cone.  

An additional motivation for eliminating rows of data based upon the Akaike statistic and the velocity and dispersion tracking method is to ensure model convergence. Since we use a likelihood maximization technique, rows with large error bars on the velocity centroids do not lead to the convergence of the best-fit model. Spatial rows that do not pass the two requirements discussed above often have very low S/N and therefore large error bars.

Third, we select only the galaxies with $n>2k$, where $k$ is the number of parameters in the bicone fit and $n$ is the number of spatial rows of data (from each individual PA) that satisfy the other two requirements described above. Typical values for $k$ and $n$ are 5-6 and 10-20, respectively. Given an average pixelscale of $0\farcs3$ pix$^{-1}$ and an average redshift of 0.05, this corresponds to emission profiles with a typical radius of 1.5-3 kpc. 

This last requirement discourages false convergence for data sets that are too small to truly constrain the geometry of a cone. We find that 18 galaxies meet these criteria. Therefore, the biconical outflows discussed in this work represent the best quality data (top 30\%) from the full sample of 61 outflow-dominated galaxies. We discuss the implications of our selection criteria in Section \ref{biases}, but in general we find that by requiring these three selection criteria, we select for galaxies that are nearby, more luminous, and more extended from the full sample of 61 outflows-dominated galaxies.

\subsection{Analytic Outflow Models}
\label{analyticModels}

\begin{figure*}
\centering
\includegraphics[scale=0.35]{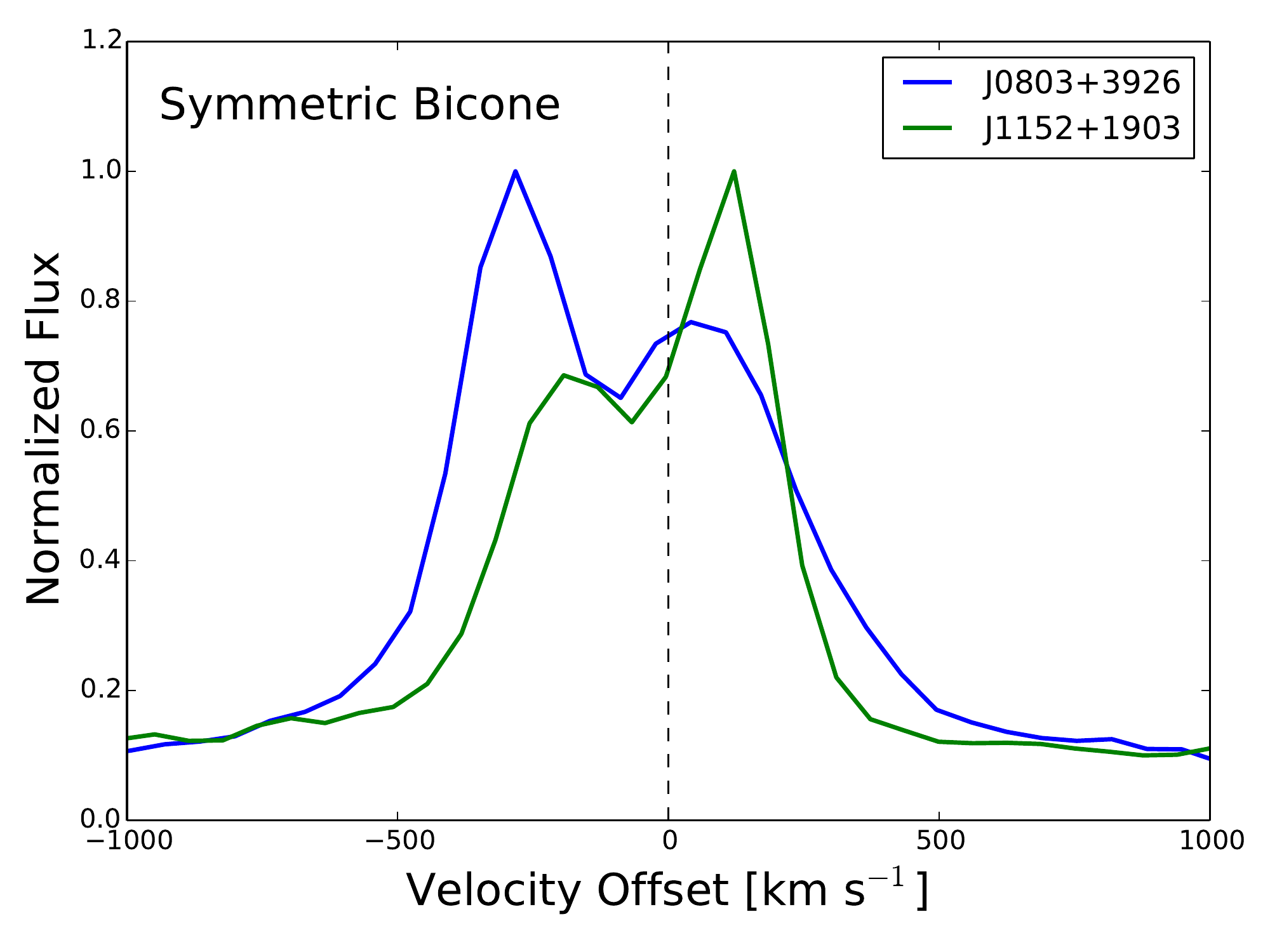}

\includegraphics[scale=0.35]{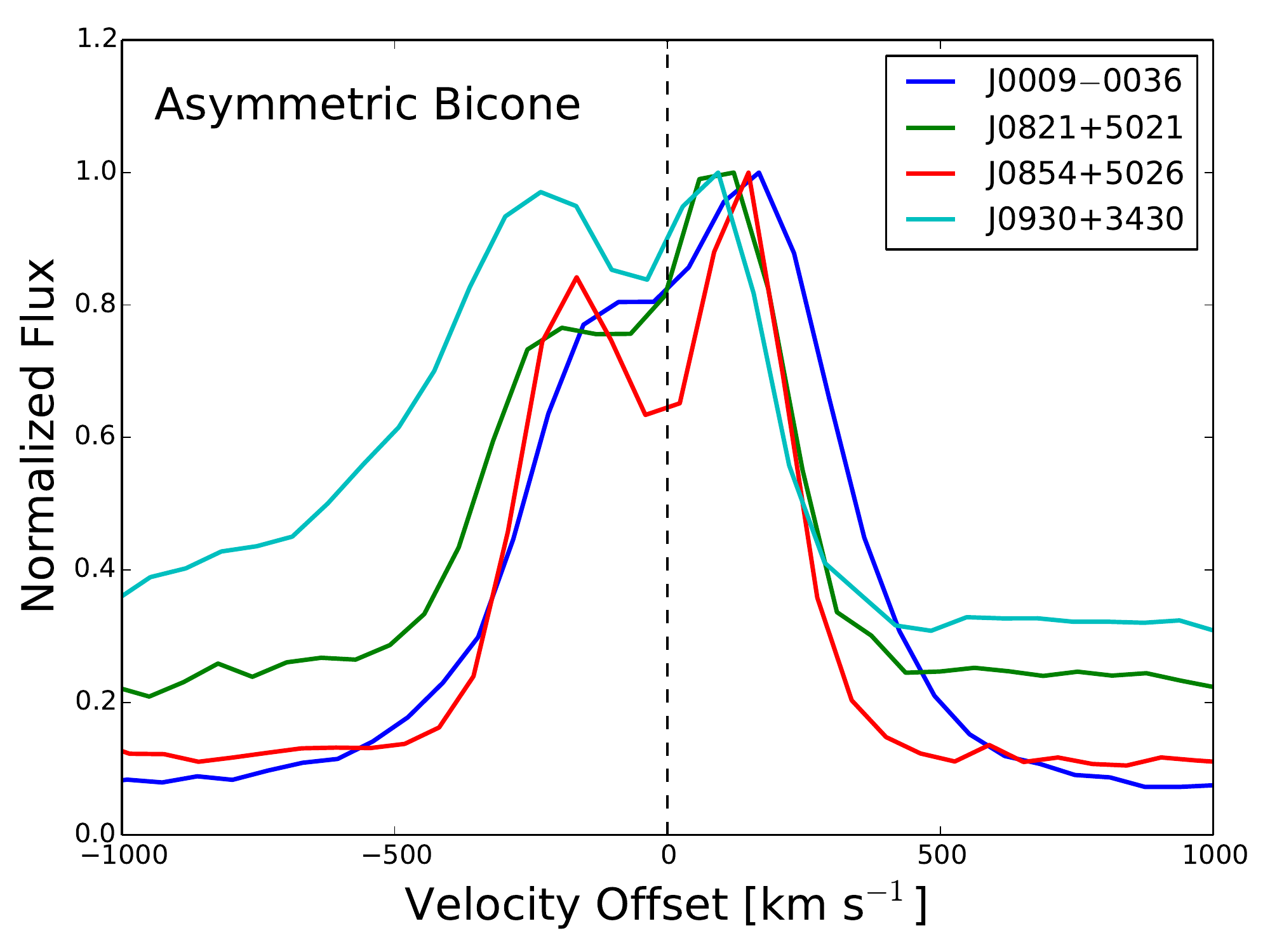}
\includegraphics[scale=0.35]{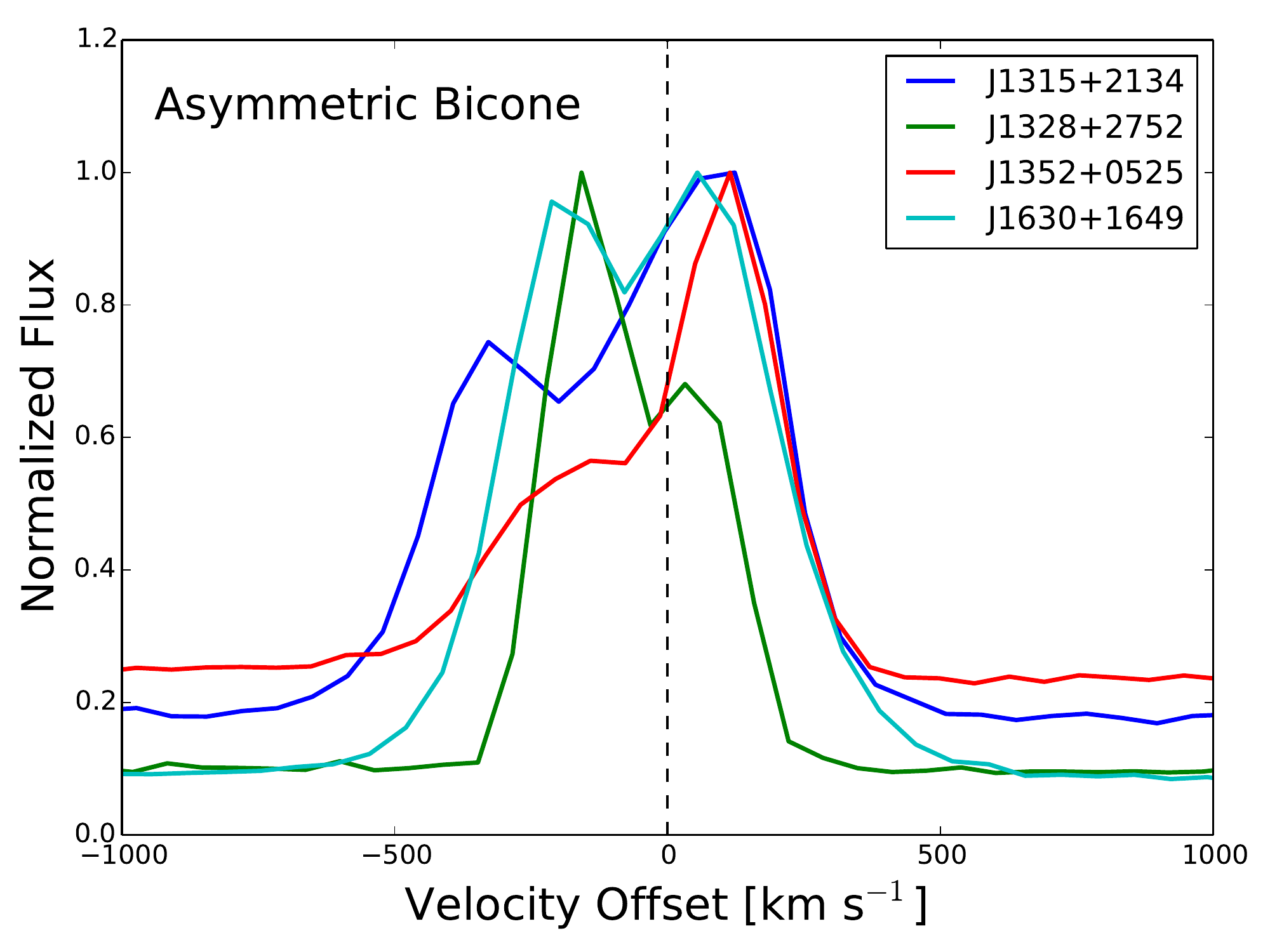}

\includegraphics[scale=0.35]{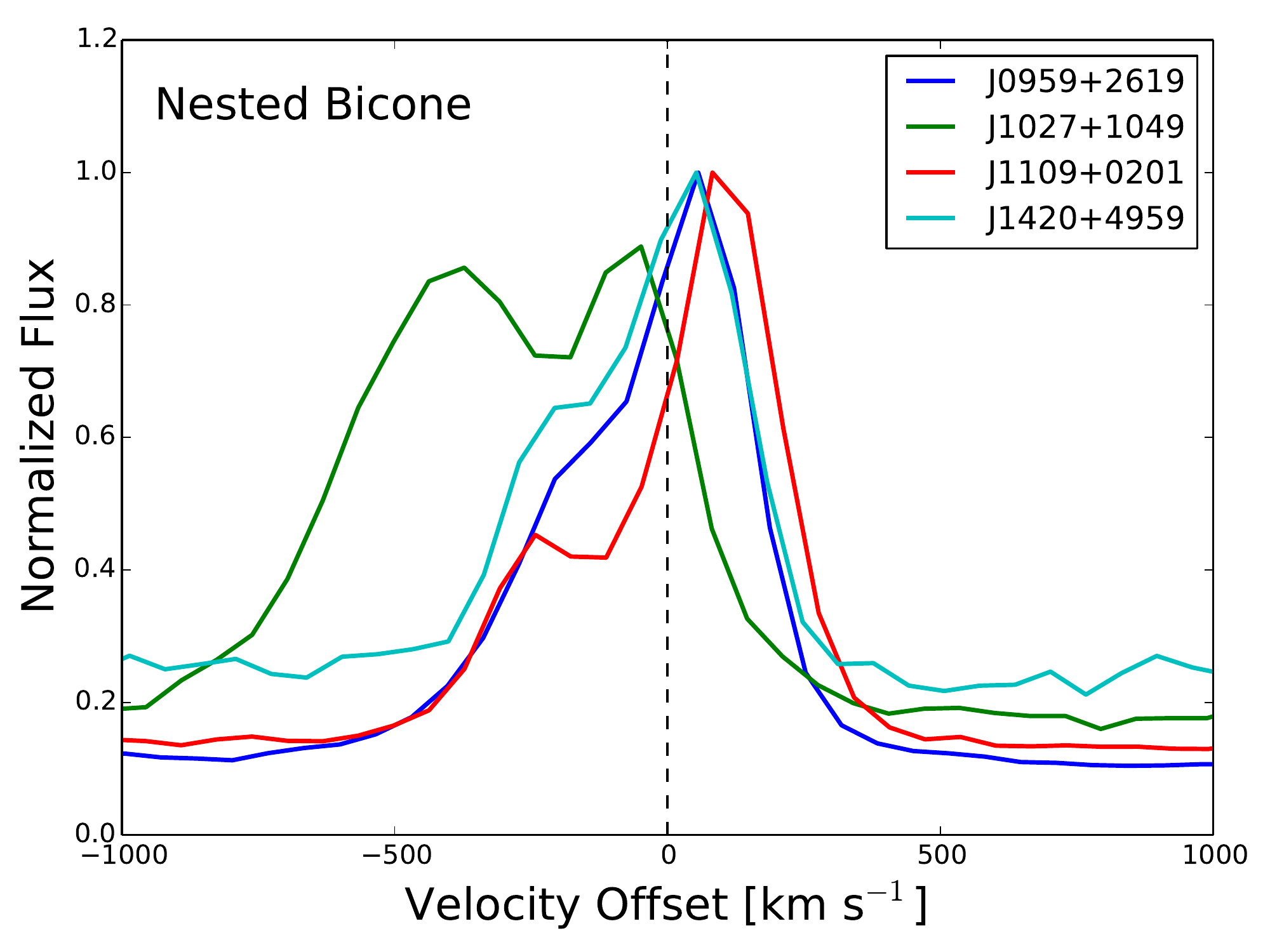}
\includegraphics[scale=0.35]{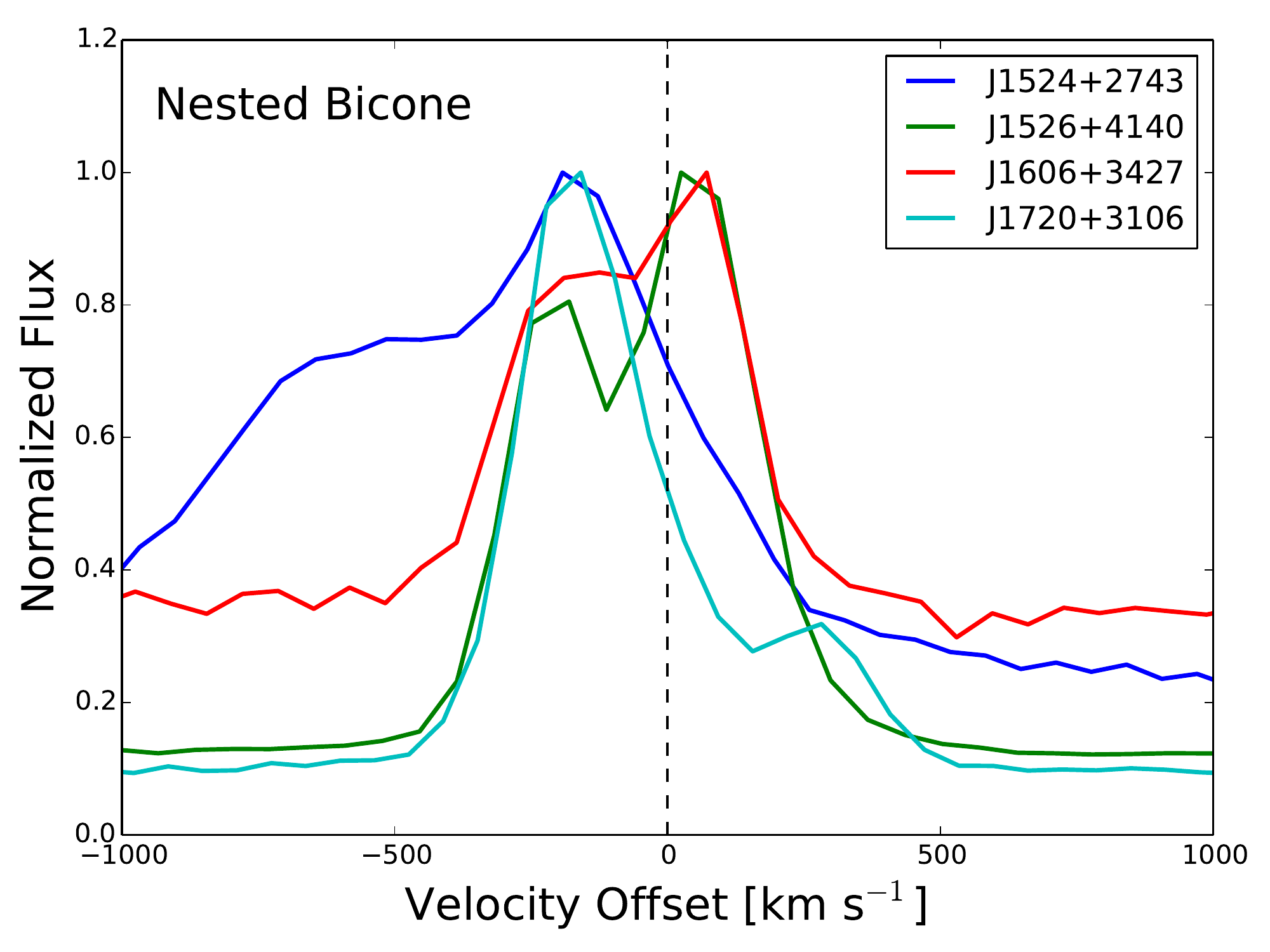}
\caption{The integrated [OIII]$\lambda$5007 SDSS profiles of the 18 galaxies modeled as biconical outflows in this work. The velocity offsets are measured relative to the host galaxy stars. Here we separate the galaxies according to the biconical outflow model that is the best fit. The symmetric bicone model with symmetric velocity centroids is the best fit model for the two galaxies in the top panel. The asymmetric bicone is the best model to describe the eight galaxies in the middle row. The nested bicone is the best model for the eight galaxies in the bottom row. Note that outflows possess distinct knots of emission that move at random velocities; therefore, profiles such as J0959+2619 (bottom left, nested bicone) may be best modeled as a nested bicone even though the redder velocity centroid is shifted redward of zero velocity in the integrated profiles. These types of profiles are better explored using spatially-resolved longslit spectra.}
\label{sdss ref}
\end{figure*}

We model the 18 galaxies selected in Section \ref{selection} as biconical outflows. A biconical model for AGN outflows is well motivated by observations and theory. From theory, a bicone model for the NLR of an outflow is expected from the unified model of AGNs; a thick torus provides the collimation necessary to produce a biconical outflow (\citealt{Antonucci1985}). Observationally, \citet{Schmitt1996} found biconical geometry in Seyfert 2 galaxies. \citet{Barbosa2014} confirm this biconical geometry for NGC 1068 with IFS. \citet{Crenshaw2000a} also find that the NLR kinematics of NGC 1068 observed with STIS longslit data are well-described by a radial biconical outflow. Other work followed to model biconical outflows with kinematic longslit and IFS data (e.g., \citealt{Crenshaw2000,Das2006,Das2007,Muller-Sanchez2011,Fischer2013,Crenshaw2015,Muller-Sanchez2016}). Although some groups model AGN outflows using a quasi-spherical geometry (e.g., \citealt{Liu2013b,Harrison2014}), our double-peaked velocity centroids are not consistent with a spherical geometry and instead suggest the inclined geometry of a bicone. Additionally, spherical shell models for bicones tend to overestimate the surface area of the outflow, and one of our main goals is to provide an accurate estimate of this parameter since it is used to estimate the kinetic energy of the outflow.

The bicone model builds off of the evacuated two-walled bicone with a front and rear wall where material begins to decelerate at a given turnover radius from \citet{Das2006}. In \citet{Das2006}, the two-walled structure is filled in between the walls. In this work, we evacuate the volume between the walls (Section \ref{discussmodels} presents the motivation for this fully evacuated bicone). The parameters for the bicone are inclination ($i$), position angle on the sky ($\mathrm{PA}_{\mathrm{bicone}}$), turnover radius for the velocity law of the outflow ($r_t$), the maximum velocity of the outflow at this turnover radius (V$_{\mathrm{max}}$), the inner half opening angle of the bicone ($\theta_{\mathrm{1,half}}$), and the outer half opening angle ($\theta_{\mathrm{2,half}}$). The height of the bicone is determined by the turnover radius ($r_t$), specifically $h=2r_t$. We measure the kinetic energy at $r_t$ to capture the bulk of the energy since the outflow decelerates beyond $r_t$.

The velocity law for the material along the wall of a biconical outflow has two phases. Exterior to $r_t$, the material decelerates linearly due to drag forces associated with the ISM (e.g., \citealt{Das2006,Das2007}). Interior to $r_t$, the bicone velocity law can be modeled with either an accelerating or constant velocity law. 

In nearby Seyferts with pc-scale resolution, observations have revealed the linear acceleration phase of AGN outflows with a turnover radius around 100 pc (e.g., \citealt{Crenshaw2000,Crenshaw2003,Das2006,Muller-Sanchez2011,Fischer2013,Crenshaw2015}). In contrast, other work with pc-scale resolution fails to find this acceleration (e.g., \citealt{Storchi-Bergmann2010}). Yet other work with larger resolution (kpc-scale) fails to find an accelerating phase and uses a constant velocity law to describe the outflowing gas (e.g., \citealt{Liu2013b,Harrison2014}).

\begin{figure*}
\includegraphics[scale=0.33]{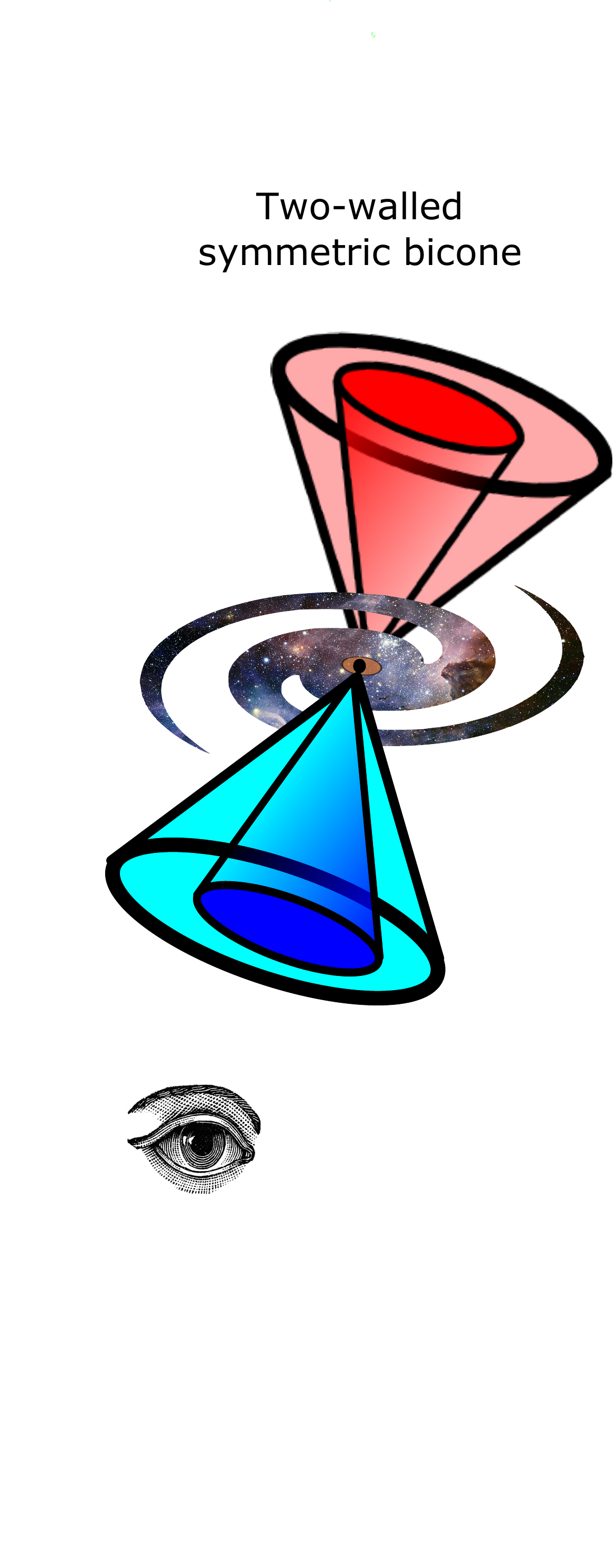}
\includegraphics[scale=0.33]{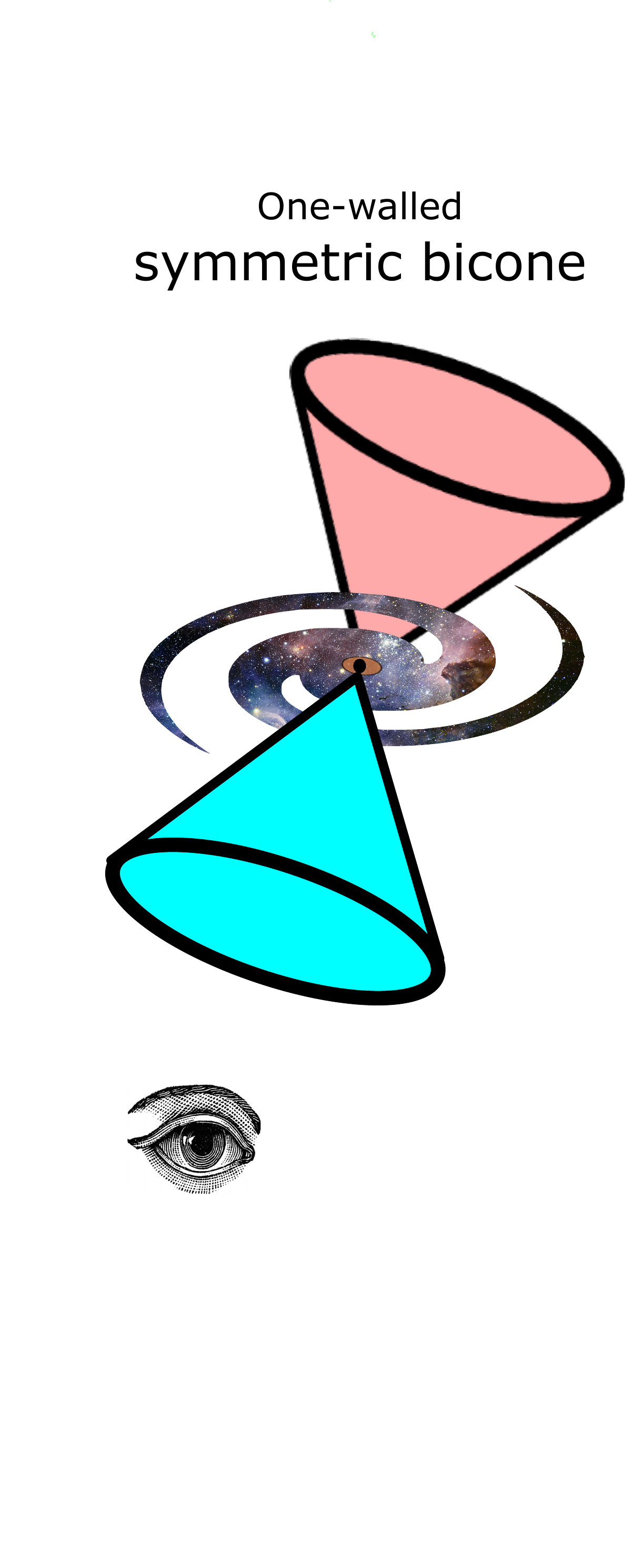}
\includegraphics[scale=0.33]{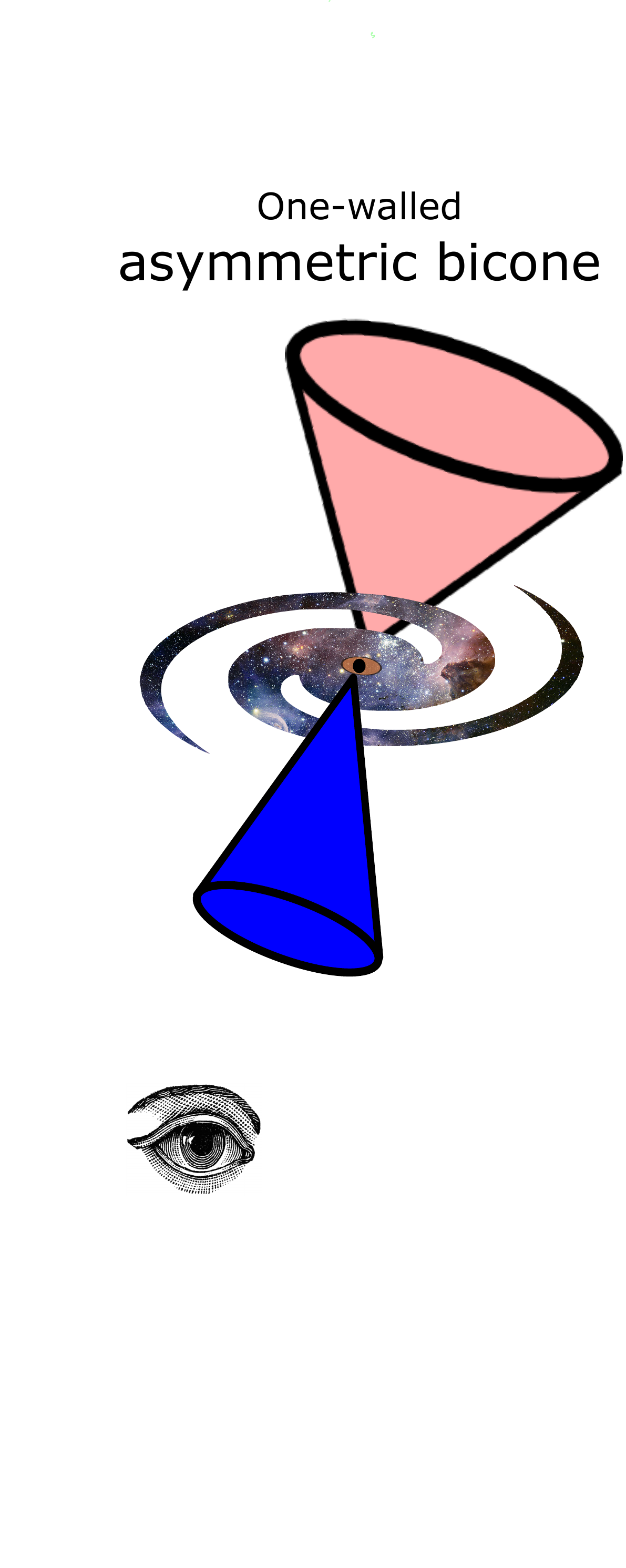}
\includegraphics[scale=0.33]{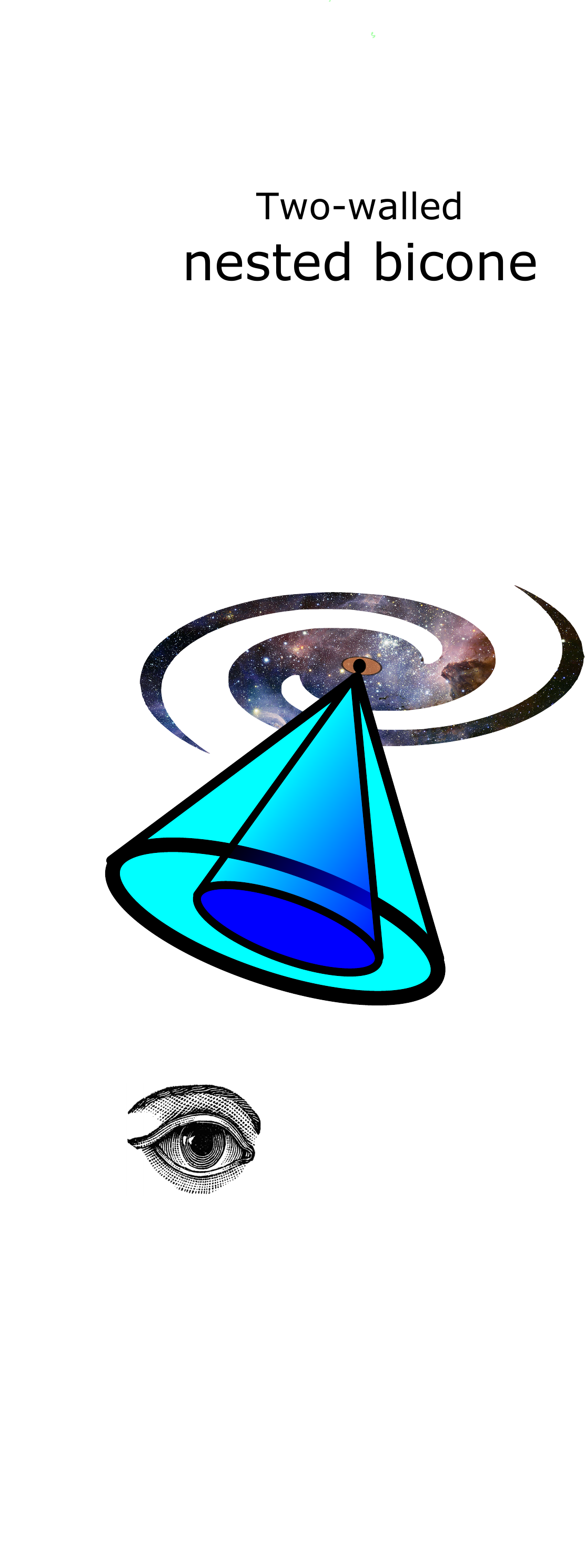}
\caption{The general structure and variations of the symmetric bicone. Only observed walls are shown, where darker colors indicate larger line of sight velocities. Here we make no distinction between illumination, obscuration, or selection effects leading to the absence of walls relative to the two-walled symmetric bicone. The general structure (left) has four total walls, all of which are aligned and described by two different opening angles. The symmetric bicone (second from left) has two cones of the same opening angle that touch apex to apex. The asymmetric bicone (second from right) has two cones of different opening angles, where the larger opening angle cone illustrates the receding emission component. We find that a larger opening angle receding cone is the case for all of the galaxies that are best modeled as an asymmetric bicone. The nested bicone (far right) has two cones, both of which are blueshifted.}
\label{anywall}
\end{figure*}

Since we are probing AGN outflows with kpc-scale resolution and also fail to observe an accelerating phase to the wind, we are well motivated to use a constant-velocity law to describe the interior regions of the outflow (prior to the deceleration phase). We note that we cannot distinguish between being unable to resolve this small-scale acceleration phase and the non-existence of this phase. We now turn to the theory of the accelerating mechanism for an outflow and how it fails to explain this acceleration phase. 

It is unclear how and where NLR winds are produced and accelerated to their observed maximum velocities of 100 to 1000 km s$^{-1}$ (\citealt{Crenshaw2005,Fischer2013,Fischer2014}). Possible proposed mechanisms for accelerating an AGN wind include thermal winds, magnetic fields, and radiative pressure (e.g., \citealt{Matzner1999,Lada1996,Das2007}). However, these mechanisms fail to explain the acceleration phase out to 100 pc observed by some work (e.g., \citealt{Everett2005, Das2007}). \citet{Everett2007} demonstrate that Parker winds (thermal winds) cannot reproduce the observed range of velocities observed. They also find that radiative pressure and magnetic fields can launch powerful winds, but these are small-scale winds that reach their terminal velocities $\sim$10 pc from the central source, the 100 pc distance as observed.

\citet{Everett2007} propose that since the various wind models fail to reproduce the observation, an already accelerated wind could be interacting with the surrounding medium. \citet{Storchi-Bergmann2010} investigate the accelerating velocity profile of the outflow in NGC 4151 observed by \citet{Das2006} and explain that since the velocity centroids probe the brightest emission, the observed accelerating structure could be produced by bright lower velocity gas entrained in the disk closer to the nucleus. Then, at greater radii in the outflow in NGC 4151, the outflowing component dominates the flux and produces the observed deceleration phase. Therefore, the observed acceleration could be attributed to a rotation-dominated component at small spatial scales.

\begin{figure*}

\includegraphics[scale=0.3,trim=0.2cm 5cm -5cm 4cm]{modify_outflow_fig_line_of_sight_symmetric_one_walled_bicone_built.pdf}
\includegraphics[scale=0.6]{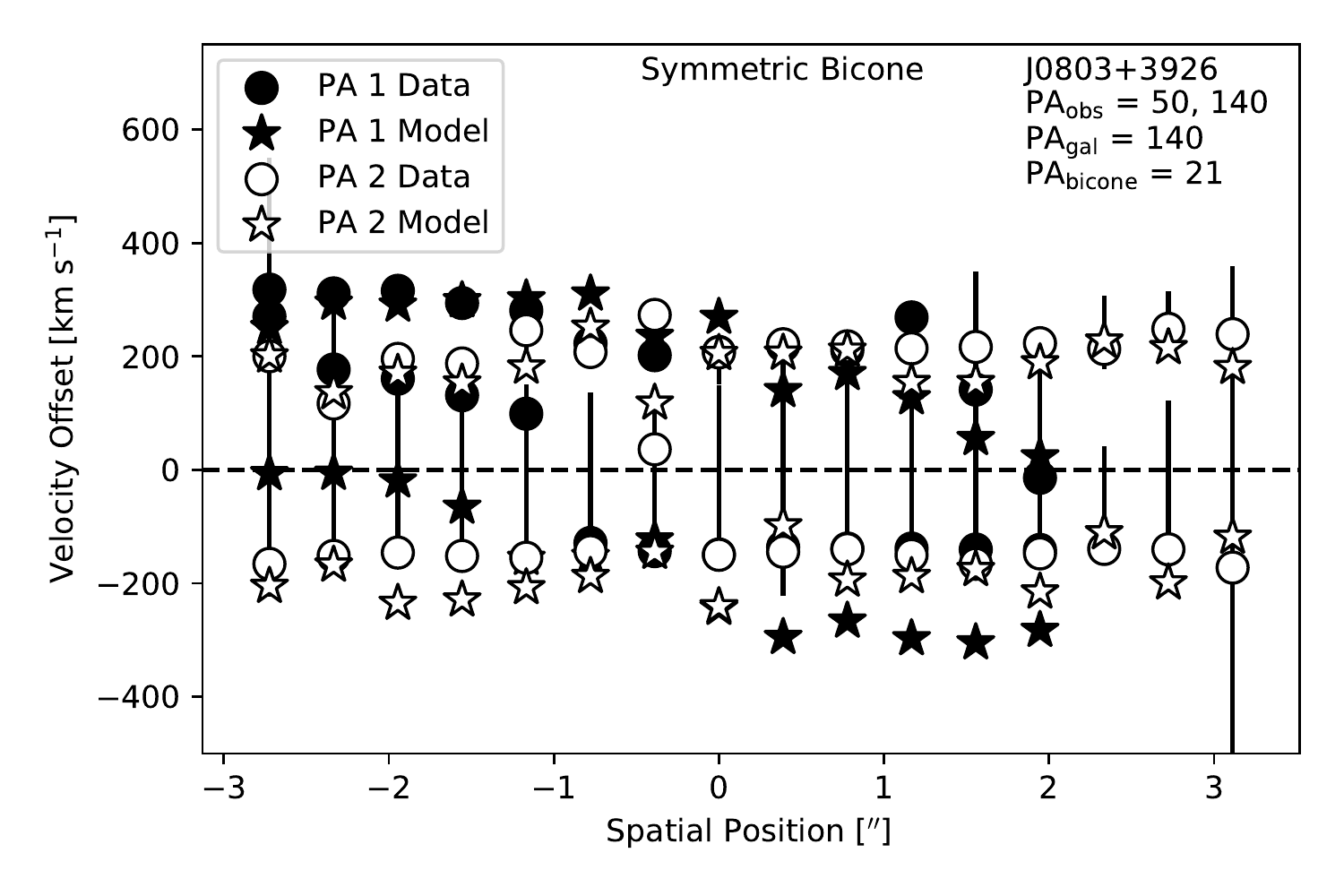}

\includegraphics[scale=0.3,trim=0.2cm 5cm -5cm 4cm]{modify_outflow_fig_line_of_sight_asymmetric_one_walled_bicone_solid.pdf}
\includegraphics[scale=0.6]{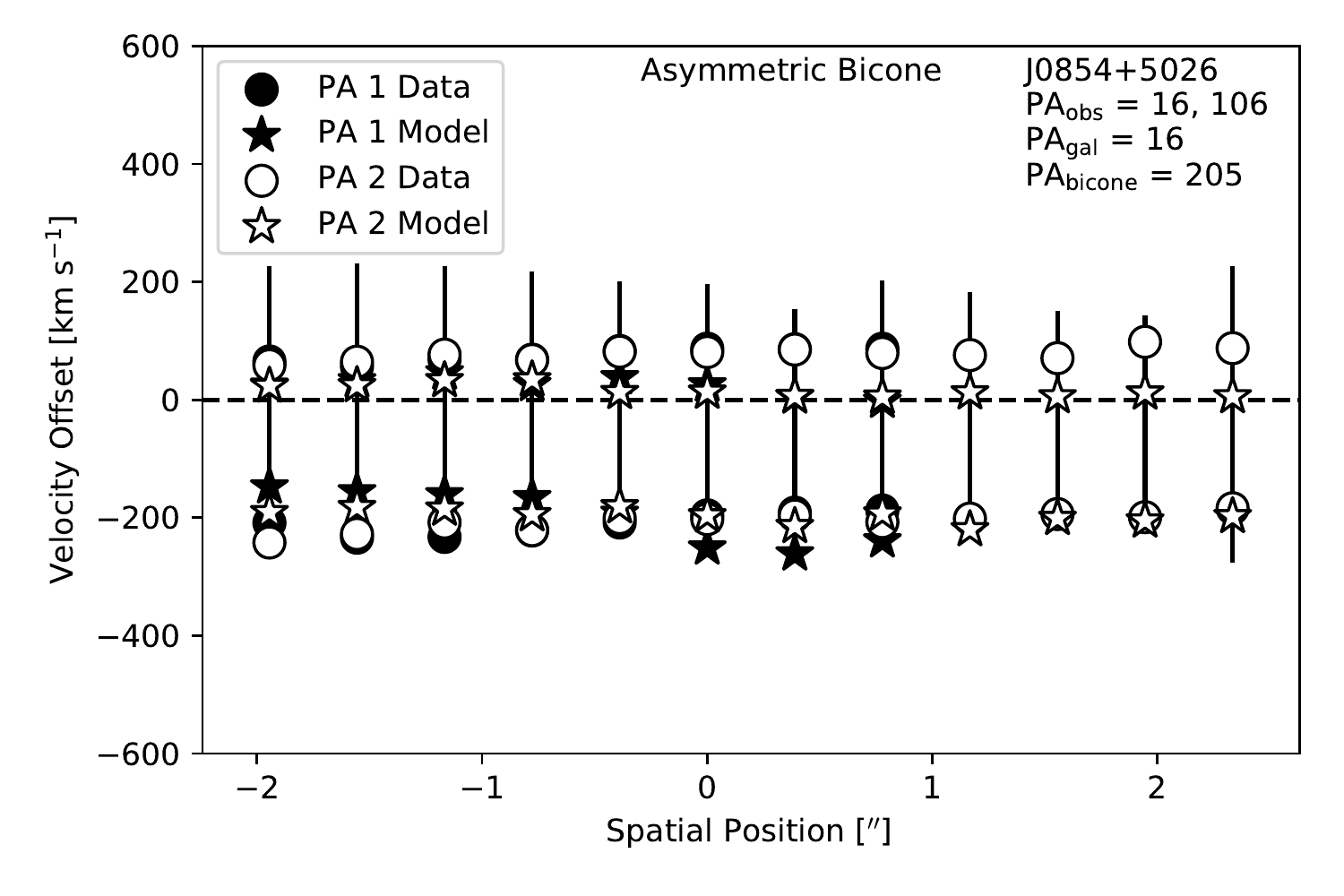}

\includegraphics[scale=0.3,trim=0.2cm 5cm -6cm 4cm]{modify_outflow_fig_line_of_sight_symmetric_two_walled_bicone_nested.pdf}
\includegraphics[scale=0.6]{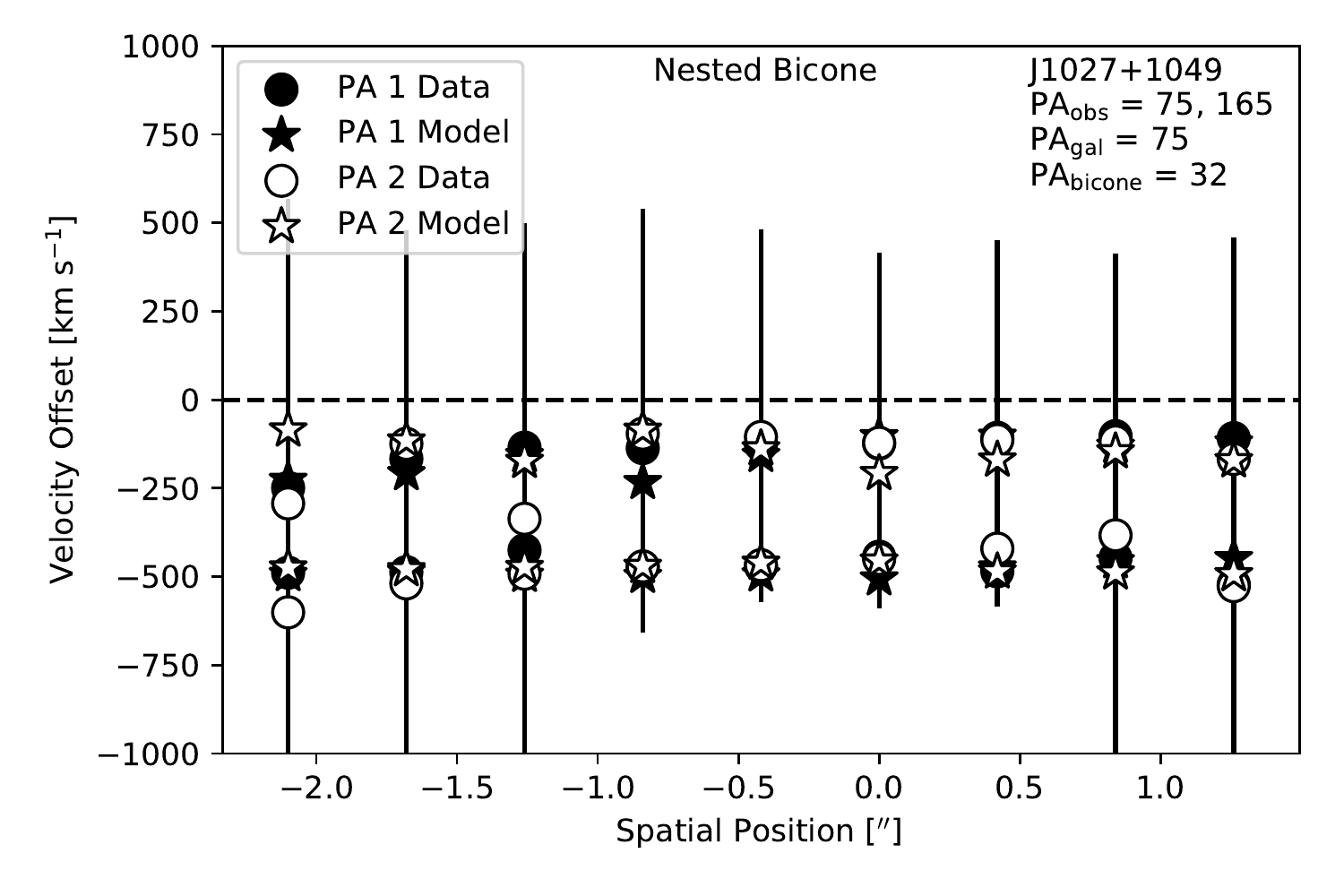}

\caption{Three variations of bicone models (left) with associated velocity profiles for example galaxies that are best fit by each model (right). We plot the velocity data in points and the best fit models with stars. The positive spatial direction corresponds to the NE direction of the slit. We plot a horizontal dashed line at zero velocity to fully demonstrate the origin of the names of the bicone models. For instance, the asymmetric bicone has velocity centroids in the spectrum that are asymmetric about zero velocity.}
\label{anywallobs}
\end{figure*}

We model the 18 AGN outflows using three biconical outflow models: the symmetric bicone, the asymmetric bicone, and the nested bicone. We are motivated to expand our models beyond that of the classical symmetric bicone due to asymmetries in the measured velocity centroids for the bluer and redder components of the two Gaussian fits to the spectra. This is apparent in both the SDSS integrated profiles (Figure \ref{sdss ref}) and the spatially resolved longslit profiles. We find that 15/18 outflows have integrated profiles that have a blueshifted overall velocity centroid. When we fit two Gaussians to the profile, 15/18 also have a mean velocity (averaging the two centroids) that is blueshifted. Motivated by the failure of a symmetric bicone model to uniformly describe the velocity centroids of all of the galaxies, we introduced the two additional analytic models. 

Each of the three bicone models can be described as a variation of a two-walled symmetric bicone; each model a total of two cone structures that are aligned with one another and produce double-peaked emission lines. We refer to a one-walled structure when there is one wall on either side of the galaxy (the one-walled symmetric bicone and one-walled asymmetric bicone). The two-walled bicones have two walls on one or both sides (the nested bicone and the general two-walled symmetric bicone, respectively). We use the same physical structure in all three model variations of the bicone, and the distinct models simply select different walls of a two-walled symmetric bicone structure. Figure \ref{anywall} shows diagrams of these three models. 

In addition to the simple case of the symmetric bicone, the asymmetric bicone and nested bicone are also motivated by observations. In a sample of SDSS Type 2 AGNs, \citet{Woo2016} find that inclination, dust obscuration, and velocity are the dominant parameters that control the modeled velocity and velocity dispersion profiles of AGNs. Their flux-weighted models demonstrate that dust can obscure the receding cone entirely when the inclination of the bicone is high, producing an observed profile similar to that of the nested bicone. 

Additionally, if the receding side of the bicone is larger than the dust plane, \citet{Woo2016} find that an asymmetric biconical structure can appear in the integrated spectrum because a wider opening angle receding cone is more favored in the flux-weighted profile. \citet{Storchi-Bergmann2010} observe very weak emission from the more inclined walls (relative to the disk) of the AGN outflow in NGC 4151. They explain that more gas is entrained at low angles to the galactic disk; an asymmetric bicone profile could originate from a wider opening angle receding cone that is close to the disk of the galaxy. \citet{Muller-Sanchez2012} find that an asymmetric bicone model is the best fit for the outflow in the Seyfert galaxy NGC 3081 with a wider opening angle receding cone.

The general symmetric two-walled bicone has a total of four cones (two receding and two approaching along the line of sight; Figure \ref{anywall}). For each of the three models, we select two cones from the symmetric two-walled bicone. Here we do not make a distinction between the existence and non-existence of various walls, obscuration effects, or illumination effects. We discuss obscuration effects, illumination effects, and lack of gas effects that may lead to these different models in Section \ref{discussmodels}. 

The symmetric one-walled bicone has two symmetric conical structures that can be described with the same opening angle on either side of the galaxy. A symmetric biconical model is constrained by five parameters ($i$, $\mathrm{PA}_{\mathrm{bicone}}$, $r_t$, $\theta_{1,\mathrm{half}}$, and V$_{\mathrm{max}}$) and produces two velocity centroids that are symmetric about zero velocity. The asymmetric bicone consists of two cones that are aligned. However, these cones can be described by two different opening angles. This model is constrained by six parameters ($i$, $\mathrm{PA}_{\mathrm{bicone}}$, $r_t$, $\theta_{1,\mathrm{half}}$, $\theta_{\mathrm{2,half}}$, and V$_{\mathrm{max}}$). Likewise, the nested bicone has six parameters ($i$, $\mathrm{PA}_{\mathrm{bicone}}$, $r_t$, $\theta_{1,\mathrm{half}}$, $\theta_{\mathrm{2,half}}$, and V$_{\mathrm{max}}$). It consists of two cones that are aligned but nested inside one another. The velocity centroids of the nested cone are both blueshifted. We show an example of a velocity profile from each of the three types of bicones in Figure \ref{anywallobs}.

Due to our $n>2k$ constraint, we are unable to assign 10 free parameters ($i$, $\mathrm{PA}_{\mathrm{bicone}}$, $r_t$, $\theta_{1,\mathrm{half}}$, V$_{\mathrm{max}}$, $i_2$, $\mathrm{PA}_{\mathrm{2,bicone}}$, $r_{2,t}$, $\theta_{2,\mathrm{half}}$, and V$_{\mathrm{2,max}}$) for the nested and asymmetric bicone models. However, past observations and asymmetric bicones in this work justify allowing just six free parameters in the nested and asymmetric bicone models, where we allow the opening angles of the two cones to vary.

First, either two different intrinsic velocities or two different opening angles could explain the asymmetric velocity centroids in the asymmetric bicone model. Observations of asymmetric bicones indicate that the redshifted wall often has a larger opening angle (e.g., \citealt{Muller-Sanchez2012,Woo2016}). Additionally, all the velocity centroids of galaxies in this work that can be described as asymmetric have a higher velocity blueshifted component and a lower velocity redshifted component. There is no physical motivation for 100\% of AGN outflows having an intrinsically lower velocity to the redshifted cone, so we favor the geometry explanation for this effect. 

We find in Section \ref{MCMC} that the choice of two distinct opening angles was merited, since the models converge on two different opening angles that are unique. Additionally, after completing the modeling, we assess the sensitivity of the parameters in Section \ref{MCMC} and find that the opening angles are the best-determined parameters. We find that the other parameters have large error bars and fitting two separate parameters for each model yields two values that are consistent with one another and therefore meaningless as separate parameters. For instance, when we fit two inclinations for each of the two sides of the asymmetric bicone, we find two values for each cone's inclination that are consistent with one another within errors. Therefore, our choice of the six free parameters for the asymmetric and nested bicones is justified by past observations of bicones as well as the limitations of our data and the sensitivity of our model (Section \ref{MCMC} and \ref{verify}).

\begin{figure*}
\centering
\includegraphics[scale=0.4]{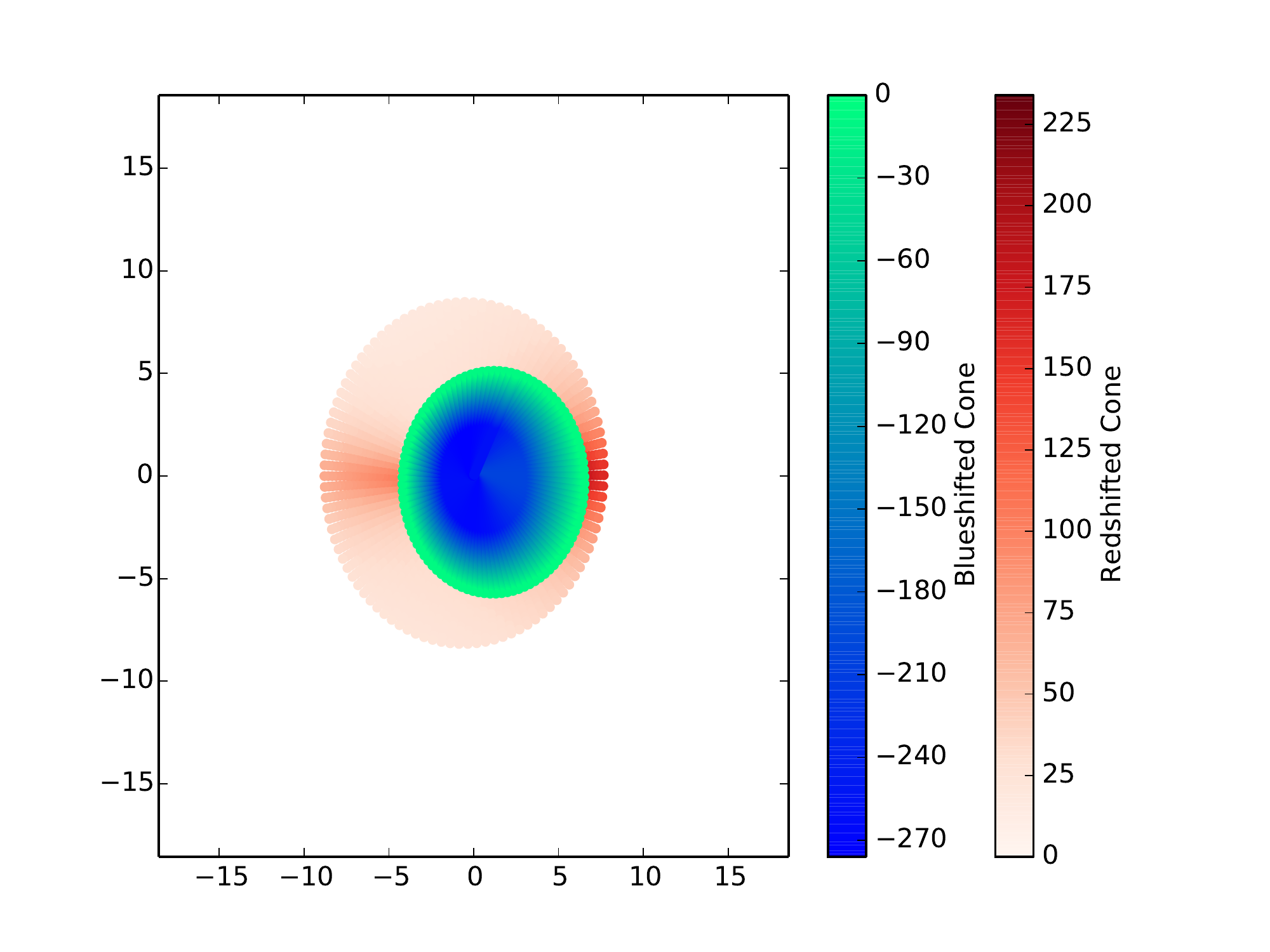}
\includegraphics[scale=0.4]{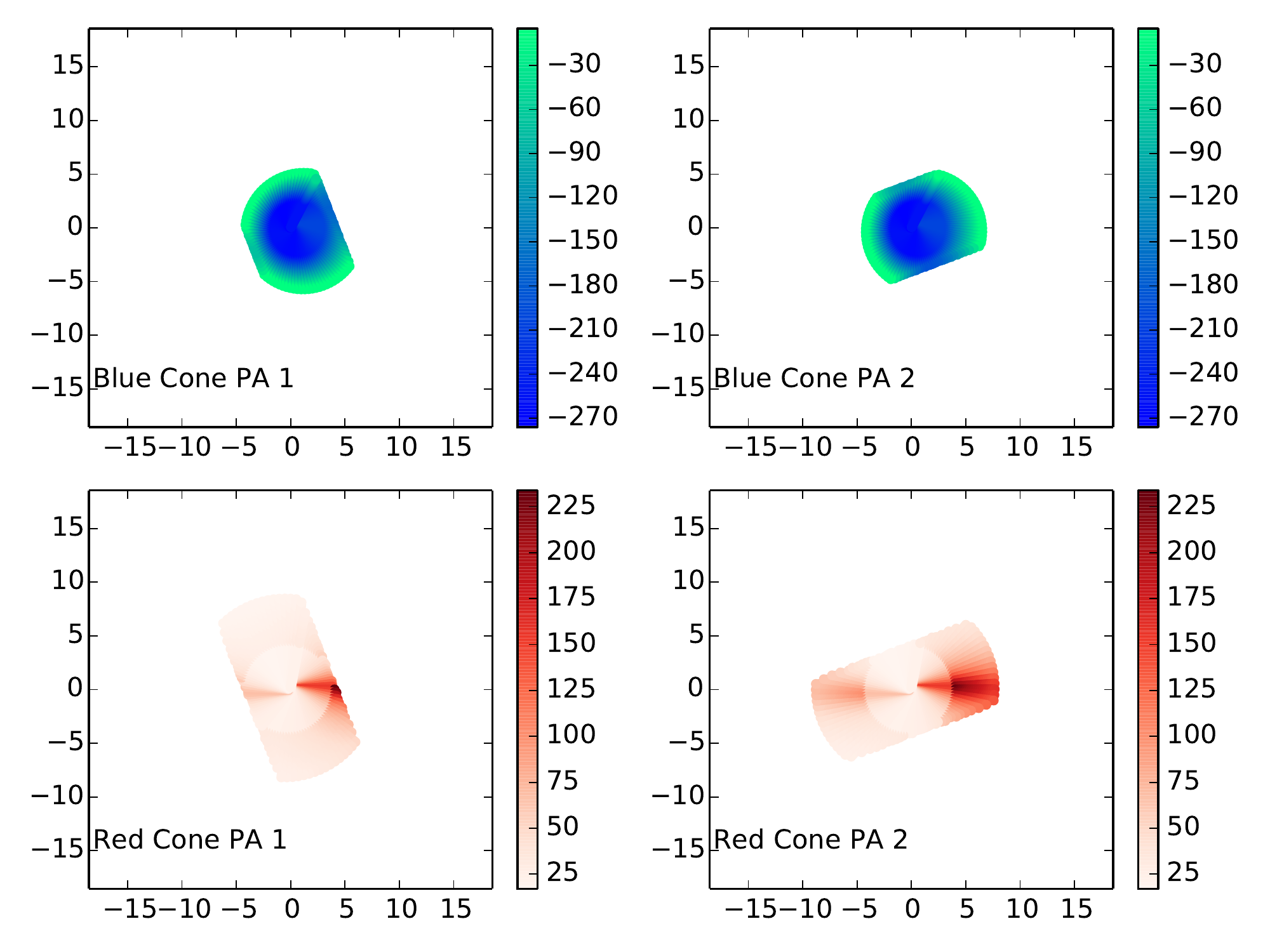}
\caption{The best fit asymmetric bicone model for the galaxy J0930+3430. Spacing on all axes corresponds to pixels (here, the pixelscale is $0\farcs389$pix$^{-1}$). Left: model of the bicone. The bicone is very inclined and the two colorbars show the line of sight velocities for both the redshifted (wider opening angle) cone and the blueshifted (smaller opening angle cone). Right: the cuts we make to the model (left) along the two observed PAs of the longslits on the sky (PA1 = 21$^{\circ}$, PA2 = 111$^{\circ}$). This takes pixelscale and slitwidth into account. We extract modeled velocity centroids for the blueshifted cone (top) and redshifted cone (bottom). Velocity color bars are in units of km s$^{-1}$.}
\label{bicone J0930}
\end{figure*}

We create a three dimensional model of each of these three cone structures. We project the velocities of our three dimensional models onto the plane of the sky and extract a line of sight velocity for all points along the observed PAs of the two slits (Figure \ref{bicone J0930}). The model accounts for the pixelscale and the slitwidth of each observation. We use a Markov Chain Monte Carlo process (described in Section \ref{MCMC}) to model all 18 galaxies using each of these three models. We find that 2/18 galaxies are best modeled as a symmetric bicone, 8/18 galaxies are best modeled as an asymmetric bicone, and 8/18 galaxies are best modeled as a nested bicone.

\subsection{Markov Chain Monte Carlo parameter estimation}
\label{MCMC}

For each of the three outflow models for each galaxy, we perform a multi-parameter Markov Chain Monte Carlo (MCMC) iterative modeling process to determine the best fit model and combination of parameters for the velocity data. In this section, we briefly outline the MCMC sampling, present the results of our modeling, discuss the methods we use to verify convergence, and discuss implications of the parameters using practical identifiability and sensitivity analysis.

We utilize the affine-invariant MCMC Ensemble sampler from \citet{Goodman2010}, conveniently packaged in the Python code {\tt emcee} (\citealt{Foreman-Mackey2013}). The MCMC method takes advantage of parallel processing to sample the posterior probability density function (PDF) for a multi-parameter space efficiently on multiple cores. Computing the PDF of a biconical outflow is computationally expensive, so we run our parallel sampling for the bicone models on the University of Colorado supercomputer JANUS. An affine-invariant sampler also performs well under all linear transformations; the sampler is insensitive to covariance between parameters. Parameters demonstrate covariance within our parameter set, so this is an advantage.

The MCMC technique minimizes the log-likelihood of the parameter space. Gaussian log-likelihood is defined as:
$$\mathrm{log \: \mathcal{L}}  = -0.5\bigg(\mathlarger{\mathlarger{\sum}} \mathrm{ln}(2 \pi \sigma_i^2)+\mathlarger{\mathlarger{\sum}} \frac{(y_i-x_i)^2}{\sigma_i^2}\bigg)$$
where $y_i$ is the data, $x_i$ is the model, and $\sigma_i$ is the associated error with each data point. Here, $\sigma_i$ includes both error on the measurement of the Gaussian centroid of each velocity component and the error inherent to the instrumental dispersion (e.g., for the instruments used in this sample, the typical error is 2.5 km s$^{-1}$). The Gaussian log $\mathcal{L}$ is related to a $\chi^2$ value, since in the Gaussian case with a normal error assumption, they are directly proportional. To compare goodness of fit between different models, we use a reduced chi-square defined as:
$$\chi^2_{\nu} = \chi^2/(n-k-1)$$
where $n$ is the number of data points and $k$ is the number of free parameters.

Due to the simplistic nature of the outflow model, the $\chi^2_{\nu}$ values are relatively high. Although we could add more parameters to make a more realistic model, the number of data points do not justify it. These models assume a continuous distribution of NLR emission along the walls of the outflow and are too simplistic for a structure with many complicated discrete knots of emission, but model the overall morphology well. Therefore, we do not expect $\chi^2_{\nu}$ values to be $\sim 1$ for these models, and we are unable to assess the absolute ``goodness of fit'' of a given model in isolation. Instead, we use the $\chi^2_{\nu}$ values only to compare between different models for an individual galaxy; the $\chi^2_{\nu}$ values are not intended to be used to compare the outflows of two different galaxies.

We find that 2/18 (11.1\%) galaxies are best modeled as a symmetric bicone, 8/18 (44.4\%) galaxies are best modeled as an asymmetric bicone, and 8/18 (44.4\%) galaxies are best modeled as a nested bicone. The $\chi^2_{\nu}$ values are reported in Table \ref{RSStable} and the parameters for the best fit model along with 1$\sigma$ error bars are reported in Table \ref{parametersbicone}, \ref{parametersassy}, and \ref{parametersnest} for the galaxies that are best fit by a symmetric bicone, asymmetric bicone, and nested bicone, respectively. We also include the mean modulus of the residuals between the observed velocities and the modeled velocities to quantify the goodness of fit. We find that this quantity is comparable to the uncertainty of the observed velocities, which indicates that the models are a good fit.

\begin{deluxetable*}{ccccccc}

\tabletypesize{\scriptsize}
\tablewidth{0pt}
\tablecolumns{5}
\tablecaption{Outflow Model Statistics}
\tablehead{
\colhead{SDSS ID} & 
\colhead{Symmetric Bicone $\chi^2_{\nu}$} & 
\colhead{Asymmetric Bicone $\chi^2_{\nu}$} &
\colhead{Nested Bicone $\chi^2_{\nu}$} &
\colhead{Best Fit Model} }

\startdata

J0009$-$0036 & 9146.3 & 243.0 & $>$10000  & Asymmetric  \\
J0803+3926& 5.1 & 18.1 & 100.3 & Symmetric\\
J0821+5021&169.1 & 68.2 & $>$10000& Asymmetric \\
J0854+5026& $>$10000 & 183.3 & 226.1 & Asymmetric \\
J0930+3430&2874.3 & 4.6 & 50.0 & Asymmetric  \\
J0959+2619& 73.2 &$>$10000  & 5.0 & Nested\\
J1027+1049&73.4 & 27.6 &4.7 &Nested   \\
J1109+0201&150.5 & 65.8 & 63.8 & Nested  \\
J1152+1903&122.8 & 139.7 & 728.9 & Symmetric \\
J1315+2134&559.5 & 14.7 & 16.0 & Asymmetric\\
J1328+2752& 10.8 & 2.3 & $>$10000& Asymmetric  \\
J1352+0525&131.7 & 87.1 & 176.3 & Asymmetric  \\
J1420+4959&32.12 & 11.87 & 1.03 & Nested \\
J1524+2743& 390.3 & 168.8 & 4.1 & Nested \\
J1526+4140 &1.8 & 0.5 & 0.3 & Nested  \\
J1606+3427& 386.2 & 21.1 & 17.0 & Nested \\
J1630+1649& 2383.9 & 1086.1 & 1591.4 & Asymmetric  \\
J1720+3106& 1240.8 & 547.4 & 235.8 & Nested

\enddata

\tablecomments{We select the best fit biconical outflow model for each galaxy by selecting the model with the lowest $\chi^2_{\nu}$ value.} 
\label{RSStable} 
\end{deluxetable*}

\begin{deluxetable*}{ccccccc}

\tabletypesize{\scriptsize}
\tablewidth{0pt}
\tablecolumns{6}
\tablecaption{Symmetric Bicone Model Parameters}
\tablehead{
\colhead{SDSS ID} & 
\colhead{$i$} & 
\colhead{$\mathrm{PA}_{\mathrm{bicone}}$} &
\colhead{$r_{t}$} &
\colhead{$\theta_{1,\mathrm{half}}$} &
\colhead{V$_{\mathrm{max}}$} &
\colhead{$<|$V$_{\mathrm{obs}}$-V$_{\mathrm{mod}}|>$}\\
 &[$^{\circ}$]  &[$^{\circ}$E of N] &[kpc] &[$^{\circ}$] & [km s$^{-1}$]  & [km s$^{-1}$] }

\startdata

J0803+3926& 
$ 40 \substack{+ 18 \\- 29 }$ & $ 20 \substack{+ 20 \\- 10 }$ & $ 9 \substack{+ 5 \\- 3 }$ & $ 53 \substack{+ 9 \\- 9 }$ & $ 430 \substack{+ 110 \\- 70 }$ & 73 \\ 
J1152+1903&
$ 42 \substack{+ 6 \\- 44 }$ & $ 40 \substack{+ 20 \\- 40 }$ & $ 6 \substack{+ 3 \\- 2 }$ & $ 60 \substack{+ 2 \\- 7 }$ & $ 370 \substack{+ 90 \\- 50 }$ &44  

\enddata

\tablecomments{Column 1: galaxy name. Column 2: outflow inclination. Column 3: position angle of the bicone axis on the sky. Column 4: turnover radius in kpc. Column 5: half opening angle. Column 6: maximum velocity. Column 7: the mean modulus of the velocity residuals.} 
\label{parametersbicone} 
\end{deluxetable*}

\begin{deluxetable*}{cccccccc}

\tabletypesize{\scriptsize}
\tablewidth{0pt}
\tablecolumns{7}
\tablecaption{Asymmetric Bicone Model Parameters}
\tablehead{
\colhead{SDSS ID} & 
\colhead{$i$} & 
\colhead{$\mathrm{PA}_{\mathrm{bicone}}$} &
\colhead{$r_{t}$} &
\colhead{$\theta_{1,\mathrm{half}}$} &
\colhead{$\theta_{2,\mathrm{half}}$} &
\colhead{V$_{\mathrm{max}}$} & 
\colhead{$<|$V$_{\mathrm{obs}}$-V$_{\mathrm{mod}}|>$}\\
&[$^{\circ}$]  &[$^{\circ}$E of N] &[kpc] &[$^{\circ}$] &[$^{\circ}$] & [km s$^{-1}$] &  [km s$^{-1}$]  }

\startdata

 J0009$-$0036&
$ 56 \substack{+ 8 \\- 6 }$ & $ 79 \substack{+ 16 \\- 17 }$ & $ 5 \substack{+ 3 \\- 1 }$ & $ 60 \substack{+ 3 \\- 4 }$ & $ 77 \substack{+ 3 \\- 2 }$ & $ 320 \substack{+ 60 \\- 80 }$ & 139 \\ 

J0821+5021&
$ 51 \substack{+ 34 \\- 8 }$ & $ 6 \substack{+ 19 \\- 7 }$ & $ 9 \substack{+ 5 \\- 3 }$ & $ 50 \substack{+ 7 \\- 11 }$ & $ 72 \substack{+ 4 \\- 6 }$ & $ 360 \substack{+ 60 \\- 100 }$ & 175\\ 
J0854+5026&
$ 60 \substack{+ 16 \\- 8 }$ & $ 205 \substack{+ 14 \\- 169 }$ & $ 7 \substack{+ 2 \\- 2 }$ & $ 42 \substack{+ 4 \\- 7 }$ & $ 75 \substack{+ 6 \\- 3 }$ & $ 290 \substack{+ 40 \\- 50 }$ & 46 \\ 
J0930+3430&
$ 80 \substack{+ 6 \\- 9 }$ & $ 75 \substack{+ 55 \\- 49 }$ & $ 10 \substack{+ 4 \\- 6 }$ & $ 37 \substack{+ 12 \\- 10 }$ & $ 67 \substack{+ 3 \\- 2 }$ & $ 290 \substack{+ 130 \\- 30 }$ &27 \\ 

J1315+2134&
$ 58 \substack{+ 7 \\- 12 }$ & $ 39 \substack{+ 14 \\- 12 }$ & $ 6 \substack{+ 2 \\- 2 }$ & $ 48 \substack{+ 4 \\- 4 }$ & $ 78 \substack{+ 3 \\- 2 }$ & $ 600 \substack{+ 130 \\- 50 }$ & 68 \\ 
J1328+2752 &
$ 78 \substack{+ 9 \\- 13 }$ & $ 52 \substack{+ 81 \\- 39 }$ & $ 6 \substack{+ 4 \\- 3 }$ & $ 48 \substack{+ 15 \\- 19 }$ & $ 81 \substack{+ 3 \\- 5 }$ & $ 230 \substack{+ 260 \\- 80 }$ & 106\\ 
J1352+0525 &
$ 43 \substack{+ 14 \\- 7 }$ & $ 43 \substack{+ 42 \\- 9 }$ & $ 6 \substack{+ 2 \\- 1 }$ & $ 61 \substack{+ 3 \\- 6 }$ & $ 78 \substack{+ 2 \\- 3 }$ & $ 440 \substack{+ 60 \\- 80 }$ & 78 \\ 

J1630+1649&
$ 83 \substack{+ 5 \\- 25 }$ & $ 38 \substack{+ 29 \\- 34 }$ & $ 6 \substack{+ 3 \\- 1 }$ & $ 40 \substack{+ 16 \\- 7 }$ & $ 82 \substack{+ 2 \\- 1 }$ & $ 290 \substack{+ 90 \\- 30 }$ & 92

\enddata

\tablecomments{Column 1: galaxy name. Column 2: outflow inclination. Column 3: position angle of the bicone axis on the sky. Column 4: turnover radius in kpc. Column 5: inner half opening angle. Column 6: outer half opening angle. Column 7: maximum velocity. Column 8: the mean modulus of the velocity residuals.} 
\label{parametersassy} 
\end{deluxetable*}

\begin{deluxetable*}{cccccccc}

\tabletypesize{\scriptsize}
\tablewidth{0pt}
\tablecolumns{7}
\tablecaption{Nested Bicone Model Parameters}
\tablehead{
\colhead{SDSS ID} & 
\colhead{$i$} & 
\colhead{$\mathrm{PA}_{\mathrm{bicone}}$} &
\colhead{$r_{t}$} &
\colhead{$\theta_{1,\mathrm{half}}$} &
\colhead{$\theta_{2,\mathrm{half}}$} &
\colhead{V$_{\mathrm{max}}$} & 
\colhead{$<|$V$_{\mathrm{obs}}$-V$_{\mathrm{mod}}|>$}\\
&[$^{\circ}$]  &[$^{\circ}$E of N] &[kpc] &[$^{\circ}$] &[$^{\circ}$] & [km s$^{-1}$] &[km s$^{-1}$]  }

\startdata

J0959+2619&
$ 78 \substack{+ 9 \\- 20 }$ & $ 13 \substack{+ 31 \\- 9 }$ & $ 7 \substack{+ 4 \\- 3 }$ & $ 49 \substack{+ 10 \\- 14 }$ & $ 71 \substack{+ 5 \\- 4 }$ & $ 280 \substack{+ 180 \\- 100 }$ & 46\\ 
J1027+1049&
$ 77 \substack{+ 9 \\- 14 }$ & $ 32 \substack{+ 81 \\- 25 }$ & $ 12 \substack{+ 6 \\- 6 }$ & $ 33 \substack{+ 9 \\- 9 }$ & $ 61 \substack{+ 5 \\- 4 }$ & $ 540 \substack{+ 100 \\- 80 }$ & 43\\ 
J1109+0201 &
$ 48 \substack{+ 5 \\- 5 }$ & $ 59 \substack{+ 16 \\- 16 }$ & $ 10 \substack{+ 4 \\- 2 }$ & $ 55 \substack{+ 5 \\- 4 }$ & $ 73 \substack{+ 3 \\- 3 }$ & $ 390 \substack{+ 50 \\- 40 }$ & 86 \\ 
J1420+4959&
$ 79 \substack{+ 7 \\- 19 }$ & $ 174 \substack{+ 98 \\- 126 }$ & $ 5 \substack{+ 4 \\- 3 }$ & $ 41 \substack{+ 10 \\- 16 }$ & $ 70 \substack{+ 6 \\- 5 }$ & $ 390 \substack{+ 170 \\- 110 }$ & 153 \\ 
J1524+2743&
$ 83 \substack{+ 5 \\- 10 }$ & $ 100 \substack{+ 29 \\- 57 }$ & $ 5 \substack{+ 5 \\- 2 }$ & $ 35 \substack{+ 7 \\- 10 }$ & $ 62 \substack{+ 2 \\- 2 }$ & $ 720 \substack{+ 100 \\- 90 }$ & 127\\ 
J1526+4140 &
$ 74 \substack{+ 12 \\- 15 }$ & $ 59 \substack{+ 66 \\- 45 }$ & $ 10 \substack{+ 7 \\- 5 }$ & $ 44 \substack{+ 15 \\- 14 }$ & $ 62 \substack{+ 11 \\- 16 }$ & $ 410 \substack{+ 270 \\- 120 }$  & 88 \\ 
J1606+3427 &
$ 81 \substack{+ 5 \\- 9 }$ & $ 114 \substack{+ 48 \\- 76 }$ & $ 9 \substack{+ 3 \\- 3 }$ & $ 37 \substack{+ 8 \\- 6 }$ & $ 74 \substack{+ 3 \\- 3 }$ & $ 370 \substack{+ 70 \\- 40 }$& 71 \\ 

J1720+3106 &
$ 84 \substack{+ 6 \\- 8 }$ & $ 34 \substack{+ 28 \\- 24 }$ & $ 14 \substack{+ 4 \\- 3 }$ & $ 18 \substack{+ 11 \\- 5 }$ & $ 30 \substack{+ 9 \\- 5 }$ & $ 300 \substack{+ 20 \\- 10 }$ & 152 

\enddata

\tablecomments{Column 1: galaxy name. Column 2: outflow inclination. Column 3: position angle of the bicone axis on the sky. Column 4: turnover radius in kpc. Column 5: inner half opening angle. Column 6: outer half opening angle. Column 7: maximum velocity. Column 8: the mean modulus of the velocity residuals. } 
\label{parametersnest} 
\end{deluxetable*}

We confirm convergence of the MCMC fit to the global minimum by assessing the acceptance fraction of the walkers and the autocorrelation function. We use the acceptance fraction of the walkers as one method to assess if the walkers have fallen into a local minimum. We ensure that the walkers are in the range of acceptance fraction (0.2-0.5) suggested by \citet{Foreman-Mackey2013}. If the acceptance fraction is less than 0.2, this implies that the walkers have fallen into a local minimum and are unable to walk their way out, and instead reject every step. An acceptance fraction that is too high would imply that the PDF is featureless, and walkers are accepting random steps across the entirety of parameter space. Our average acceptance fraction is in the 0.2-0.5 range due to our careful selection of data that are high enough in quality and quantity as discussed in Section \ref{selection}. 

Another method for quality assurance of the fit is to determine the ``burn-in'' period and ensure that the MCMC process iterates for at least this long. The code {\tt emcee} provides an estimation of the autocorrelation time, which is defined as the time lag that drives the value of the autocovariance function of a time series to zero. When the autocovariance is zero, the chain has fully sampled the probability space. For our parameters, the typical autocorrelation time is 50-60 steps, indicating that it takes 50-60 steps for the walkers to converge upon the true value. We run all chains for 200 steps to ensure that the runs extend for at least twice the maximum autocorrelation time. On average, our chains run for four autocorrelation times.

\begin{figure*}
\centering
\includegraphics[scale=0.8, angle=0,trim=0cm 0cm -0cm 0cm]{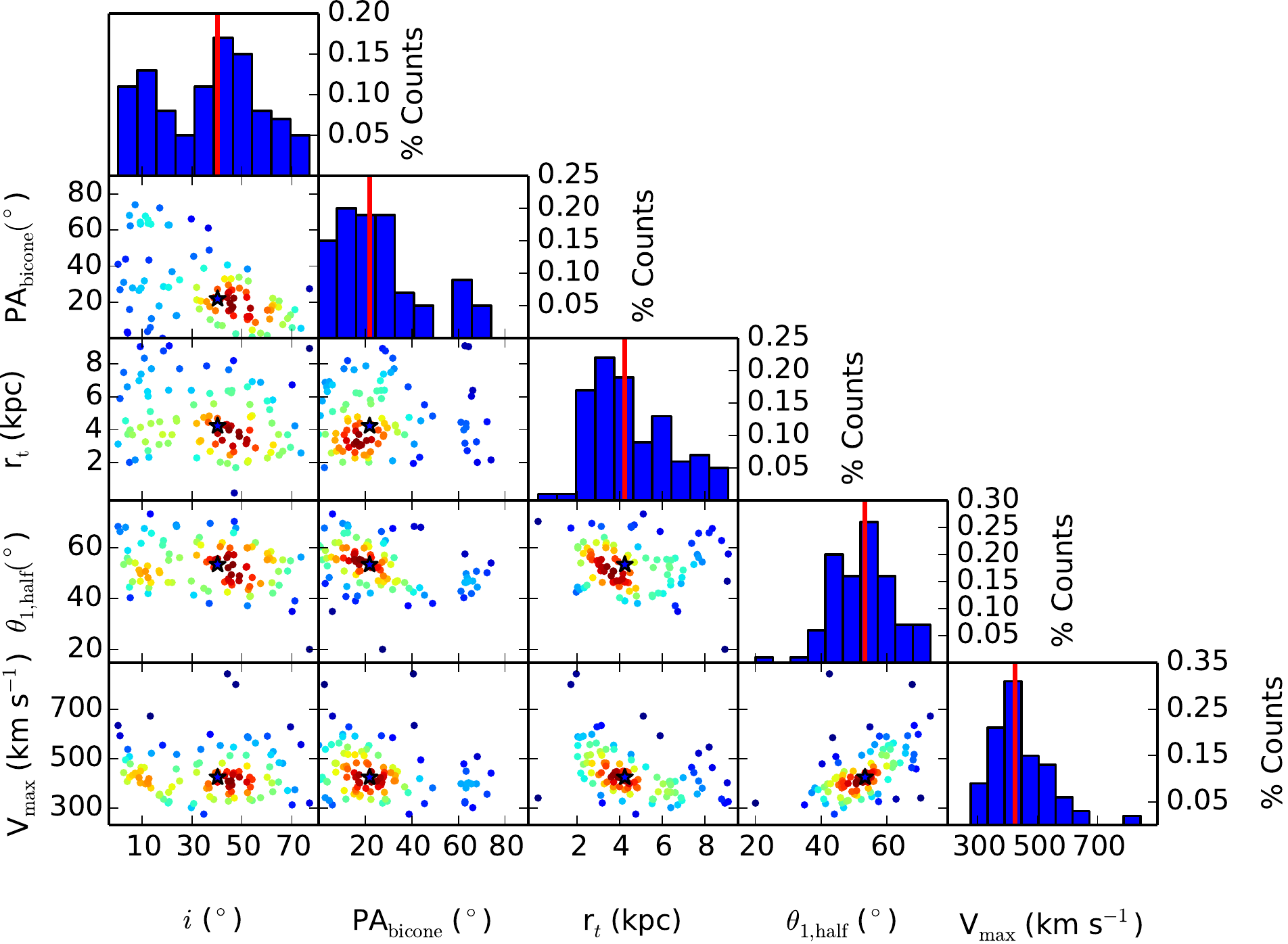}
\caption{Triangle plot for the best fit model parameters for the symmetric bicone model for the galaxy J0803+3926. Each column and row represents one of the five parameters for which the MCMC method estimates a best value. From left to right these parameters are: Inclination $i$, position angle on the sky ($\mathrm{PA}_{\mathrm{bicone}}$), turnover radius $r_t$, half opening angle of the cone ($\theta_{1,\mathrm{half}}$), and maximum velocity (V$_{\mathrm{max}}$). The top plot of each column gives the histogram of final values for each parameter, where the red line represents the median value. We demonstrate that our method successfully returns normally-distributed histograms for the value of each parameter. In the interior plots, walker final locations are colored according to point density on each plot, where red represents the densest clustering of points. The black star illustrates the best fit values provided by the median value of each parameter histogram.}
\label{triangleJ0930}
\end{figure*}

We next investigate the relative sensitivity of the six outflow parameters and their practical identifiability. We determine the best fit parameter values and their associated $1\sigma$ errors from the marginalized distributions (Figure \ref{triangleJ0930}). The PDF of each parameter is constructed from the final position of all 100 walkers in parameter space. One advantage of MCMC sampling is that the final parameter distributions are not restricted to symmetric errors. Instead, we use the shape and width of these distributions to determine how the various parameters affect the modeling process.

The first parameter-related check that we perform is a test of practical identifiability. \citet{Rothenberg1971} define lack of identification as the lack of sufficient information to distinguish between alternative structures or models based upon the data. There are different types of identifiability; here we discuss practical identifiability. Lack of identifiability where the data may not uniquely identify a model could either be a structural problem with the model itself or a problem that arises due to noisy data (\citealt{Campbell2013}). We assess the practical identifiability of the model by constructing a synthetic data set based upon the symmetric bicone model with a pixelscale and spectral resolution typical of the longslit data for the 18 galaxies. We then create two different realizations of this synthetic data; one where the error is equal to that of the data ($\sigma \approx 10\%$) and one where the error is inflated ($\sigma \approx 100 \%$).

The synthetic bicone has parameters typical of the galaxies we model here: $i$ = 0; $\mathrm{PA}_{\mathrm{bicone}}$ = 15$^{\circ}$; $r_t$ = 5 (pixels); $\theta_{1,\mathrm{half}}$ = 55$^{\circ}$; and V$_{\mathrm{max}}$ = 500 km s$^{-1}$. We run the synthetic model through {\tt emcee} starting at an intentionally incorrect starting point for the parameters. The goal is to determine if the {\tt emcee} process returns the correct set of parameters and examine the posterior probability of the output. 

When examining the walkers from both runs, it becomes apparent that the walkers better converge upon the true parameter values for the run with smaller, more representative error bars. This is unsurprising, as it indicates that the probability space is well-defined for the smaller errors. For the larger errors the likelihood is smooth and featureless (less conducive to convergence). We have demonstrated practical identifiability for our data, which has errors of order 10\%. However, extreme caution should be taken when attempting to identify models with velocity measurement errors on the order of 100\% of the value of the velocity.

We also assess the sensitivity of the six parameters involved in the biconical outflow models. For instance, from the parameter error bars reported in Tables \ref{parametersbicone}, \ref{parametersassy}, and \ref{parametersnest}, it is apparent that $\mathrm{PA}_{\mathrm{bicone}}$ is not well determined; the modeling process is not particularly sensitive to this parameter. We use one-factor-at-a-time (OFAT) sensitivity analysis to investigate the relative sensitivity of all parameters. We keep all parameters but the one in question at their baseline (nominal) values and compute the change in likelihood produced by varying the parameter in question through the full range of allowed values. For example, for an OFAT sensitivity analysis of the inclination, we vary the inclination between 0 and 90$^{\circ}$. We quantify the change in likelihood by computing $\Delta \chi^2_{\nu}$ between the best fit $\chi^2_{\nu}$ value and the largest $\chi^2_{\nu}$ value within this parameter range.

As an example, we discuss the OFAT sensitivity analysis for J0930+3430, which is best fit by an asymmetric bicone. We find the following $\Delta \chi^2_{\nu}$ values for the parameters listed in order of increasing sensitivity: (1.2, 256.2, 262.3, 613.2, 1051.1, 1335.3) for ($\mathrm{PA}_{\mathrm{bicone}}$, V$_{\mathrm{max}}$, $r_t$, $i$, $\theta_{1,\mathrm{half}}$, $\theta_{\mathrm{2,half}}$). As expected from the large error bars on $\mathrm{PA}_{\mathrm{bicone}}$, $\mathrm{PA}_{\mathrm{bicone}}$ is the least sensitive parameter. This is explained by our finding that the AGN outflows are biased to have large inclinations in Section \ref{biases}. A large inclination outflow is face-on, which causes the line of sight velocities to only change slightly over a range of values of $\mathrm{PA}_{\mathrm{bicone}}$. We calculate energy diagnostics for each outflow (Section \ref{energetics}) using V$_{\mathrm{max}}$, $r_t$, $\theta_{1,\mathrm{half}}$, and $\theta_{\mathrm{2,half}}$, which are among the most well-determined parameters in our models.

\subsection{Verifying the Models}
\label{verify}

\begin{figure*}
\centering
\includegraphics[scale=1, trim=0cm 1cm 0cm 0.5cm]{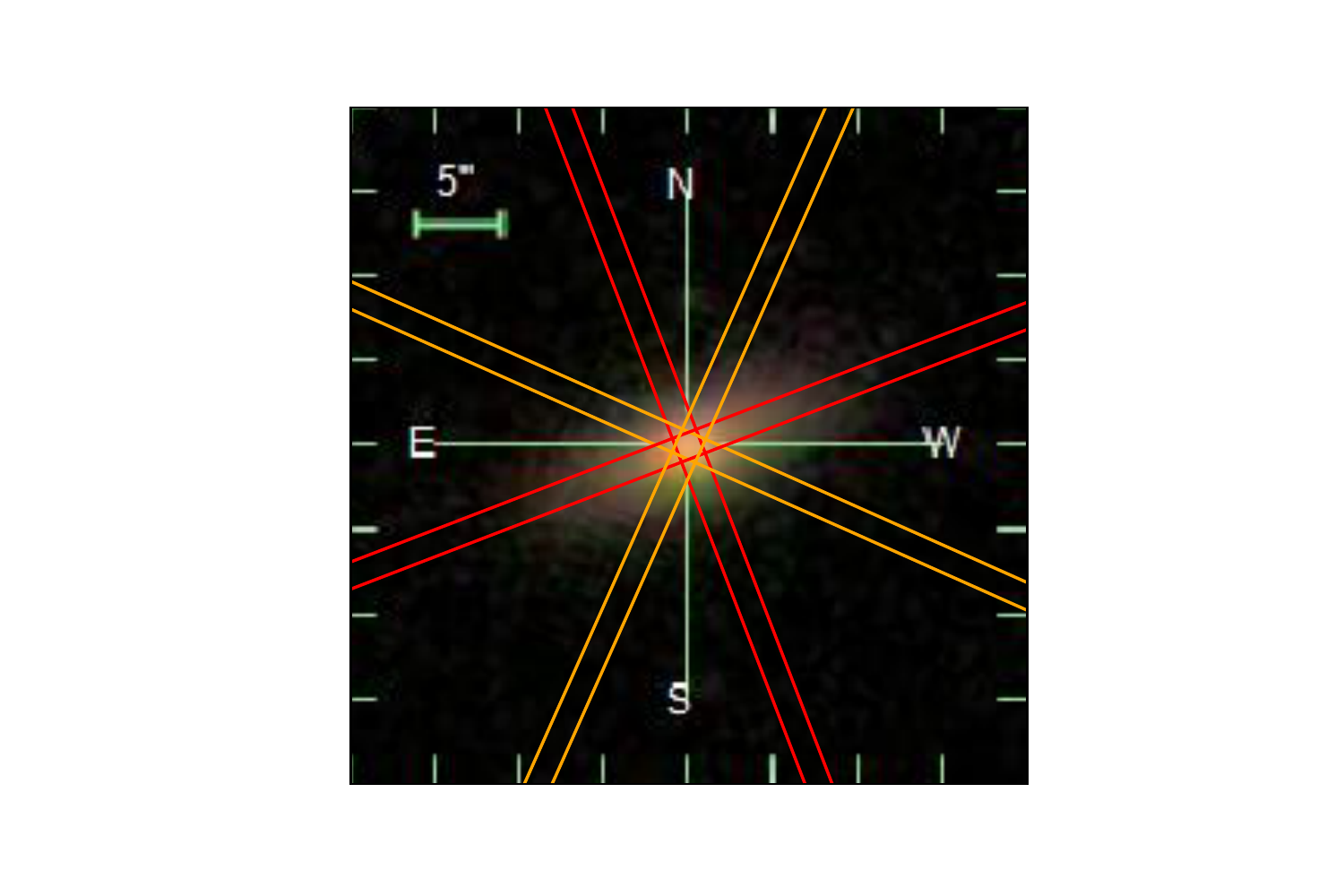}
\caption{Four observed PAs for J0930+3430 overplotted on the SDSS $gri$ image. We combine them to construct a pseudo-IFS map. The PAs are E of N: 21, 66, 111, and 156 degrees. PA 66 and 156 are the new observations, shown in orange, obtained from APO. The original two PAs of 21 and 111 degrees, shown in red, were obtained from Palomar  (\citealt{Nevin2016}). The slitwidth is 1.5$^{\prime \prime}$. }

\label{fullspatial}
\end{figure*}

We note that we are limited to two PAs of longslit data, with a limited number of data points for each. Therefore, we use two tests to verify that the models are converging and that we have enough longslit data points to converge upon the parameters of the bicone. For the first test, in this section we examine a galaxy for which we have obtained additional longslit data. We are motivated by \citet{Fischer2017}, who find that their original longslit observations of the biconical outflow in Mrk 573 (\citealt{Fischer2010}) were insufficient to return the correct parameters for their bicone and disk models for the kinematics of the galaxy. Returning to gather IFS observations in \citet{Fischer2017}, they find a different model for Mrk 573 that indicates an outflow is occurring on small scales and that rotation-dominated kinematics dominate at large scales. We will address the concern of rotation-dominated kinematics in Appendix \ref{APa}. Here, we examine the limitations of our longslit data.  

To address this concern for the longslit observations in this paper, we present additional observations for the galaxy J0930+3430. We choose to further investigate this galaxy because the best-fit PA for the bicone is between the two originally observed PAs. Additionally, the original best-fit model was one of the largest and most energetic of the 18 galaxies. The original two PAs observed and presented in \citet{Nevin2016} are 21 and 111 degrees, and we found that the best-fit bicone model is oriented along PA 75. These two position angles were observed with the Palomar Blue Channel Spectrograph (pixelscale $0\farcs389$ pixel$^{-1}$). We obtain two new PAs at 66 and 156 degrees with the Dual Imaging Spectrograph at APO (pixelscale $0\farcs42$ pixel$^{-1}$). In Figure \ref{fullspatial} we overplot all of the observed PAs.

We have modified our bicone modeling code to incorporate four PAs. We have rerun the the \texttt{emcee} code for all possible combinations of two PAs as well as the four PA run (pseudo-IFS) in Table \ref{parametersassystart}. We use the asymmetric bicone model for J0930+3430, and we find that all parameters agree within the 1$\sigma$ error interval for each run. Additionally, the model with four PAs does not significantly reduce the error interval although it provides better spatial coverage (Figure \ref{fullspatial}). 

For the second test, in Appendix \ref{APa} we examine a galaxy for which we have IFS data. This galaxy was not included in the 18 galaxy sample for this work. It was classified as an `Outflow Composite' in \citet{Nevin2016} but excluded from the modeling in this paper due to a third component at small scales and the lack of data points at one PA.

\begin{deluxetable*}{ccccccc}

\tabletypesize{\scriptsize}
\tablewidth{0pt}
\tablecolumns{6}
\tablecaption{J0930+3430 Outflow Model Parameters}
\tablehead{
\colhead{Modeled PAs} & 
\colhead{$i$} & 
\colhead{$\mathrm{PA}_{\mathrm{bicone}}$} &
\colhead{$r_{t}$} &
\colhead{$\theta_{1,\mathrm{half}}$} &
\colhead{$\theta_{2,\mathrm{half}}$} &
\colhead{V$_{\mathrm{max}}$} \\
&[$^{\circ}$]  &[$^{\circ}$E of N] &[kpc] &[$^{\circ}$] &[$^{\circ}$] & [km s$^{-1}$]   }

\startdata

Pseudo-IFS & $ 62 \substack{+ 16 \\- 22 }$ & $ 75 \substack{+ 121 \\- 53 }$ & $ 11 \substack{+ 5 \\- 6 }$ & $ 38 \substack{+ 23 \\- 27 }$ & $ 81 \substack{+ 5 \\- 23 }$ & $ 320 \substack{+ 290 \\- 170 }$  \\ 

21, 66 & $ 68 \substack{+ 15 \\- 29 }$ & $ 92 \substack{+ 105 \\- 45 }$ & $ 10 \substack{+ 6 \\- 6 }$ & $ 34 \substack{+ 28 \\- 27 }$ & $ 75 \substack{+ 10 \\- 14 }$ & $ 330 \substack{+ 240 \\- 170 }$ \\ 

21, 111 & $ 80 \substack{+ 6 \\- 9 }$ & $ 75 \substack{+ 55 \\- 49 }$ & $ 10 \substack{+ 4 \\- 6 }$ & $ 37 \substack{+ 12 \\- 10 }$ & $ 67 \substack{+ 3 \\- 2 }$ & $ 290 \substack{+ 130 \\- 30 }$\\

21, 156 & $ 72 \substack{+ 12 \\- 25 }$ & $ 85 \substack{+ 100 \\- 63 }$ & $ 10 \substack{+ 6 \\- 7 }$ & $ 30 \substack{+ 23 \\- 23 }$ & $ 79 \substack{+ 5 \\- 14 }$ & $ 350 \substack{+ 270 \\- 170 }$ \\ 

66, 156 & $ 55 \substack{+ 26 \\- 35 }$ & $ 128 \substack{+ 119 \\- 91 }$ & $ 11 \substack{+ 8 \\- 8 }$ & $ 23 \substack{+ 24 \\- 17 }$ & $ 63 \substack{+ 18 \\- 28 }$ & $ 360 \substack{+ 300 \\- 190 }$ \\

66, 111 & $ 69 \substack{+ 15 \\- 20 }$ & $ 90 \substack{+ 98 \\- 56 }$ & $ 10 \substack{+ 7 \\- 7 }$ & $ 37 \substack{+ 23 \\- 19 }$ & $ 70 \substack{+ 13 \\- 14 }$ & $ 330 \substack{+ 300 \\- 180 }$ \\ 

111, 156 & $ 69 \substack{+ 14 \\- 33 }$ & $ 105 \substack{+ 105 \\- 70 }$ & $ 11 \substack{+ 5\\- 9 }$ & $ 29 \substack{+ 25 \\- 19 }$ & $ 76 \substack{+ 6 \\- 20 }$ & $ 390 \substack{+ 330 \\- 210 }$

\enddata

\tablecomments{Best-fit parameters for all observed PAs of J0930+3430 with 1$\sigma$ errors. Column 1: PAs used in model. `Pseudo-IFS' indicates that all four PAs were used. Column 2: outflow inclination. Column 3: position angle of the bicone axis on the sky. Column 4: turnover radius in kpc. Column 5: inner half opening angle. Column 6: outer half opening angle. Column 7: maximum velocity.} 
\label{parametersassystart} 
\end{deluxetable*}

\section{Results}
\label{results}
\subsection{Bicone and Disk Orientation}
\label{resultorientation}
For each of the 18 galaxies, we compare the orientation of the bicone with the orientation of the photometric major axis of the galaxy. The orientation statistics for this sample of modeled outflows enable us to comment both on biconical outflow theory with the orientation of a theoretical collimating structure and on how the ionized outflows may affect the ISM and therefore drive feedback in the host galaxy; Section \ref{discussalignment} and Section \ref{energydiscuss}.

The position angle of the photometric major axis of the galaxy (PA$_{\mathrm{gal}}$) is the photometric major axis of the galaxy in the SDSS $r$-band. In this work, we determine the alignment of the bicone axis and the opening angle of the bicone structure from the analytic models ($\mathrm{PA}_{\mathrm{bicone}}$ and $\theta_{\mathrm{2,half}}$, respectively). We use $\theta_{\mathrm{1,half}}$ for the galaxies that are best fit with a symmetric bicone. We report these two position angles and the half opening angles in Table \ref{Alignment} and determine if the biconical structure is aligned with the photometric major axis, if it intersects the photometric major axis, and if it is perpendicular to the photometric major axis for all 18 galaxies. Alignment is defined as PA$_{\mathrm{gal}} = $ PA$_{\mathrm{bicone}}$ within a 1$\sigma$ error margin.

We find that a significant portion of outflows are aligned with the photometric major axis of their host galaxy (10/18, 55.6\%). However, we also find that a significant portion of outflows are aligned with a position angle that is perpendicular to the photometric major axis of the galaxy (10/18, 55.6\%). We find that six galaxies are included in both of these groups and that this is a reflection of the uncertainty of the PA$_{\mathrm{bicone}}$ parameter. When we remove these overlapping galaxies we find that 4/18, or 22.2\% of the outflows are aligned with the photometric major axis of the galaxy. The 95\% binomial confidence interval on the measured alignment fraction is 3.0\% to 41.4\%. 

The typical 1$\sigma$ error margin on the measurement of the position angle of the bicone is $\sim$20$^{\circ}$. If the bicone axes were randomly oriented, 22.2\% of all outflows should be measured to be within 20$^{\circ}$ of the photometric major axis. We derive the 22.2\% random orientation percentage from the total error margin, 40$^{\circ}$, divided by the total possibility of orientations on the sky (180$^{\circ}$). As a result, the percentage of outflows that have a biconical outflow axis that is aligned with the photometric major axis of the host galaxy is consistent with the percentage expected for a randomly oriented bicone. We discuss the implications of this result in Section \ref{discussalignment}.

\begin{turnpage}

\begin{deluxetable*}{lccccccccc}

\tablecolumns{9}
\tablecaption{Galaxy Alignment}

\tablehead{\colhead{SDSS ID} & 
\colhead{Slit PA 1} & 
\colhead{Slit PA 2} &
\colhead{PA$_{\mathrm{gal}}^{a}$} &
\colhead{$\mathrm{PA}_{\mathrm{bicone}}$} &
\colhead{$\theta_{\mathrm{2,half}}^{b}$}&
\colhead{Intersect?}&
\colhead{Aligned with PA$_{\mathrm{gal}}$?}&
\colhead{Perpendicular to PA$_{\mathrm{gal}}$?}\\
& [$^{\circ}$E of N]  & [$^{\circ}$E of N]  & [$^{\circ}$E of N]  & [$^{\circ}$E of N] & [$^{\circ}$]&  PA$_{\mathrm{gal}} $ = $\mathrm{PA}_{\mathrm{bicone}} \pm \theta_{\mathrm{2,half}}$ & PA$_{\mathrm{gal}} $ = $\mathrm{PA}_{\mathrm{bicone}}$ & PA$_{\mathrm{gal}} +90^{\circ} $ = $\mathrm{PA}_{\mathrm{bicone}}$}

\startdata

J0009$-$0036 &23&67& $65 $ & $ 79 \substack{+ 16 \\- 17 }$ & $77\substack{+ 3 \\- 2 }$  & yes&yes & no\\  
J0803+3926&50&140& $140$  &$ 22 \substack{+ 22 \\- 13 }$ & $53\substack{+ 9 \\- 9 }$ &  yes & no &yes\\
J0821+5021&43&133& $133$ & $ 6 \substack{+ 19 \\- 7 }$& $72\substack{+ 4 \\- 6 }$  & yes & no &no\\
J0854+5026&16&106& $ 16 $ &  $ 25 \substack{+ 14 \\- 169 }$& $75\substack{+ 6 \\- 3 }$  & yes & yes &yes \\
J0930+3430&21&111& $110 $  & $ 75 \substack{+ 55 \\- 49 }$ & $67\substack{+ 3 \\- 2 }$& yes & yes &yes\\
J0959+2619&28&118& $152  $  & $ 13 \substack{+ 31 \\- 9 }$& $71\substack{+ 5 \\- 4 }$  & yes & no &no\\
J1027+1049&75&165& $75$ &  $ 32 \substack{+ 81 \\- 25 }$& $61\substack{+ 5 \\- 4 }$  & yes & yes &no\\
J1109+0201&31&121& $31$ &  $ 59 \substack{+ 16 \\- 16 }$ & $73\substack{+ 3 \\- 3 }$& yes & no &no\\
J1152+1903&17&107& $37 $ &  $ 40 \substack{+ 25 \\- 36 }$  & $60\substack{+ 2 \\- 7 }$& yes & yes &no\\
J1315+2134&74&164& $164 $  & $ 39 \substack{+ 14 \\- 12 }$ & $78\substack{+ 3 \\- 2 }$ & yes & no&no\\
J1328+2752&39&129& $39  $ & $ 52 \substack{+ 81 \\- 39 }$& $81\substack{+ 3 \\- 4 }$ & yes & yes &yes\\
J1352+0525&162&252& $ 162  $ &  $ 43 \substack{+ 42 \\- 9 }$ & $78\substack{+ 2 \\- 3 }$ &yes & no &yes\\
J1420+4959&79&169& $169   $&$ 174 \substack{+ 98 \\- 126 }$ & $70\substack{+ 6 \\- 5 }$& yes & yes &yes\\
J1524+2743&12&102& $102  $ & $ 101 \substack{+ 102 \\- 63 }$& $62\substack{+ 2 \\- 2 }$  & yes & yes &yes\\
J1526+4140 &38&128& $38  $ &$ 59 \substack{+ 66 \\- 45 }$ & $62\substack{+ 11 \\- 16 }$ & yes & yes&yes\\
J1606+3427&17&107& $17  $ &  $ 114 \substack{+ 48 \\- 76 }$& $74\substack{+ 3 \\- 3 }$  &yes & no &yes\\
J1630+1649&30&120&  $30  $ &  $ 38 \substack{+ 29 \\- 34 }$& $82\substack{+ 2 \\- 1 }$  & yes & yes&no\\
J1720+3106&152&242& $153$ &  $ 34 \substack{+ 28 \\- 24 }$& $30\substack{+ 9 \\- 5 }$ & yes & no &yes

\enddata

\tablecomments{Column 1: galaxy name. Column 2 and 3: the observed position angles. Column 4: the position angle of the photometric major axis from the $r$-band from the SDSS. Column 5: the position angle of the bicone from the best fit model for each galaxy. Column 6: the larger half opening angle from the best fit model for each galaxy. Column 7: we determine if the bicone intersects the photometric major axis using the requirement PA$_{\mathrm{gal}}$ = PA$_{\mathrm{bicone}} \pm \theta_{\mathrm{2,half}}$ within the 1$\sigma$ error margin for all parameters. Column 8: we determine if the bicone axis is aligned with the photometric major axis using the requirement PA$_{\mathrm{gal}}$= PA$_{\mathrm{bicone}}$ within the 1$\sigma$ error margin. Column 9: we determine if the bicone axis is perpendicular to the photometric major axis using PA$_{\mathrm{gal}} +90^{\circ} $ = $\mathrm{PA}_{\mathrm{bicone}}$, again with a 1$\sigma$ error margin.
 } 
 \tablenotetext{a}{Associated 1$\sigma$ errors are $\sim$7$^{\circ}$.} 
  \tablenotetext{b}{We use $\theta_{1,\mathrm{half}}$ for the galaxies that are best modeled as a symmetric bicone, J0803+3926 and J1152+1903.} 
\label{Alignment} 
\end{deluxetable*}

\end{turnpage}

\subsection{Energy Diagnostics}
\label{energetics}

After constraining the geometry of the NLR outflows in the analytic modeling process, we derive energy diagnostics for the biconical outflows. By constraining the kinetic luminosity of the momentum-driven outflows, we can determine the ratio of the kinetic luminosity to the total radiated luminosity. This diagnostic enables us to make observational comparisons to the theoretically predicted 0.5\% threshold, which has been quoted as the ratio of L$_{\mathrm{KE}}/\mathrm{L}_{\mathrm{bol}}$ necessary to evacuate cold molecular gas from the inner regions of a galaxy and suppress star formation in the ISM (\citealt{Hopkins2010a}). 

To determine the mass outflow rate of the wind and the kinetic luminosity, we use the best fit parameters from our biconical models as well as density and temperature diagnostics from emission line ratios in the integrated SDSS spectra. We use the SDSS DR7 value-added catalogues (OSSY) to obtain information on integrated spectral lines (\citealt{Oh2011}). 

The mass outflow rate is defined as:

$$\dot{\mathrm{M}} =  m_p n_e \mathrm{V}_{\mathrm{max}} f ( \mathrm{A}_{1} + \mathrm{A}_{2})$$
where $m_p$ is the proton mass, $n_e$ is the electron density of the NLR, V$_{\mathrm{max}}$ is the maximum velocity of the outflow, $f$ is the filling factor, and A$_{1}$ and A$_{2}$ are the lateral surface areas of each cone in the bicone. A$_{1}$ is the smaller opening angle cone and A$_{2}$ is the larger opening angle cone. 

To calculate the electron density, $ n_e$, we calculate the intensity ratios of [OII]$\lambda 3729 /\mathrm{[OII]} \lambda 3726$ and [SII]$ \lambda 6716 / \mathrm{[SII]}\lambda 6731$, which are sensitive to density (\citealt{Osterbrock2006}). We find mean values of $\sim 0.82$ and $\sim 1.18$, respectively, for these intensity ratios for the 18 galaxy sample. Typical temperatures are in the range $(1-2)\times 10^4$ K in the NLRs of AGNs (\citealt{Osterbrock2006}), and thus the corresponding electron density is $\sim (1-5)\times 10^2$ cm$^{-3}$. Thus, we verify that  $10^2 < n_e $ (cm$^{-3}$) $< 10^3 $, which is typical for the NLR (\citealt{Taylor2003}). We use a density of 100 cm$^{-3}$ in our calculations. This assumption is consistent with previous work that finds an electron density of 100 cm$^{-3}$ exterior to 1 kpc from the AGN (\citealt{Karouzos2016}). With our spatial resolution, we always resolve gas at $>1$ kpc.

The filling factor represents the proportion of the bicone surface that contains ionized NLR clouds. It scales inversely with the electron density, $n_e$: $n_e \propto f^{-1/2}$ (\citealt{Oliva1997}) and has values in the range of $0.01<f<0.1$ for the NLR  (\citealt{Storchi-Bergmann2010}). We adopt a value of 0.01, which is a conservative lower limit based upon the literature. Some authors choose to adopt $f=0.1$, but we choose to be conservative in our calculation of outflowing energy and adopt $f=0.01$, representing a bicone in which 1/100 of the bicone surface contains ionized NLR clouds.

The lateral surface area, A, which is the generalized form of A$_{1}$ and A$_{2}$, is:
$$\mathrm{A}=\pi r \sqrt{h^2+r^2} $$
where $h$ is the height, which we define as the turnover radius for energy calculation purposes, and $r$ is the deprojected radius, determined by the half opening angle of the cone:
$$r=r_t \sin (\theta_{\mathrm{half}})$$
where $r_t$ is the turnover radius.

We determine the turnover radius, the opening angle, and the maximum velocity for the 18 galaxies using the best fit analytic models (Section \ref{MCMC}). As we discuss in Section \ref{discussmodels}, various additional walls of these three biconical outflow models could be obscured or not illuminated so this calculation is a lower limit for energy outflow rate.

Once we have derived the mass outflow rate of the biconical outflows, we calculate the kinetic luminosity:

$$\mathrm{L}_{\mathrm{KE}} = \frac{1}{2} \dot{\mathrm{M}} \mathrm{V}_{\mathrm{max}}^2$$

We compare the kinetic luminosity to the AGN bolometric luminosity, which is calculated from the dereddened [OIII]$\lambda 5007$ luminosity from the SDSS DR7 value-added catalogues in \citet{Nevin2016}.

We report the lateral surface area, mass outflow rate, kinetic luminosity, maximum outflow velocity, half opening angle, turnover radius, AGN bolometric luminosity, and ratio of kinetic to AGN bolometric luminosity for the 18 galaxies in Table \ref{energy}. 

We find that 16/18 (88.9\%) of the galaxies have a L$_{\mathrm{KE}}/\mathrm{L}_{\mathrm{bol}}$ ratio that is above the 0.5\% threshold value to drive two-staged feedback. Of these galaxies, 100\% have a bicone that intersects the photometric major axis of the host galaxy. We discuss the implications of these results in Section \ref{energydiscuss}.

\begin{deluxetable*}{lccccccccc}

\tabletypesize{\scriptsize}
\tablewidth{0pt}
\tablecolumns{10}
\tablecaption{Energy Diagnostics}
\tablehead{
\colhead{SDSS ID} & 
\colhead{$r_t$} &
\colhead{$\theta_{1,\mathrm{half}}$} & 
\colhead{$\theta_{2,\mathrm{half}}$} & 
\colhead{V$_{\mathrm{max}}$} &
\colhead{A$_{1}$ + A$_{2}$} & 
\colhead{$\dot{\mathrm{M}}$} &
\colhead{L$_{\mathrm{KE}}$} & 
\colhead{L$_{\mathrm{bol}}$} & 
\colhead{L$_{\mathrm{KE}}$/L$_{\mathrm{bol}}$} \\
& [kpc] &[deg] &[deg] & [km s$^{-1}$] & [kpc$^2$] &  [M$_{\odot}$ yr$^{-1}$] & [erg s$^{-1}$] & [erg s$^{-1}$]  & Lower Limit
}

\startdata

J0009$-$0036  & $ 2.0 \substack{+ 1.1 \\- 0.3 }$ & $ 60 \substack{+ 3 \\- 4 }$ & $ 77 \substack{+ 3 \\- 2 }$ & $ 320 \substack{+ 60 \\- 80 }$ & $ 30.1 \substack{+ 46.4 \\- 9.1 }$ & $ 250 \substack{+ 260 \\- 84 }$ & $( 7.9 \substack{+ 8.7 \\- 3.4 }) \times 10^{ 42 }$ & $( 2.0178 \pm  0.1764 ) \times 10^{ 45 }$ &  $ 0.002 $  \\ 
J0803+3926  & $ 4.2 \substack{+ 2.4 \\- 1.6 }$ & $ 53 \substack{+ 9 \\- 9 }$ & $ 77 \substack{+ 0 \\- 0 }$ & $ 430 \substack{+ 110 \\- 70 }$ & $ 102.1 \substack{+ 163.4 \\- 53.5 }$ & $ 1098 \substack{+ 1537 \\- 536 }$ & $( 6.7 \substack{+ 9.1 \\- 3.0 }) \times 10^{ 43 }$ & $( 3.1736 \pm  0.2616 ) \times 10^{ 45 }$ &  $ 0.011 $  \\ 
J0821+5021  & $ 6.4 \substack{+ 3.3 \\- 1.8 }$ & $ 50 \substack{+ 7 \\- 11 }$ & $ 72 \substack{+ 4 \\- 6 }$ & $ 360 \substack{+ 60 \\- 100 }$ & $ 286.6 \substack{+ 313.1 \\- 130.1 }$ & $ 2504 \substack{+ 2912 \\- 1246 }$ & $( 9.9 \substack{+ 13.0 \\- 5.8 }) \times 10^{ 43 }$ & $( 1.181 \pm  0.188 ) \times 10^{ 44 }$ &  $ 0.320 $  \\ 
J0854+5026  & $ 4.5 \substack{+ 1.1 \\- 1.2 }$ & $ 42 \substack{+ 4 \\- 7 }$ & $ 75 \substack{+ 6 \\- 3 }$ & $ 290 \substack{+ 40 \\- 50 }$ & $ 133.4 \substack{+ 63.4 \\- 60.6 }$ & $ 1019 \substack{+ 385 \\- 506 }$ & $( 2.6 \substack{+ 1.2 \\- 1.6 }) \times 10^{ 43 }$ & $( 6.44 \pm  0.99 ) \times 10^{ 43 }$ &  $ 0.168 $  \\ 
J0930+3430  & $ 4.6 \substack{+ 2.0 \\- 2.6 }$ & $ 37 \substack{+ 12 \\- 10 }$ & $ 67 \substack{+ 3 \\- 2 }$ & $ 290 \substack{+ 130 \\- 30 }$ & $ 129.3 \substack{+ 127.9 \\- 105.3 }$ & $ 1019 \substack{+ 1205 \\- 821 }$ & $( 2.8 \substack{+ 4.3 \\- 2.2 }) \times 10^{ 43 }$ & $( 1.175 \pm  0.37 ) \times 10^{ 44 }$ &  $ 0.054 $  \\ 
J0959+2619  & $ 3.3 \substack{+ 1.8 \\- 1.3 }$ & $ 49 \substack{+ 10 \\- 14 }$ & $ 71 \substack{+ 5 \\- 4 }$ & $ 280 \substack{+ 180 \\- 100 }$ & $ 75.6 \substack{+ 100.9 \\- 46.0 }$ & $ 468 \substack{+ 765 \\- 237 }$ & $( 1.1 \substack{+ 5.0 \\- 0.84 }) \times 10^{ 43 }$ & $( 8.14 \pm  0.88 ) \times 10^{ 43 }$ &  $ 0.038 $  \\ 
J1027+1049  & $ 6.2 \substack{+ 3.0 \\- 3.1 }$ & $ 33 \substack{+ 9 \\- 9 }$ & $ 61 \substack{+ 5 \\- 4 }$ & $ 540 \substack{+ 100 \\- 80 }$ & $ 215.0 \substack{+ 244.7 \\- 162.9 }$ & $ 3059 \substack{+ 3181 \\- 2191 }$ & $( 2.4 \substack{+ 4.5 \\- 1.7 }) \times 10^{ 44 }$ & $( 9.09 \pm  1.54 ) \times 10^{ 43 }$ &  $ 0.943 $  \\ 
J1109+0201  & $ 4.9 \substack{+ 1.9 \\- 0.8 }$ & $ 55 \substack{+ 5 \\- 4 }$ & $ 73 \substack{+ 3 \\- 3 }$ & $ 390 \substack{+ 50 \\- 40 }$ & $ 181.8 \substack{+ 139.0 \\- 56.9 }$ & $ 1918 \substack{+ 940 \\- 651 }$ & $( 8.9 \substack{+ 4.7 \\- 3.1 }) \times 10^{ 43 }$ & $( 1.023 \pm  0.142 ) \times 10^{ 44 }$ &  $ 0.562 $  \\ 
J1152+1903  & $ 4.5 \substack{+ 2.1 \\- 1.7 }$ & $ 60 \substack{+ 2 \\- 7 }$ & $ 73 \substack{+ 0 \\- 0 }$ & $ 370 \substack{+ 90 \\- 50 }$ & $ 148.4 \substack{+ 165.0 \\- 101.0 }$ & $ 1567 \substack{+ 1053 \\- 1053 }$ & $( 6.4 \substack{+ 4.1 \\- 4.0 }) \times 10^{ 43 }$ & $( 6.66e \pm  1.21 ) \times 10^{ 44 }$ &  $ 0.039 $  \\ 
J1315+2134  & $ 3.2 \substack{+ 1.3 \\- 0.9 }$ & $ 48 \substack{+ 4 \\- 4 }$ & $ 78 \substack{+ 3 \\- 2 }$ & $ 600 \substack{+ 130 \\- 50 }$ & $ 70.8 \substack{+ 70.9 \\- 31.0 }$ & $ 1131 \substack{+ 988 \\- 494 }$ & $( 1.3 \substack{+ 1.2 \\- 0.61 }) \times 10^{ 44 }$ & $( 8.835 \pm  0.914 ) \times 10^{ 44 }$ &  $ 0.089 $  \\ 
J1328+2752  & $ 4.0 \substack{+ 2.7 \\- 1.9 }$ & $ 48 \substack{+ 15 \\- 19 }$ & $ 81 \substack{+ 3 \\- 5 }$ & $ 230 \substack{+ 260 \\- 80 }$ & $ 116.4 \substack{+ 170.3 \\- 80.5 }$ & $ 746 \substack{+ 890 \\- 474 }$ & $( 1.1 \substack{+ 4.4 \\- 0.83 }) \times 10^{ 43 }$ & $( 7.47 \pm  1.21 ) \times 10^{ 43 }$ &  $ 0.047 $  \\ 
J1352+0525  & $ 1.1 \substack{+ 0.3 \\- 0.2 }$ & $ 61 \substack{+ 3 \\- 6 }$ & $ 78 \substack{+ 2 \\- 3 }$ & $ 440 \substack{+ 60 \\- 80 }$ & $ 10.2 \substack{+ 5.8 \\- 2.9 }$ & $ 111 \substack{+ 52 \\- 28 }$ & $( 7.0 \substack{+ 3.4 \\- 3.5 }) \times 10^{ 42 }$ & $( 9.92 \pm  1.93 ) \times 10^{ 43 }$ &  $ 0.034 $  \\ 
J1420+4959  & $ 2.7 \substack{+ 2.0 \\- 1.5 }$ & $ 41 \substack{+ 10 \\- 16 }$ & $ 70 \substack{+ 6 \\- 5 }$ & $ 390 \substack{+ 170 \\- 110 }$ & $ 47.0 \substack{+ 92.7 \\- 37.7 }$ & $ 451 \substack{+ 642 \\- 333 }$ & $( 2.0 \substack{+ 3.5 \\- 1.5 }) \times 10^{ 43 }$ & $( 4.576 \pm  1.445 ) \times 10^{ 44 }$ &  $ 0.011 $  \\ 
J1524+2743  & $ 2.8 \substack{+ 3.0 \\- 1.0 }$ & $ 35 \substack{+ 7 \\- 10 }$ & $ 62 \substack{+ 2 \\- 2 }$ & $ 720 \substack{+ 100 \\- 90 }$ & $ 44.7 \substack{+ 151.0 \\- 24.3 }$ & $ 788 \substack{+ 2154 \\- 403 }$ & $( 1.2 \substack{+ 2.2 \\- 0.55 }) \times 10^{ 44 }$ & $( 1.262 \pm  0.122 ) \times 10^{ 44 }$ &  $ 0.505 $  \\ 
J1526+4140  & $ 0.8 \substack{+ 0.5 \\- 0.4 }$ & $ 44 \substack{+ 15 \\- 14 }$ & $ 62 \substack{+ 11 \\- 16 }$ & $ 410 \substack{+ 270 \\- 120 }$ & $ 3.2 \substack{+ 4.6 \\- 2.3 }$ & $ 33 \substack{+ 40 \\- 24 }$ & $( 1.7 \substack{+ 4.7 \\- 1.2 }) \times 10^{ 42 }$ & $( 8.2 \pm  0.3 ) \times 10^{ 42 }$ &  $ 0.051 $  \\ 
J1606+3427  & $ 1.1 \substack{+ 0.3 \\- 0.3 }$ & $ 37 \substack{+ 8 \\- 6 }$ & $ 74 \substack{+ 3 \\- 3 }$ & $ 370 \substack{+ 70 \\- 40 }$ & $ 7.7 \substack{+ 6.1 \\- 3.7 }$ & $ 79 \substack{+ 40 \\- 37 }$ & $( 3.3 \substack{+ 2.2 \\- 1.3 }) \times 10^{ 42 }$ & $( 1.83 \pm  0.45 ) \times 10^{ 43 }$ &  $ 0.111 $  \\ 
J1630+1649  & $ 0.5 \substack{+ 0.3 \\- 0.1 }$ & $ 40 \substack{+ 16 \\- 7 }$ & $ 82 \substack{+ 2 \\- 1 }$ & $ 290 \substack{+ 90 \\- 30 }$ & $ 1.5 \substack{+ 2.9 \\- 0.5 }$ & $ 11 \substack{+ 25 \\- 4 }$ & $( 2.7 \substack{+ 16.0 \\- 0.89 }) \times 10^{ 41 }$ & $( 4.77 \pm  0.25 ) \times 10^{ 43 }$ &  $ 0.004 $  \\ 
J1720+3106  & $ 2.9 \substack{+ 0.8 \\- 0.7 }$ & $ 18 \substack{+ 11 \\- 5 }$ & $ 30 \substack{+ 9 \\- 5 }$ & $ 300 \substack{+ 20 \\- 10 }$ & $ 21.9 \substack{+ 18.3 \\- 5.3 }$ & $ 167 \substack{+ 151 \\- 39 }$ & $( 4.7 \substack{+ 4.8 \\- 1.0 }) \times 10^{ 42 }$ & $( 1.347 \pm  0.228 ) \times 10^{ 44 }$ &  $ 0.026 $

\enddata

\tablecomments{Column 1: galaxy name. Column 2: turnover radius in kpc. Column 3: inner half opening angle or half opening angle for the symmetric bicone galaxies. Column 4: outer half opening angle for the nested and asymmetric bicone galaxies. Column 5: maximum intrinsic velocity. Column 6: lateral surface area of the bicones. Column 7: mass outflow rate. Column 8: kinetic luminosity. Column 9: AGN bolometric luminosity from \citet{Nevin2016}. Column 10: lower limit calculated from the asymmetric energy ratio distribution. We take the median value from the distribution and subtract the 1$\sigma$, or 34th percentile of the distribution (L$_{\mathrm{KE}}$/L$_{\mathrm{bol}}$ - $\sigma_{\mathrm{lower}}$). We use this ratio to assess if the energy ratio is above the 0.5\% threshold value.} 
\label{energy} 
\end{deluxetable*}

\section{Discussion}
\label{discuss}

\subsection{This sample of biconical outflows is biased to be very large and energetic}
\label{biases}

Before we analyze the energetics of our outflows, we first must understand the observational biases. In this section, we discuss the selection criteria that bias our sample towards larger and more energetic outflows. We also discuss a theoretical `minimum energy bicone' that corresponds to a bicone with the smallest possible surface area that it is possible to recover from our sample given the sample biases.

First, the galaxies in this sample were selected from the SDSS for their double-peaked emission lines. The average velocity separation of the double peaked narrow lines of the integrated spectra of the 18 galaxy sample is $\sim$300 km s$^{-1}$. Second, these 18 galaxies also have large spatial extents of emission (the average extent of emission is 6.8 kpc) by our selection criteria that requires that the number of rows of statistically significant emission be greater than twice the number of parameters. Third, the average pixelscale of the instruments in this sample is $0\farcs3$ pix$^{-1}$ which biases our sample towards larger (kpc-scale) outflows.

These three factors have several effects on the best fitting bicone models. First, the large separation in velocity space produces biconical outflows with preferentially large opening angles and higher intrinsic velocities. A higher intrinsic velocity along the walls of the cone produces a larger observed velocity separation between the velocity centroids, regardless of the orientation of the structure, while a larger opening angle cone's geometry can produce this same effect.

Second, the requirement of many statistically significant rows of emission produces bicone structures with larger turnover radii, larger inclinations, and/or larger opening angles. The average pixelscale of the instruments ($0\farcs3$ pix$^{-1}$) corresponds to a physical distance of $\sim$0.3 kpc at $z=0.05$, which is the typical redshift. If we require that the observed bicone structure cover 5 spatial rows from the center of the galaxy (10 total; $n=2k$ where $k=5$ for the symmetric bicone) at each PA, for example, the bicone structure will extend out to a radius of 1.5 kpc. This distance is the full extent of the measured bicone in our data and the turnover radius is interior to this point. However, for our sample the turnover radius is close to the full extent because we observe very little deceleration in the bicones modeled here. Therefore, we are biased towards finding larger bicone structures that tend to also have a larger turnover radius. The spatial row selection also ensures that we observe double peaked emission at both position angles. This selects for more inclined bicones with larger opening angles that can open up along both position angles, producing the full spatial coverage of both orthogonal position angles. 

We are therefore selecting for AGN outflows that are more energetic (larger sizes, opening angles, and velocities) and that have a greater chance of intersecting the photometric major axis of their host galaxies with their larger opening angles.

We create a theoretical minimum energy bicone using the limitations of the sample to characterize the selection biases towards higher energy bicones. This helps us place a lower limit on the mass outflow rates of the galaxies in this sample. To produce the minimum energy bicone, we minimize the turnover radius, opening angle, and maximum velocity. We first use the smallest resolvable turnover radius of one pixel at a representative pixelscale of $0\farcs3$pix$^{-1}$ (the average pixelscale of the collection of instruments used here). We use the typical redshift for this sample of 0.05 which corresponds to a conversion factor of $\sim$1 kpc/$\prime \prime$. This yields a turnover radius of 0.3 kpc. This is slightly smaller than the smallest modeled turnover radius in the sample, which is 0.48 kpc for J1630+1649. 

We use the representative separation of double peaks of 300 km s$^{-1}$ from our sample. This is an observed velocity and does not directly correspond to intrinsic velocity. However, for a randomly oriented sample of outflows where inclinations and half opening angles are distributed between 0 and 90$^{\circ}$, the observed velocity could range between 0 and 600 km s$^{-1}$ for an intrinsic velocity of 300 km s$^{-1}$. Therefore, 300 km s$^{-1}$ is a fair average intrinsic velocity. It is also slightly larger than the smallest measured intrinsic velocity from our sample, which is 281.5 km s$^{-1}$ for J0959+2619.

The fact that we observe velocity separations at both observed orthogonal PAs places limits on the possible range of values for the combination of opening angles and inclination. For example, if the inclination is zero, the half opening angles are constrained to be at least 45$^{\circ}$ so that the bicone is observable at both orthogonal PAs. For the nested bicone, the inclination must be high so that the bicone walls are observed across the plane of the sky. When we consider inclination and opening angle at the same time, this requires that the combination of inclination and $\theta_{1,\mathrm{half}}$ be greater than 90$^{\circ}$. For instance, if the nested bicone axis is inclined at 45$^{\circ}$, the half opening angle of the inner cone must be at least 45$^{\circ}$ so we observe it at all spatial positions. For the asymmetric bicone, the same rules apply.

For all three models it is possible that the bicone structure is inclined exactly 90$^{\circ}$ relative to the line of sight with a small opening angle bicone. However, this is very unlikely given that our sample selects for high inclinations but few are greater than 80$^{\circ}$. We choose to use the average inclination of 66$^{\circ}$, which requires that the half opening angle be $>$ 24$^{\circ}$ for the bicone to intersect both observed PAs at all spatial positions. We use an inclination of 66$^{\circ}$ and a half opening angle of 24$^{\circ}$.

Using the combination of these minimized parameters, we find a kinetic luminosity of $5.1 \times 10^{40}$ erg s$^{-1}$ for our minimum energy bicone case. This is roughly an order of magnitude below our lowest measured kinetic luminosity ($2.7 \times 10^{41}$ erg s$^{-1}$ for J1630+1649). The corresponding minimum mass outflow rate is 1.8 M$_{\odot}$ yr$^{-1}$. We use a horizontal line in Figure \ref{grandcompare} to compare the mass outflow rate of the minimum energy bicone to the rest of our sample.

\subsection{The biconical outflows in this sample are large and energetic}
\label{comparesize}
For the 18 galaxies we model, we find that the average intrinsic maximum velocity is 370 $\pm$ 146 km s$^{-1}$, the average inner half opening angle is 44.5 $\pm$ 11.8$^{\circ}$, the average outer half opening angle (for those galaxies that were best fit as nested bicones or asymmetric bicones) is 69.5 $\pm$ 12.4$^{\circ}$, and the average turnover radius is 3.4 $\pm$ 1.8 kpc. Our sample of moderate-luminosity AGNs (42 $<$ log L$_{\mathrm{bol}}$ (erg s$^{-1}$) $<$ 46) have large surface geometries due to their large half opening angles and turnover radii. This leads to large mass outflow rates (1 $<$ log $\dot{\mathrm{M}}$ (M$_{\odot}$ yr$^{-1}$) $<$ 3), large kinetic luminosities (41 $<$ log L$_{\mathrm{KE}}$ (erg s$^{-1}$) $<$ 45), and therefore large kinetic to total AGN bolometric luminosity ratios (0.001 $<$ L$_{\mathrm{KE}}$/L$_{\mathrm{bol}}$ $<$ 1.0). In this section we compare these findings to previous work and examine assumptions made in mass outflow rate estimates in the literature. 

\begin{figure*}
\includegraphics[scale=0.6]{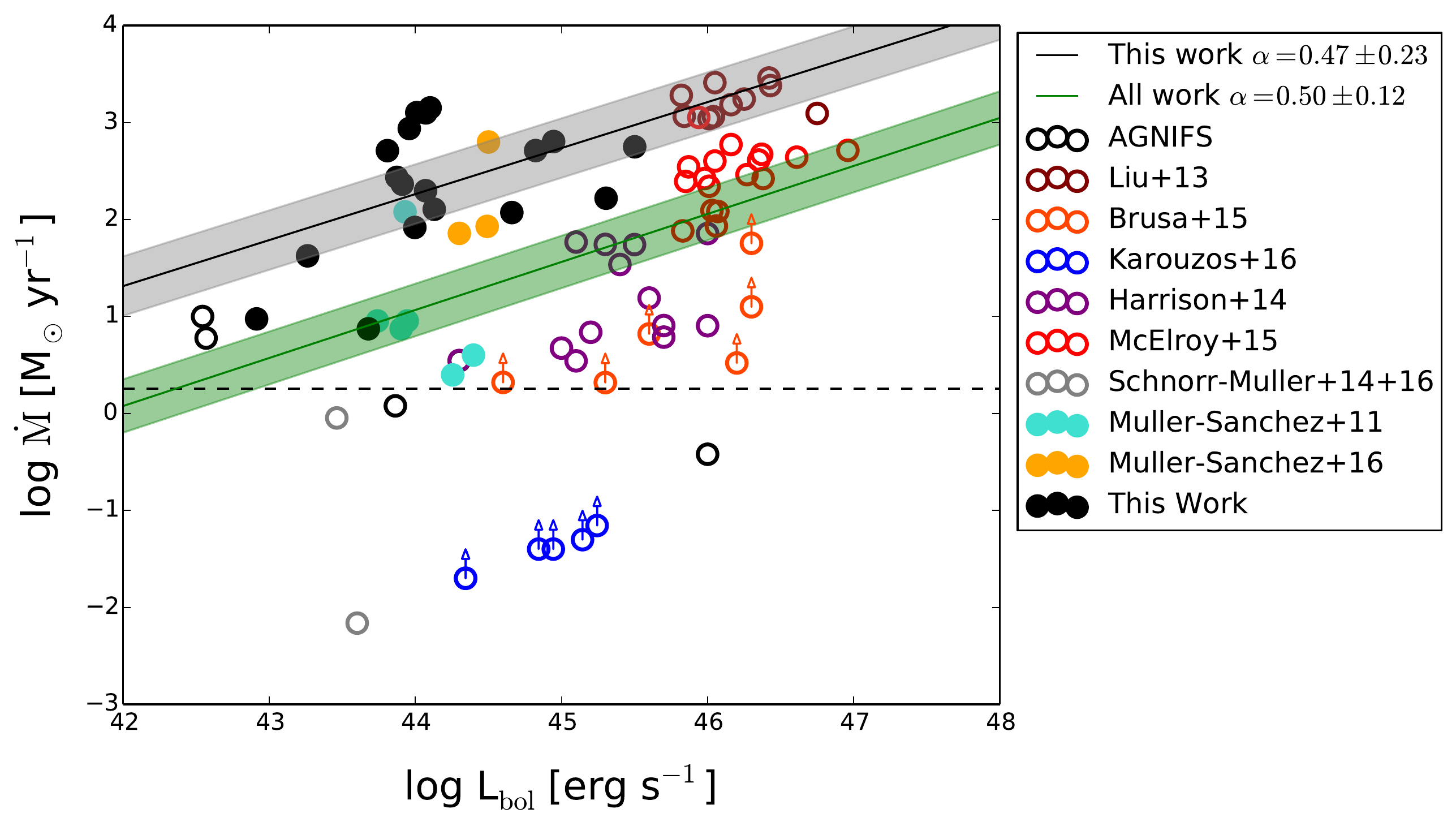}
\caption{The measured mass outflow rates for AGN outflows in the literature and this work plotted against AGN bolometric luminosity. The filled circles utilize biconical kinematic models to constrain the parameters to measure mass outflow rate. The open circles assume either a spherical geometry or a biconical geometry for their outflow and emission line fluxes to estimate the material swept up in the shell. These geometries for the open circles are not kinematically constrained. We include arrows to indicate underestimation of the mass outflow rates according to the discussion in Section \ref{comparesize}. We plot a horizontal line representing the mass outflow rate ($\sim$ 1.8 M$_{\odot}$ yr$^{-1}$) associated with the minimum energy bicone discussed in Section \ref{biases}. We plot a black line ($\alpha =0.47$) and a green line ($\alpha = 0.50$) for the best fit slopes associated with this sample and all data points plotted, respectively. We overplot the confidence intervals on both lines. References include: AGNIFS (\citealt{Storchi-Bergmann2010,Riffel2011a,Riffel2011b,Riffel2013,Riffel2015,Schonell2014}); \citealt{Liu2013a,Liu2013b,Brusa2015,Karouzos2016,Harrison2014,McElroy2015,Schnorr-Muller2014,Schnorr-Muller2016,Muller-Sanchez2011,Muller-Sanchez2016}. }
\label{grandcompare}
\end{figure*}

\begin{deluxetable*}{ccccccc}

\tabletypesize{\scriptsize}
\tablewidth{0pt}
\tablecolumns{5}
\tablecaption{Mass Outflow Rates in the Literature}
\tablehead{
\colhead{Group} & 
\colhead{$\dot{\mathrm{M}}$ Range} &
\colhead{Average $\dot{\mathrm{M}}$ } &  
\colhead{Technique Used} & 
\colhead{n$_e$ Assumed}\\
& [M$_{\odot}$ yr$^{-1}$] & [M$_{\odot}$ yr$^{-1}$] & &[cm$^{-3}$] }

\startdata

This work & 10$^{0.5}$ - 10$^{3.5}$ & 10$^{2.7}$&  [OIII] kinematic modeling of a bicone  & 100 \\
AGNIFS Group & 10$^{-1}$ - 10$^1$& 10$^{0.5}$& HII luminosity of a bicone &500  \\
\citet{Liu2013b} & 10$^{2.5}$ - 10$^{3.5}$& 10$^{3.2}$& H$\alpha$ and H$\beta$ luminosity of a sphere& 100\\
\citet{Brusa2015} & 10$^{1.5}$ - 10$^3$& 10$^{1.1}$ & [OIII] luminosity of a sphere & 100\\
\citet{Karouzos2016} & 10$^{-2}$ - 10$^{-1}$ & 10$^{-1.4}$ &[OIII] luminosity of a sphere &200-800 \\
\citet{Harrison2014} & 10$^0$ - 10$^2$ & 10$^{1.4}$& H$\beta$ luminosity of a sphere & 100\\
\citet{McElroy2015} & 10$^{1.5}$ - 10$^{3}$ & 10$^{2.5}$ &H$\alpha$ and H$\beta$ luminosity of a sphere& 100 \\
\citet{Schnorr-Muller2014,Schnorr-Muller2016} & 10$^{-2.5}$ - 10$^0$& 10$^{-0.3}$ & H$\alpha$ luminosity of a bicone & 1350 \\
\citet{Muller-Sanchez2011} & 10$^0$ - 10$^2$& 10$^{1.4}$&CLR kinematic modeling of a bicone &5000  \\
\citet{Muller-Sanchez2016} & 10$^{1.5}$ - 10$^3$& 10$^{2.4}$ & Pa$\alpha$ kinematic modeling of a bicone&100

\enddata

\label{tablegraph} 
\end{deluxetable*}

In Figure \ref{grandcompare}, we plot the mass outflow rates and AGN bolometric luminosities of the galaxies in this sample against other estimated mass outflow rates from AGN-driven outflows in the literature. We also compile the mass outflow rates from the literature in Table \ref{tablegraph}. 

Our galaxies span a wide range of AGN bolometric luminosities, overlapping with low-luminosity local Seyfert galaxies (e.g., \citealt{Muller-Sanchez2011,Schnorr-Muller2014}) as well as high-luminosity quasars (e.g., \citealt{Liu2013b,McElroy2015}). The average mass outflow rate for the 18 galaxies modeled in this work is $\sim$10$^{2.7}$ M$_{\odot}$ yr$^{-1}$. The mass outflow rates we derive agree with the mass outflow rates of high-luminosity AGNs and have some overlap with the mass outflow rates of moderate-luminosity AGNs. They are greater than that of samples of low-luminosity AGN outflows. 

\citet{Liu2013b} and \citet{McElroy2015} measure the mass outflow rates of samples of high-luminosity AGN outflows and find averages of 10$^{3.2}$ and 10$^{2.5}$ M$_{\odot}$ yr$^{-1}$, respectively. Both of these studies use non-kinematic estimation techniques; they use a spherical shell assumption and H$\beta$ and H$\alpha$ luminosities to estimate a total mass outflow rate. Although they assume similar number densities for the NLR ($\sim$100 cm$^{-3}$), the non-kinematic technique can underestimate the mass entrained in the outflow. However, the use of H$\beta$ and H$\alpha$ could lead to an overestimate of the mass outflow rate, since H$\alpha$ and H$\beta$ can also trace gas associated with the disk of the galaxy. \citet{Karouzos2016} find that 60\% of the kinetic energy calculated using H$\alpha$ may be unrelated to the outflow. It is unclear if the combined effect of hydrogen tracers and a luminosity-based technique result in an overestimate or an underestimate of the mass outflow rate.

\citet{Harrison2014}, \citet{Brusa2015}, \citet{Karouzos2016}, and \citet{Muller-Sanchez2016} measure the mass outflow rates for moderate-luminosity AGN outflows and find averages of 10$^{1.4}$, 10$^{1.1}$, 10$^{-1.4}$, and 10$^{2.4}$ M$_{\odot}$ yr$^{-1}$, respectively. Again, the majority of these samples (\citealt{Harrison2014,Brusa2015,Karouzos2016}) use non-kinematic models. \citet{Muller-Sanchez2016} use the same kinematic technique as this work and therefore agrees most closely with the mass outflow rates estimated here. \citet{Harrison2014}, \citet{Brusa2015}, and \citet{Karouzos2016} use a spherical geometry and the luminosity of the H$\beta$, [OIII], and [OIII] emission lines, respectively. Again, a luminosity-based technique in combination with a hydrogen tracer as in \citet{Harrison2014} has an unknown effect on the estimate mass outflow rate. \citet{Karouzos2016} and \citet{Brusa2015} use a luminosity-based technique with [OIII] as a tracer. This significantly underestimates the mass outflow rate of the outflow. Additionally, \citet{Karouzos2016} use an electron density of 200-800 cm$^{-3}$, which further drives down the estimate of the mass outflow rate. We calculate a mass outflow rate using the [OIII] luminosities of the galaxies in our sample as in \citet{Karouzos2016} and find a rate that is a large underestimate of the mass outflow rate.

The AGNIFS group (\citealt{Storchi-Bergmann2010,Riffel2011a,Riffel2011b,Riffel2013,Riffel2015,Schonell2014}), \citet{Schnorr-Muller2016}, and \citet{Muller-Sanchez2011} find mass outflow rates for their samples of low-luminosity AGN outflows of 10$^{0.5}$, 10$^{-0.3}$, and 10$^{1.4}$ M$_{\odot}$ yr$^{-1}$, respectively. \citet{Schnorr-Muller2016} use a biconical geometry, a higher electron density of 1350 cm$^{-3}$, and H$\alpha$ tracers, which have an unknown combined effect on the estimate of the mass outflow rate. \citet{Muller-Sanchez2011} use lines with higher ionization potential in the CLR to trace a fast outflow in a higher ionization phase. They also find the same ionization cones for the NLR outflows. The AGNIFS group use an electron density of 500 cm$^{-3}$ and a similar biconical geometry to this work with HII gas to trace the outflow. This has an unknown effect on the mass outflow rate. 

Interestingly, some of the AGN outflows in other work with high mass outflow rates also have double-peaked narrow emission line profiles. For instance, 19\% of the galaxies in \citet{Harrison2014} and 41\% of the galaxies in \citet{McElroy2015} have double-peaked profiles. Additionally, \citet{Liu2013b} fit multiple Gaussian components to their [OIII]$\lambda$5007 profiles. \citet{Muller-Sanchez2016} also find significant velocity offsets in the emission lines of the three AGN outflows in their sample, so while these profiles cannot be characterized as double-peaked, they are also selected to be highly energetic outflows by selecting for a significant velocity offset in the spectral lines. While these AGN outflows are selected in a variety of ways, double-peaked or offset line profiles indicate a wide separation in velocity space and select for highly energetic outflows with large mass outflow rates as discussed in Section \ref{biases}.

Despite the potpourri of different estimation techniques and therefore the large scatter in mass outflow estimates, overall, the mass outflow rate of AGN outflows increases with AGN bolometric luminosity. Therefore, the ratio of L$_{\mathrm{KE}}$/L$_{\mathrm{bol}}$ remains constant over a large range of AGN luminosities. We fit a line to all the data and find a log-log slope of $\alpha$ = 0.50$\pm$0.12. When we fit a line to only the 18 galaxies from this work, we find a consistent slope of $\alpha$ = 0.47$\pm$0.23. We confirm that this relationship is statistically significant using the t-statistic. We find a p-value for this statistic of 0.05, which indicates that we can reject the null hypothesis that the slope is equal to zero at 95\% confidence.

This positive slope indicates a trend of increased mass outflow rate with increased AGN bolometric luminosities. An additional danger of using a non-kinematic AGN outflow model to measure mass outflow rate is that this creates an artificial positive correlation for this relationship if L$_{\mathrm{[OIII]}}$ is included in both the calculation of mass outflow rate and L$_{\mathrm{bol}}$. We repeat the slope measurement by excluding the two studies from the sample that use this technique and find a consistent slope of $\alpha$ = 0.50$\pm$0.10. Therefore, since only two studies included in Figure \ref{grandcompare} use [OIII] as a probe of mass outflow rate, the measured positive correlation is real and unrelated to artificial correlations from this technique.

Overall, while we find that the mass outflow rates of the galaxies in our sample are biased towards larger values (Section \ref{biases}), they are broadly consistent with other AGN outflows in the literature and follow the same trend of increased mass outflow rates with increased AGN bolometric luminosities. An implication of this trend is that lower luminosity AGNs still have the potential to exceed the critical value of the energy ratio required to expel gas. This is reflected in this work; we find that the majority of biconical outflows in our work, regardless of AGN bolometric luminosity, exceed the 0.5\% threshold.

\subsection{Selection, illumination, and obscuration effects explain the best fit models}
\label{discussmodels}
We find that 2/18 (11.1\%) galaxies are best modeled as a symmetric bicone, 8/18 (44.4\%) galaxies are best modeled as an asymmetric bicone, and 8/18 (44.4\%) galaxies are best modeled as a nested bicone. The relative percentages of best fit models as well as the nature of the bicone walls in each model can be explained by invoking a combination of obscuration, illumination, lack of gas, and selection effects. First, we discuss that very few biconical outflows are best fit by the symmetric bicone model. Second, we consider the nature of the observed walls in the asymmetric and nested models and the implications for the structure of the general bicone model.

Only two of the 18 total galaxies are best fit by a symmetric bicone model. The lack of symmetric bicone models in this work could be explained by a more general interpretation of biconical outflows as four-walled structures with two opening angles. Some studies model a biconical outflow using a four-wall filled structure where the outer and inner walls can be described by distinct opening angles (e.g., \citealt{Das2006,Muller-Sanchez2011,Crenshaw2015}). Our bicones may be better described with a four-walled evacuated structure with different amounts of gas and/or illumination on various walls. We first describe the evidence for an evacuated general model of a bicone in this work. Then, we discuss other work that provides evidence of illumination and/or obscuration effects that could explain the lack of symmetric bicone models in this work.

\citet{Das2006}, \citet{Muller-Sanchez2011}, and \citet{Crenshaw2015} model the bicone using an averaging process to approximate the velocity of material between the two different opening angle walls. In other words, they use a filled bicone structure. This is an acceptable approximation because the outer and inner opening angles seldom differ by more than $\sim$ 10-15$^{\circ}$, so the average velocity is always close to the velocity of material along the walls of the bicone. However, in our case, $\theta_{1,\mathrm{half}}$ and $\theta_{\mathrm{2,half}}$ often differ by $>$ 20$^{\circ}$, so this approximation is no longer valid. Instead, although the four-walled structure may exist, the selection bias that leads to large opening angle outflows and the double-peaked profiles that trace two walls necessitate a different overall model of a bicone. Therefore, we created individual models to describe bicones that have illuminated material along distinct opening angle walls. It is no longer adequate to describe a `filled bicone' structure by averaging the velocity of the two opening angles because the opening angles now differ considerably. 

The location of the filled material within the walls has never been investigated, although it has been hypothesized that it is very clumpy and sparsely distributed (e.g., \citealt{Nenkova2008,Mor2009}). Our models indicate that although the material still has a large velocity dispersion, the material is not evenly distributed between the two walls (\citealt{Nevin2016}). Instead, it seems to be clumped in distinct velocities around various walls of the more general four-walled structure. Additionally, since this sample includes large opening angle cones, the material is positioned at distinct lines of sight from the ionizing source (the AGN) and therefore can experience different amounts of illumination. These effects can lead to alternate bicone structures such as an asymmetric or nested bicone.

In many studies, a symmetric geometry has been preferred for kinematic modeling (e.g., \citealt{Das2006,Muller-Sanchez2011,Crenshaw2015}). However, other work indicates that the canonical symmetric biconical outflow structure may not be the best description for all AGN-driven outflows. For instance, \citet{Muller-Sanchez2012} find that an asymmetric bicone model is the best fit for the outflow in NGC 3081. \citet{Storchi-Bergmann2010} model NGC 4151 using a two-walled cone structure with weaker emission from the steeply inclined wall of the cone and stronger emission from the walls that are closer to the photometric major axis. This produces an asymmetric bicone structure if the wider opening angle posterior wall is brighter. \citet{Woo2016} use a larger sample of 39,000 Type 2 AGN outflows and find that the amount of dust extinction is a main driver of the observed velocity profile. They use models of outflows with obscuration effects in the photometric major axis of the galaxy and produce a large fraction of nested and asymmetric types of biconical outflows with asymmetric integrated line profiles.

The illuminated walls in the asymmetric and nested biconical outflows can be explained by selection effects, different amounts of illumination, lack of gas, and obscuration effects. First, by selecting for double-peaked narrow emission line profiles in the SDSS spectra, we are selecting for two walls of illumination at two distinct velocities in our bicone structure. Although the general structure of a symmetric biconical outflow (Figure \ref{anywall}) has four illuminated surfaces, we are only able to select for cones in which two of these are illuminated and/or not obscured. This selection effect does not eliminate the possibility of a structure inherently having more than two walls. For instance, the third and fourth walls could be much fainter due to illumination or obscuration effects. These lower flux components are much harder to detect and could be swamped out by the emission of the brighter walls. We are therefore limited to a model of a two-walled structure due to the flux limit of the longslit data.

Second, we can analyze the best fit models to make conclusions about the obscuration of walls in our biconical outflows. Since we observe only nested cones with blueshifted walls (we observe no nested cones with two redshifted walls), this indicates that obscuration plays a key role in the observed walls of the bicone. Obscuring dust in the disk of the galaxy could be preferentially allowing us to observe the anterior walls while obscuring the two posterior walls. While it is difficult to distinguish between this obscuration scenario and a complete absence of the posterior two-walled structure associated with the general structure of a bicone (four total walls), we find evidence that the obscuration scenario is most likely. If the absence of a side of the bicone is the better explanation, then we would expect to see equal numbers of redshifted and blueshifted nested bicones. Since we see only blueshifted bicones, obscuration is a likelier explanation than preferential illumination or lack of structure in one direction.

Third, we observe asymmetric bicones with a preferential orientation. The posterior cone always has a larger opening angle; we observe this for all eight asymmetric bicones. We can rule out obscuration effects as solely responsible for this observation. If obscuration effects were involved, we would expect to see the anterior wider opening angle wall and not the posterior wider opening angle wall. 

Illumination effects explain the relative brightnesses of the velocity components of our biconical outflows but they are not the complete explanation. We expect the material in walls with lower inclinations relative to the galaxy to be brighter. \citet{Storchi-Bergmann2010} witness this illumination effect in the outflow in NGC 4151 and hypothesize that the bicone walls with high inclination relative to the line of sight are fainter. By examining the relative fluxes of the two emission components in our asymmetric bicones, we find that the majority (7/8) have a brighter integrated redshifted component. Additionally, 7/8 of the nested bicones have a brighter velocity component nearer to zero velocity (the wall with a lower inclination relative to the line of sight). Therefore, since the blueshifted wider opening angle wall has a low inclination relative to the line of sight, it is expected to be very bright. However, since we do not observe this to be the case, different illumination of various walls relative to the line of sight is not the full explanation for the lack of a blueshifted wider opening angle wall. 

We can also rule out a lack of material in this blueshifted wall as the sole explanation since it is unlikely that the material is preferentially more clumpy on the side of the galaxy facing us. Therefore, the most probable explanation is a combination of an obscuration effect, an illumination effect, a lack of material effect, and a selection effect. First, the lower inclination walls are the brightest. Second, the faint high inclination redshifted wall is most likely to be totally obscured by dust in the plane of the galaxy. Third, we select for galaxies with only two velocity peaks. This explains why we observe equal numbers of asymmetric and nested bicones. Fourth and finally, the anterior and posterior low inclination walls must be absent to observe both nested and asymmetric bicones.

 Overall, by selecting for double-peaked narrow emission line profiles, we eliminated the possibility of observing more than two distinct velocity peaks corresponding to more than two walls. The presence of various walls of a general four-walled structure can be explained by obscuration effects, illumination effects, and/or lack of material in various walls.

\subsection{The outflows have random orientations}
\label{discussalignment}

We determine that 4/18 (22.2\%) of the galaxies have a bicone axis that is aligned with the photometric major axis of the galaxy in Table \ref{Alignment}. The bicones are randomly oriented with respect to the photometric major axis of the galaxy. We also determine that 100\% of galaxies have a bicone structure that intersects the photometric major axis of the galaxy; this measurement takes into account the large half opening angles of the bicones, which is on average 68$^{\circ}$. A bicone with a half opening angle of 68$^{\circ}$ will cover 272$^{\circ}$, which is 76\% of the plane of the sky. Therefore, the 100\% intersection percentage is unsurprising and we discuss its effect on feedback in Section \ref{energydiscuss}. Here we focus instead on the implications of the orientations of the bicones in this sample both in terms of previous observations and the theory of a collimating torus.

Observations of biconical NLRs necessitate an optically thick, collimating torus that exists at parsec scales (e.g., \citealt{Antonucci1985,Mulchaey1996}), but the orientation and structure of this theoretical torus remain uncertain. If Type 1 and Type 2 AGNs are to be explained by the orientation of the torus alone, the relative fraction of observed Type 1 and Type 2 AGNs require that the torus be geometrically thick ($H/R \sim 1$, $H$ is the height and $R$ is the radius of the torus), covering an angle of 65$^{\circ}$ as seen from the central source (\citealt{Risaliti1999}). However, theory has shown that it is difficult to maintain a geometrically thick cold rotating structure even if the torus is clumpier (e.g., \citealt{Krolik1988,Krolik2007}). Alternately, \citet{Ramos2011} fit a model of a clumpy torus to the spectral energy distributions (SEDs) of seven Seyfert galaxies and find that the torus has no preferential orientation with respect to Seyfert 1s and Seyfert 2s. Additionally, they find that a clumpy, or somewhat transparent torus is the only explanation for the observations of a BLR in NGC 7469, which has an edge-on torus. The parsec extent of the torus makes it difficult to resolve, but we can probe both the degree of clumpiness and the orientation of the torus using the alignment and opening angles of our large-scale outflows.

Observational and theoretical work has found a range of NLR outflow alignments; they are not preferentially aligned with the photometric major axis, aligned with the photometric major axis, and aligned perpendicular to the photometric major axis. For instance, some theoretical work has shown that AGN-driven outflows tend to follow the `path of least resistance', emerging perpendicular to the photometric major axis of the host galaxy (e.g., \citealt{Gabor2014a}). However, other work finds no preferential orientation. \citet{Muller-Sanchez2011} and \citet{Fischer2013} find no alignment between the inclinations of a sample of Seyfert galaxies with biconical outflows and the photometric major axes of their host galaxies. Other work finds equatorial outflows that are aligned parallel to the photometric major axis of the host galaxy (e.g., \citealt{Elitzur2006,Riffel2014,Ricci2015}). 

Our findings agree with \citet{Muller-Sanchez2011} and \citet{Fischer2013}, which is one of the only other large statistical studies of AGN outflows. A randomly oriented bicone structure has important implications for the theory of a collimating torus. If the torus is fully collimating and optically thick, this implies that the torus is randomly oriented with respect to the photometric major axis of the galaxy. We cannot rule out this possibility. However, since our biconical outflows have such large opening angles, this would imply a very wide opening angle for the thick molecular torus. The widest half opening angle is 82$^{\circ}$ which implies $H/R \sim 1/7$. This is an unphysically thin torus or a torus with an unphysically large radius, so this implies that it is more likely that the torus is clumpy as opposed to thin. 

\subsection{Type 1 vs Type 2 AGNs}
In addition to the orientation of the bicones with relation to the plane of the host galaxies, we also discuss the orientation of the bicone structures with relation to the line of sight (LOS) and the implications for the unification of Type 1 and Type 2 AGNs. Since the 18 bicones modeled in this sample have large inclinations and opening angles (Section \ref{biases}), 100\% satisfy the condition $\lvert{}i\rvert{}$+$\theta_{1,\mathrm{half}} > 90^{\circ}$. We choose $\theta_{1,\mathrm{half}}$ as opposed to the larger $\theta_{2,\mathrm{half}}$ because the walls of the smaller inner cone more tightly constrain the LOS to the BLR. If the classification of Type 1 and Type 2 AGNs depends only on the orientation of the collimating torus, and therefore the bicone inclination and opening angle, this would imply that these are all Type 1 objects with a direct view to the BLR.  However, these objects are all classified as Type 2 AGNs in SDSS and have no observed BLRs. We discuss three possible explanations for this apparent discrepancy. 

If we account for the large errors on the half opening angles and inclinations measured for these bicones (the uncertainties of the parameters are discussed in Section \ref{MCMC}), within a 3$\sigma$ error margin, 50\% of the bicones are consistent with $\lvert{}i\rvert{}$+$\theta_{1,\mathrm{half}} < 90^{\circ}$, and therefore the bicone and/or collimating torus obscure the BLR from view. The entire sample is consistent with $\lvert{}i\rvert{}$+$\theta_{1,\mathrm{half}} < 106^{\circ}$, which means that the bicone walls are consistent with being within $\sim10^{\circ}$ of the LOS (90$^{\circ}$) for the entire sample. Since we assume optically thick walls for the bicone that have a finite thickness, a LOS along the edge of a wall could obscure the BLR. It is realistic to assume that both of these parameters ($i$ and $\theta_{1,\mathrm{half}}$) are overestimated in the modeling for 76\% and 60\% of galaxies, respectively, due to asymmetric error bars. The lower limit is larger for the inclination for 76\% of the 18 galaxies and 60\% of the inner opening angle. However, it is unrealistic to assume that the large error bars are solely responsible for the lack of visible BLRs for all of the galaxies, so we turn towards physical explanations.

The classification of Type 1 or Type 2 objects may depend more on the intrinsic properties of the torus rather than on orientation effects related just to inclination and opening angle. \citet{Ramos2009} do not find a clear trend in the inclination of the torus with Seyfert 1s and Seyfert 2s with their SED models and find that the intrinsic properties of the torus for Type 1 and Type 2 AGNs may be different. \citet{Ramos2011} find that the Type 2 torii in their sample have larger geometric covering factors, and therefore a smaller probability of having a direct view of the BLR. They also find that Type 2 torii are intrinsically clumpier with a higher density of clouds closer to the nucleus. The clumpy torus model is supported by other work (e.g., \citealt{Krolik1988,Krolik2007}) as well as our finding that a clumpy torus is required for the large opening angles of the bicones in this work. With a clumpy torus model, there is a finite possibility of seeing the central source at any inclination through the clumpy material (\citealt{Netzer2015}). As a result, some Type 1 AGNs might have high inclination angles (edge-on line of sight to the central source), while obscuration by a large cloud could lead to a Type 2 AGN classification (e.g., \citealt{Ramos2011}).  

Some work finds that different areas of the torus require different physical models for their structure. \citet{Davies2015} find that the inner boundary of the geometrically thick torus may be decoupled from the outer regions. The region that provides collimation of the outflow may be distinct from the region that allows a direct LOS to the BLR. This is similar to the findings of \citet{Ramos2011} that there is a higher density of clouds towards the center of the torus structure.

Finally, obscuration of the BLR can occur at a variety of spatial scales. The obscuration of the BLR need not come only from pc-scale structures such as the torus. \citet{Bianchi2012} and references therein discuss that obscuration on $\sim$100 pc scales in a host galaxy can contribute to AGN obscuration. For instance, optically selected AGN samples are biased against edge-on galaxies due to dust in the plane of the galaxy (e.g., \citealt{Maiolino1995,Lagos2011}). Additionally, interferometric maps of molecular gas show evidence of a large amount of dense gas in the 100 pc regions surrounding the AGN (e.g., \citealt{Schinnerer2000,Boone2011,Krips2011}). Other work has confirmed larger-scale obscuring structures that are aligned with the host galaxy plane (e.g., \citealt{Gelbord2004}).

While it is beyond the scope of this work to fully explore AGN unification by delving into physical modeling of the torus, BLR variability, and/or intrinsic Type 2 AGNs, it is apparent that the large half opening angles and inclinations of the biconical outflows in the sample are not consistent with a thick fully collimating torus alone (e.g., \citealt{Antonucci1985}). Instead, we find that a combination of large uncertainties on our modeled parameters as well as physical structures (a clumpier torus with obscuring material on a variety of spatial scales) can explain why our galaxies are classified as Type 2 AGNs.

\begin{deluxetable*}{lcccccc}

\tabletypesize{\scriptsize}
\tablewidth{0pt}
\tablecolumns{7}
\tablecaption{Quenched Galaxies - A Matched Sample}
\tablehead{
\colhead{SDSS ID} & 
\colhead{Num. in matched control sample} &
\colhead{$z$} &
\colhead{log M$_{*}$} & 
\colhead{L$_{\mathrm{bol}}$} & 
\colhead{$g$-$r$ color} & 
\colhead{log sSFR} \\
& & & [M$_{\odot}$]& [10$^{42} $erg s$^{-1}$]& [Magnitude]&[ yr$^{-1}$ ]
}

\startdata

J0821+5021
& & $ 0.095 $ & $ 10.96 $ & $ 130 \pm 21 $ & $ 0.95  $ Consistent  &  $ -10.6 $ Consistent  \\  & 131 & $ 0.096 \pm 0.011 $ & $ 10.94 \pm 0.05 $ & $ 98 \pm 18 $ & $ 0.88 \substack{+ 0.09 \\- 0.10 }$ & $ -10.5 \substack{+ 10.2 \\- 10.7 }$ \\ 
J0854+5026
& & $ 0.096 $ & $ 10.91 $ & $ 69 \pm 11 $ & $ 0.86  $ Consistent  &  $ -10.4 $ Consistent  \\  & 254 & $ 0.096 \pm 0.011 $ & $ 10.89 \pm 0.05 $ & $ 63 \pm 15 $ & $ 0.85 \substack{+ 0.10 \\- 0.09 }$ & $ -10.5 \substack{+ 10.4 \\- 10.6 }$ \\ 
J0930+3430
& & $ 0.061 $ & $ 10.74 $ & $ 110 \pm 32 $ & $ 0.92  $ Redder  &  $ -11.0 $ Consistent  \\  & 54 & $ 0.064 \pm 0.006 $ & $ 10.73 \pm 0.05 $ & $ 93 \pm 22 $ & $ 0.78 \substack{+ 0.12 \\- 0.09 }$ & $ -10.5 \substack{+ 10.3 \\- 10.7 }$ \\ 
J0959+2619
& & $ 0.051 $ & $ 10.58 $ & $ 89 \pm 10 $ & $ 0.74  $ Consistent  &  $ -10.6 $ Consistent  \\  & 29 & $ 0.049 \pm 0.005 $ & $ 10.56 \pm 0.06 $ & $ 73 \pm 18 $ & $ 0.71 \substack{+ 0.12 \\- 0.10 }$ & $ -10.2 \substack{+ 10.1 \\- 10.5 }$ \\ 
J1027+1049
& & $ 0.065 $ & $ 10.29 $ & $ 88 \pm 14 $ & $ 0.74  $ Consistent  &  $ -10.1 $ Consistent  \\  & 39 & $ 0.067 \pm 0.007 $ & $ 10.32 \pm 0.05 $ & $ 70 \pm 17 $ & $ 0.75 \substack{+ 0.08 \\- 0.13 }$ & $ -10.2 \substack{+ 10.1 \\- 10.4 }$ \\ 
J1109+0201
& & $ 0.063 $ & $ 11.00 $ & $ 106 \pm 8 $ & $ 0.86  $ Consistent  &  $ -11.6 $ Quenched  \\  & 59 & $ 0.066 \pm 0.007 $ & $ 10.98 \pm 0.04 $ & $ 97 \pm 22 $ & $ 0.83 \substack{+ 0.09 \\- 0.08 }$ & $ -10.8 \substack{+ 10.4 \\- 10.9 }$ \\ 
J1152+1903
& & $ 0.097 $ & $ 11.08 $ & $ 728 \pm 170 $ & $ 0.98  $ Consistent  &  $ -11.1 $ Consistent  \\  & 44 & $ 0.101 \pm 0.011 $ & $ 11.04 \pm 0.04 $ & $ 589 \pm 145 $ & $ 0.88 \substack{+ 0.12 \\- 0.08 }$ & $ -10.5 \substack{+ 10.0 \\- 10.6 }$ \\ 
J1315+2134
& & $ 0.07 $ & $ 10.95 $ & $ 892 \pm 75 $ & $ 0.81  $ Consistent  &  $ -10.1 $ Consistent  \\  & 21 & $ 0.071 \pm 0.008 $ & $ 10.95 \pm 0.05 $ & $ 749 \pm 196 $ & $ 0.8 \substack{+ 0.08 \\- 0.12 }$ & $ -10.4 \substack{+ 10.0 \\- 10.6 }$ \\ 
J1328+2752
& & $ 0.091 $ & $ 10.76 $ & $ 84 \pm 14 $ & $ 0.83  $ Consistent  &  $ -10.7 $ Consistent  \\  & 228 & $ 0.092 \pm 0.011 $ & $ 10.75 \pm 0.05 $ & $ 72 \pm 18 $ & $ 0.83 \substack{+ 0.10 \\- 0.10 }$ & $ -10.3 \substack{+ 10.3 \\- 10.5 }$ \\ 
J1420+4959
& & $ 0.063 $ & $ 10.62 $ & $ 125 \pm 39 $ & $ 0.81  $ Consistent  &  $ -10.6 $ Consistent  \\  & 41 & $ 0.065 \pm 0.007 $ & $ 10.61 \pm 0.05 $ & $ 93 \pm 18 $ & $ 0.74 \substack{+ 0.10 \\- 0.08 }$ & $ -10.4 \substack{+ 10.4 \\- 10.6 }$ \\ 
J1524+2743
& & $ 0.068 $ & $ 10.99 $ & $ 128 \pm 11 $ & $ 0.83  $ Consistent  &  $ -10.5 $ Consistent  \\  & 62 & $ 0.071 \pm 0.007 $ & $ 10.97 \pm 0.04 $ & $ 99 \pm 18 $ & $ 0.82 \substack{+ 0.09 \\- 0.06 }$ & $ -10.8 \substack{+ 10.3 \\- 10.9 }$ \\ 
J1606+3427
& & $ 0.055 $ & $ 10.51 $ & $ 22 \pm 6 $ & $ 0.90  $ Redder  &  $ -11.8 $ Quenched  \\  & 95 & $ 0.057 \pm 0.006 $ & $ 10.50 \pm 0.05 $ & $ 20 \pm 5 $ & $ 0.78 \substack{+ 0.08 \\- 0.09 }$ & $ -10.6 \substack{+ 10.4 \\- 10.8 }$ \\ 
J1630+1649
& & $ 0.034 $ & $ 10.40 $ & $ 47 \pm 3 $ & $ 0.71  $ Consistent  &  $ -10.7 $ Consistent  \\  & 13 & $ 0.034 \pm 0.003 $ & $ 10.39 \pm 0.06 $ & $ 36 \pm 11 $ & $ 0.75 \substack{+ 0.08 \\- 0.08 }$ & $ -10.5 \substack{+ 10.5 \\- 10.7 }$ 
\enddata

\tablecomments{Column 1: galaxy name. Column 2: number of galaxies in the matched sample. Column 3: redshift of the galaxy, and mean redshift and standard deviation of the matched sample of galaxies. Column 4: stellar mass of the galaxy, and mean stellar mass and standard deviation of the matched sample of galaxies. Column 5: AGN bolometric luminosity and error, and mean AGN bolometric luminosity and standard deviation of the matched sample of AGNs. Column 6: color index, and mean color index and standard deviation for the matched sample. If the galaxy's $g-r$ color is $>1\sigma$ larger than the mean color of the matched sample, we label the galaxy `Redder', if it is consistent, the galaxy is `Consistent', and if it $>1\sigma$ smaller than the mean color of the matched sample, the galaxy is `Bluer'. Column 7: specific star formation rate, and mean specific star formation rate and standard deviation for the matched sample. If galaxy's sSFR is $>1\sigma$ smaller than the mean sSFR of the matched sample, we label the galaxy `Quenched', if it is consistent, the galaxy is `Consistent', and if it is $>1\sigma$ larger than the mean sSFR of the matched sample, the galaxy is `Enhanced'. } 
\label{reddead} 
\end{deluxetable*}

\begin{figure*}
\center
\includegraphics[scale=0.75]{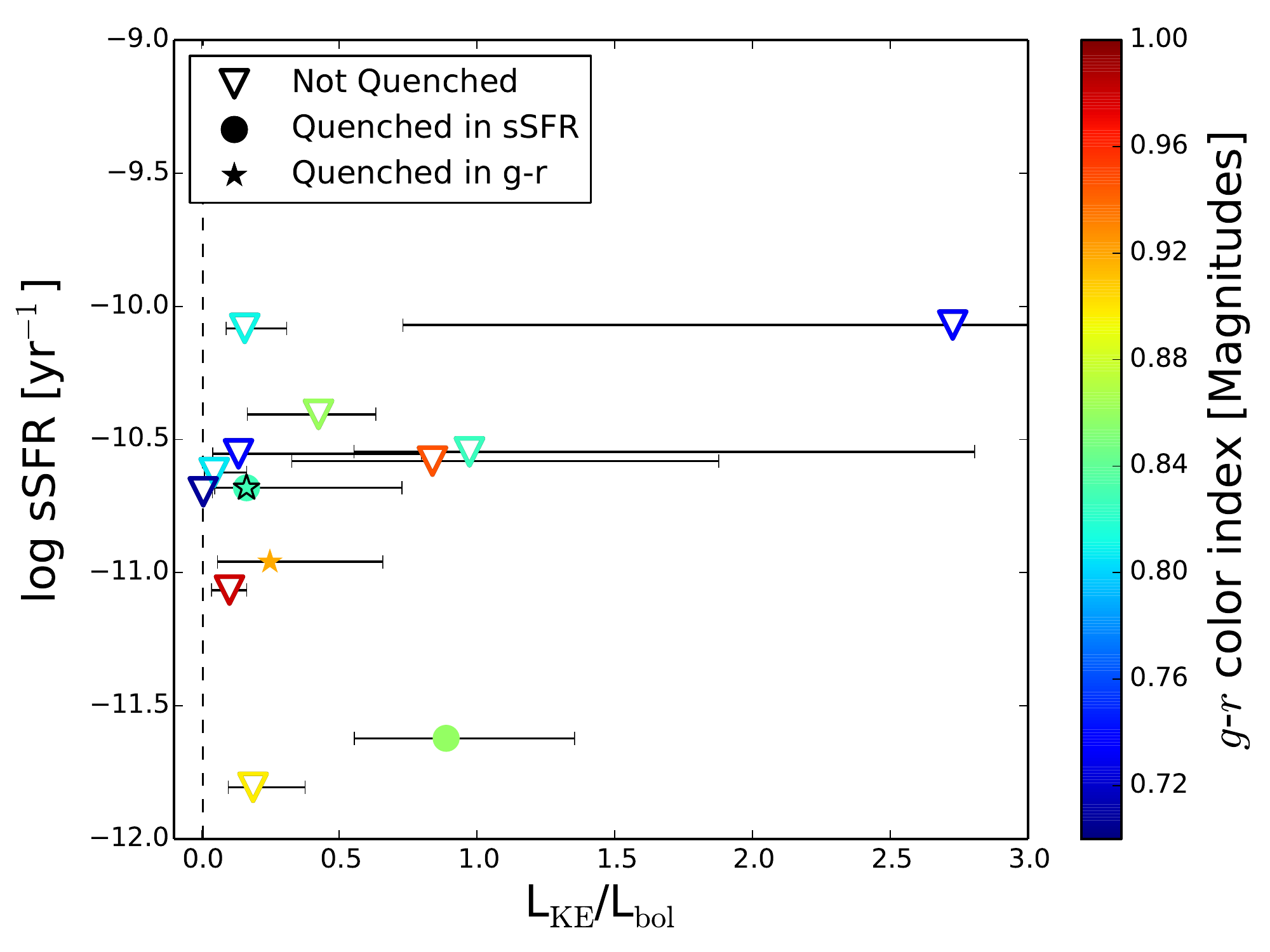} 
\caption{sSFR compared to kinetic to AGN bolometric luminosity ratio for 13 outflow host galaxies that have $>$10 comparison galaxies matched in stellar mass, redshift, and AGN bolometric luminosity (Table \ref{reddead}). Filled stars denote galaxies that are quenched according to $g-r$ color, while filled circles denote galaxies that are quenched according to sSFR. Open triangles represent galaxies that are not quenched. The colorbar is the $g-r$ color index. The vertical dashed line is the 0.5\% energy threshold from \citet{Hopkins2010a} that is required to drive a two-stage feedback process.}
\label{reddeadplot}
\end{figure*}

\subsection{The outflows are energetic enough to drive feedback}
\label{energydiscuss}

We find that 16/18 of the biconical outflows in this sample are above the 0.5\% energy threshold necessary to drive a two-staged feedback process in their host galaxy (Section \ref{energetics}). To investigate the implications of this further, we first discuss their outflow geometry and how they may interact with the ISM, and then we discuss indications of positive or negative feedback in the host galaxies.

As discussed in Section \ref{discussalignment}, 100\% of the galaxies have a biconical structure that intersects the photometric major axis of the host galaxy. This has important implications for outflow interactions with the ISM and feedback in the host galaxy. Since these outflows have the geometrical alignment to interact with the gas in the galactic disk, the \citet{Hopkins2010a} threshold has more physical implications. To determine if these outflows actually affect star formation in the host galaxies, we examine the star formation rates and colors of the host galaxies.

To place these galaxies in the context of AGNs in the local universe, we compare each outflow host galaxy to a control sample of SDSS galaxies that are matched in stellar mass, redshift, and AGN bolometric luminosity using the MPA-JHU (\citealt{Kauffmann2003}) and OSSY (\citealt{Oh2011}) value-added catalogs. To build statistically significant ($>$10) control samples for most of our AGNs, we use thresholds of 20\%, 20\% and 50\%, respectively, for stellar mass, redshift, and AGN bolometric luminosity matching. We remove four outflow host galaxies (J0009-0036, J0803+3926, J1352+0525, and J1526+4140) from this analysis due to matching control samples that were too small ($<$10 matches) and one galaxy (J1720+3106), for which the specific star formation rate (sSFR) could not be measured, due to artifacts in the SDSS spectrum. When we adjust the comparison thresholds by 10\% in either direction to test for consistency with changing sample sizes of the comparison sample, we find the same results. Only when we increase all thresholds by more than 10\% do we find that a single galaxy changes its classification from not quenched to quenched. We list the 13 galaxies that we use for the matched sample comparison of sSFR and $g$-$r$ color index in Table \ref{reddead}.

We define a galaxy as quenched if its $g-r$ color is more than one standard deviation above the mean $g-r$ color of the comparison sample, and/or if its sSFR is more than one standard deviation below the mean sSFR of the comparison sample. We find that 3/13 (23.1\%) galaxies are quenched according to either $g-r$ color or sSFR. One of these galaxies is also quenched according to both of these criteria. Of the 13 galaxies, 12 have an energy ratio that exceeds 0.5\%, so 3/12 of the galaxies with an energy ratio that exceeds the threshold value are quenched in this sample. We plot the energy ratio and sSFR of these 13 galaxies in Figure \ref{reddeadplot} and find that there is no correlation between energy ratio and quenching. Nine galaxies have a high energy ratio but have colors and sSFRs that are consistent (within 1$\sigma$) with those of the comparison samples. Importantly, none of the galaxies in this sample have enhanced sSFR or bluer color indices relative to their comparison samples. Therefore, our analysis favors the negative feedback scenario over the positive feedback scenario for these galaxies. 

For comparison, \citet{Wylezalek2016} find a negative correlation between outflow strength and sSFR for a sample of 132 AGN. This is consistent with AGN feedback operating on the host galaxies in the sample, and is most apparent for galaxies that are gas rich with high SFRs. None of the galaxies in this work have a SFR $>100$ M$_{\odot}$ yr$^{-1}$, which is the cutoff for the galaxies with the most detectable negative correlation in \citet{Wylezalek2016}. Since the galaxies in this work have less star formation and therefore less gas to couple with the AGN outflow, the effects of feedback may be less pronounced for the galaxies in this sample. 

Overall, we find preliminary evidence for negative feedback in the galaxies in this sample. However, to fully confirm the negative feedback in these galaxies, we require detailed star formation histories or maps of the gas, e.g., with ALMA.

\section{Conclusion}
\label{conclude}
We model 18 SDSS galaxies with double-peaked narrow emission lines as AGN-driven biconical outflows using three models: a symmetric bicone, an asymmetric bicone, and a nested bicone. We find that 8/18 are best fit as asymmetric bicones, 8/18 are best fit as nested bicones, and 2/18 are best fit as symmetric bicones. These results inform us that obscuration, illumination, and our sample selection of double-peaked NLR profiles dictate the type of bicone structure observed. The results of the analytic modeling also yield the geometry and energetics associated with the ionized outflows. Based upon these results, we find that:
\begin{enumerate}
\item Our bicones have large opening angles (average $\theta_{1,\mathrm{half}}$ = 44.5$^{\circ}$ and average $\theta_{2,\mathrm{half}}$ = 69.5$^{\circ}$), large turnover radii (average 3.4 kpc), and fast intrinsic velocities (average 370 km s$^{-1}$). Since these galaxies have double-peaked narrow emission lines, they are biased to have larger opening angles than similar moderate-luminosity AGN outflows, and as a result they also have larger kinetic energies.

\item Using the geometry of the bicone structures, we find that the bicone axes have random orientations with respect to the photometric major axis of the galaxy. This implies that the torus responsible for producing the bicone structure is randomly oriented and clumpy, where clumpiness enables radiation to escape along the plane of the torus.

\item We find that 16/18 (88.9\%) of galaxies exceed the kinetic luminosity to AGN bolometric luminosity threshold value from \citet{Hopkins2010a} of 0.5\%, which means that they have the potential to drive a two-staged feedback process.

\item Of the outflows that exceed the 0.5\% energy threshold, 100\% intersect the photometric major axis of the galaxy on kpc-scales coincident with the location of circumnuclear star formation. They have the potential to directly deliver energy to the ISM of the galaxy.

\end{enumerate}

While we can make tentative conclusions that these galaxies are quenched and thus potentially experiencing negative feedback as a result of the AGN-driven outflows, we cannot make any definitive conclusions without observations of the molecular gas. In the future, we could pursue ALMA as a means to observe the direct effect of feedback in these galaxies.

The sample of AGN outflows in this work demonstrates that moderate luminosity AGNs have the potential to drive feedback in their host galaxies. Since moderate luminosity AGNs are common in the local universe (10\% of the AGN population, whereas high-luminosity AGNs are only 1\% of the population), this indicates that they may play an important role in driving galaxy-SMBH co-evolution as well.

\section*{Acknowledgements}

R.N. is supported by an National Science Foundation (NSF) Graduate Research Fellowship Program (GRFP) Fellowship. The observations reported here were obtained at the Apache Point Observatory 3.5m telescope, which is owned and operated by the Astrophysical Research Consortium; the Hale Telescope, Palomar Observatory as part of a continuing collaboration between the California Institute of Technology, NASA/JPL, and Cornell University; the MMT Observatory, a joint facility of the University of Arizona and the Smithsonian Institution; and the W.M. Keck Observatory, which is operated as a scientific partnership among the California Institute of Technology, the University of California and the National Aeronautics and Space Administration. The W.M. Keck Observatory was made possible by the generous financial support of the W.M. Keck Foundation. 

The authors wish to recognize and acknowledge the very significant cultural role and reverence that the summit of Mauna Kea has always had within the indigenous Hawaiian community. We are most fortunate to have the opportunity to conduct observations from this mountain.

This work utilized the Janus supercomputer, which is supported by the National Science Foundation (award number CNS-0821794) and the University of Colorado Boulder. The Janus supercomputer is a joint effort of the University of Colorado Boulder, the University of Colorado Denver and the National Center for Atmospheric Research.

R.N. thanks Brian Zahartos, Jordan Mirocha, Evan Tilton, and Allison Youngblood for their help with the MCMC Hammer.

{\it Facilities:} APO (Dual Imaging spectrograph), Keck (DEep Imaging spectrograph), MMT (Blue Channel spectrograph), Palomar (Double spectrograph)

Funding for the Sloan Digital Sky Survey IV has been provided by the Alfred P. Sloan Foundation, the U.S. Department of Energy Office of Science, and the Participating Institutions. SDSS-IV acknowledges
support and resources from the Center for High-Performance Computing at
the University of Utah. The SDSS web site is www.sdss.org.

SDSS-IV is managed by the Astrophysical Research Consortium for the 
Participating Institutions of the SDSS Collaboration including the 
Brazilian Participation Group, the Carnegie Institution for Science, 
Carnegie Mellon University, the Chilean Participation Group, the French Participation Group, Harvard-Smithsonian Center for Astrophysics, 
Instituto de Astrof\'isica de Canarias, The Johns Hopkins University, 
Kavli Institute for the Physics and Mathematics of the Universe (IPMU) / 
University of Tokyo, Lawrence Berkeley National Laboratory, 
Leibniz Institut f\"ur Astrophysik Potsdam (AIP),  
Max-Planck-Institut f\"ur Astronomie (MPIA Heidelberg), 
Max-Planck-Institut f\"ur Astrophysik (MPA Garching), 
Max-Planck-Institut f\"ur Extraterrestrische Physik (MPE), 
National Astronomical Observatories of China, New Mexico State University, 
New York University, University of Notre Dame, 
Observat\'ario Nacional / MCTI, The Ohio State University, 
Pennsylvania State University, Shanghai Astronomical Observatory, 
United Kingdom Participation Group,
Universidad Nacional Aut\'onoma de M\'exico, University of Arizona, 
University of Colorado Boulder, University of Oxford, University of Portsmouth, 
University of Utah, University of Virginia, University of Washington, University of Wisconsin, 
Vanderbilt University, and Yale University.

\bibliographystyle{apj}
\bibliography{library6}

\appendix
\section{Investigating Outflow vs Rotational Kinematics}
\label{APa}

In this Appendix we discuss the outflow-dominated kinematics of this sample. We compare these AGN outflows to other galaxies that demonstrate rotation-dominated kinematics on large scales (e.g., \citealt{Fischer2017}) to explain why we are motivated to model only outflowing components for our sample. Additionally, we present a case study of a galaxy from \citet{Nevin2016} that was not included in this work. This galaxy offers insight into rotational and outflow kinematics and sheds light on the nature of the sample of 18 galaxies in this work.

We first discuss the previous work on the full sample of 71 AGNs in \citet{Nevin2016} as well as other general properties of this sample of AGNs that lead to our conclusion that the 18 galaxies in this work are dominated by outflow  kinematics, not large-scale rotation. In \citet{Nevin2016}, we kinematically classify the full sample of 71 double-peaked AGNs as either outflow-dominated or rotation-dominated. The outflow-dominated AGNs are further classified as outflow or outflow composite. Outflow composite AGNs are best fit with $>$ 2 kinematic components and outflow AGNs are best fit with 2 kinematic components. The outflow-dominated classification requires that one of the kinematic components have a velocity dispersion $\sigma > $ 500 km s$^{-1}$ or a velocity offset $v >$ 400 km s$^{-1}$ for at least one spatial position. While outflows can also have lower velocity dispersions or velocity offsets, these cutoffs are meant to conservatively eliminate rotation-dominated kinematics as the origin of the kinematic components.

For this work, we carefully select 18 of the outflow-dominated galaxies from \citet{Nevin2016} with the best quality data to model as biconical outflows. We describe this selection in Section \ref{selection}. The classification in \citet{Nevin2016} previously identified these 18 AGNs as those with outflow-dominated kinematics apparent at one or more spatial positions in the galaxy. The selection in this work further required that these AGNs have double-peaked profiles consistent with a bicone at all of the modeled spatial positions. Therefore, we are selecting for AGN outflows to model that have outflow-dominated kinematics and that are double-peaked on large spatial scales. 

Additionally, the double-peaked nature of the [OIII]$\lambda$5007 profiles at all spatial positions indicates that the 18 galaxies in this work are indeed dominated by outflow kinematics because rotation cannot explain double-peaked profiles at all spatial scales. This was one of the main outcomes of \citet{Nevin2016}. Unlike in some modeled Seyfert galaxies, where the profiles are double peaked towards the center, but single-peaked and rotation-dominated at spatial extremes (e.g., \citealt{Fischer2017}), we found profiles for the 18 galaxies that were double-peaked at all spatial positions.

The 18 galaxies are outflow-dominated and not rotation-dominated, but as another test we examine how well a rotating structure can fit the gas velocity in each galaxy. We create spatially-resolved velocity maps for each galaxy by superimposing each of the two observed longslit PAs and fitting a single Gaussian peak to the [OIII]$\lambda$5007 profile at each position along each slit. We determine whether the spatially-resolved velocity map for each galaxy is rotation-dominated using two criteria: The velocity map must be centered on zero velocity (systemic velocity for the galaxy) and it must demonstrate symmetric rotation, where if one extreme of an observed PA is redshifted, the other must be blueshifted by the same amount. We find that while three AGNs potentially demonstrate disk-like rotation on large spatial scales (J0009$-$0036, J0803+3926, and J1152+1903), the other 15 do not have large-scale rotation. 

These results are unsurprising given Figure \ref{sdss ref}, which plots the double-peaked [OIII]$\lambda$5007 profile of each galaxy from the SDSS spectra. If these galaxies were entirely dominated by a disk on a large scale and if this disk has a bright integrated flux, we would expect the integrated profiles to be centered on zero velocity. However, that is not the case for most galaxies in our sample and only two galaxies (J0803+3926 and J1152+1903) are classified and modeled as a `Symmetric Bicone' because their double-peaked velocity profiles are centered at zero velocity. These two galaxies are also two of the galaxies that show potential disk-like rotation in the spatially-resolved velocity maps. The third galaxy with disk-like rotation is J0009$-$0036, which is best described as an `Asymmetric Bicone', but it does have velocities that are centered around zero velocity. 

\begin{figure*}
\hspace{-0.2in}
\includegraphics[scale=0.65]{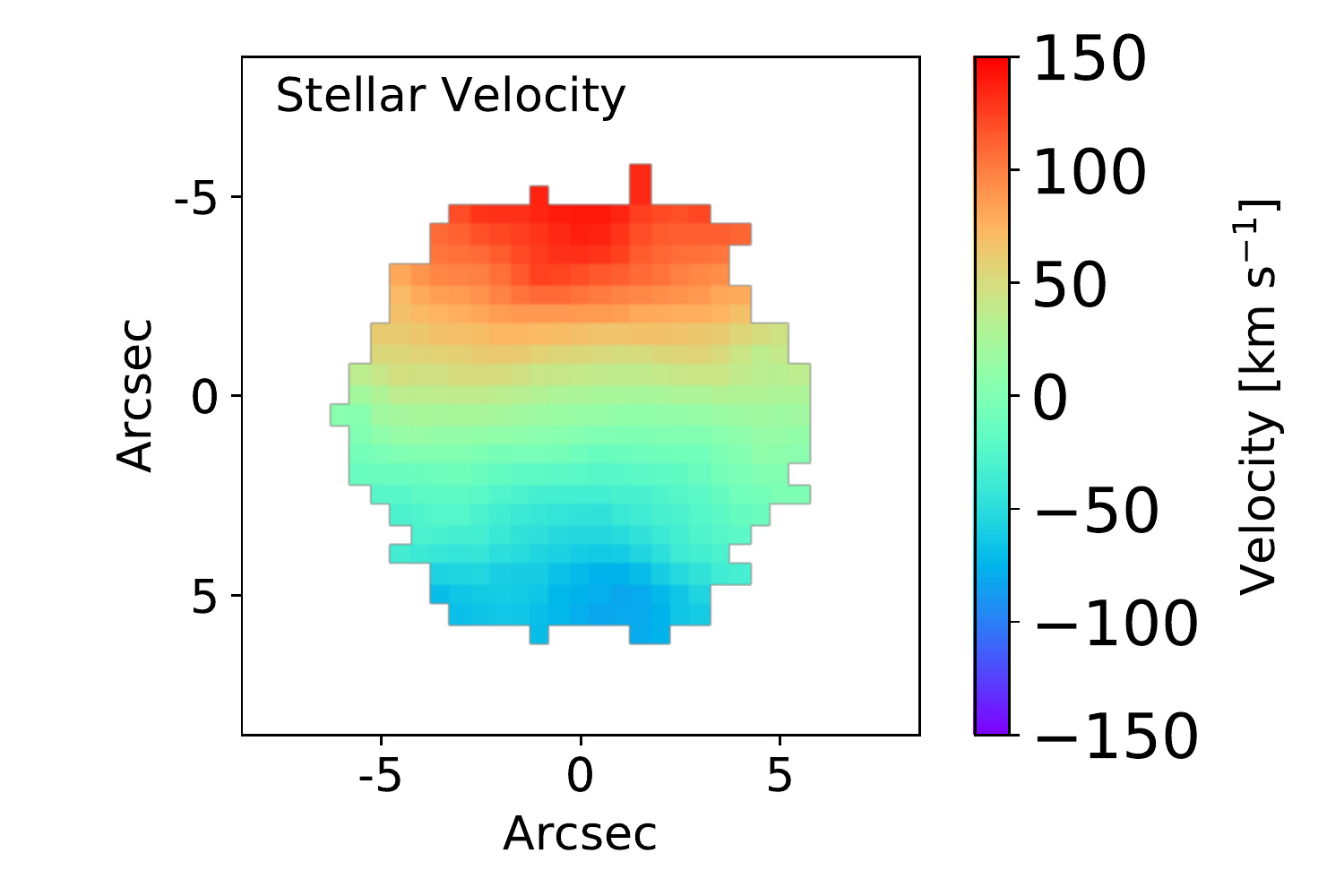}
 \hspace{-0.47in}
  \includegraphics[scale=0.65]{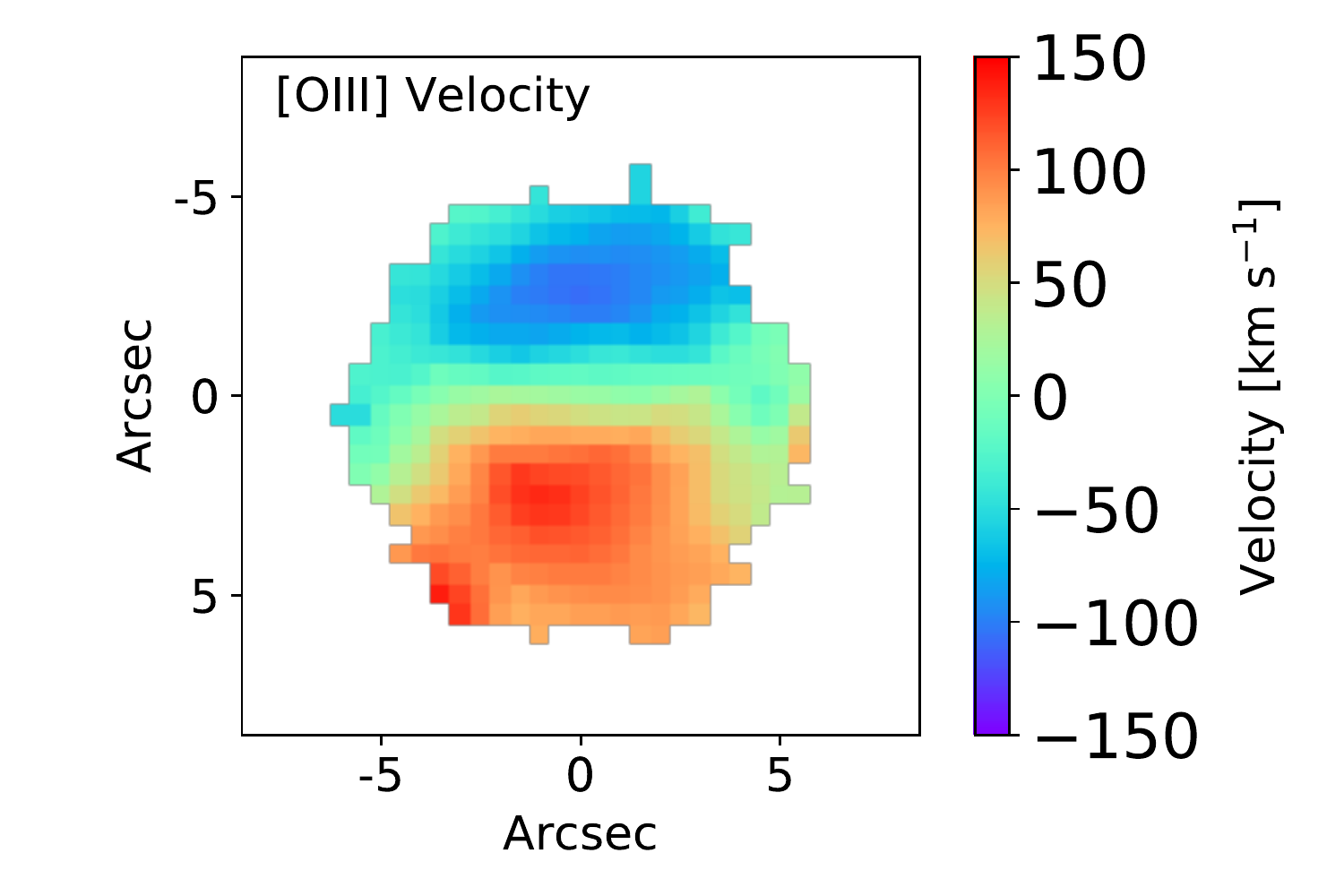}
 \hspace{-0.4in}
 
 \hspace{-0.4in}
 \includegraphics[scale=0.45]{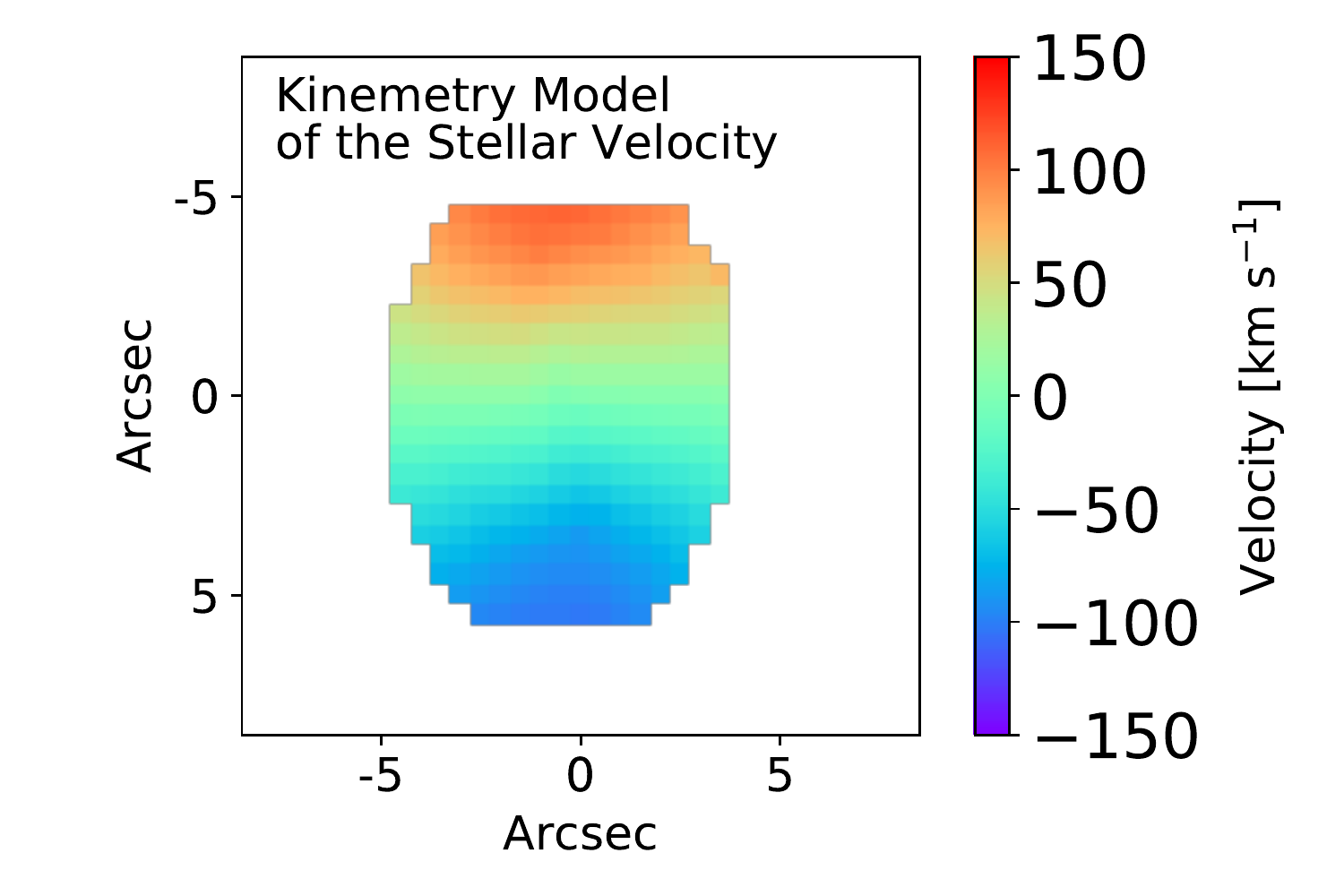}
 \hspace{-0.34in}
 \includegraphics[scale=0.45]{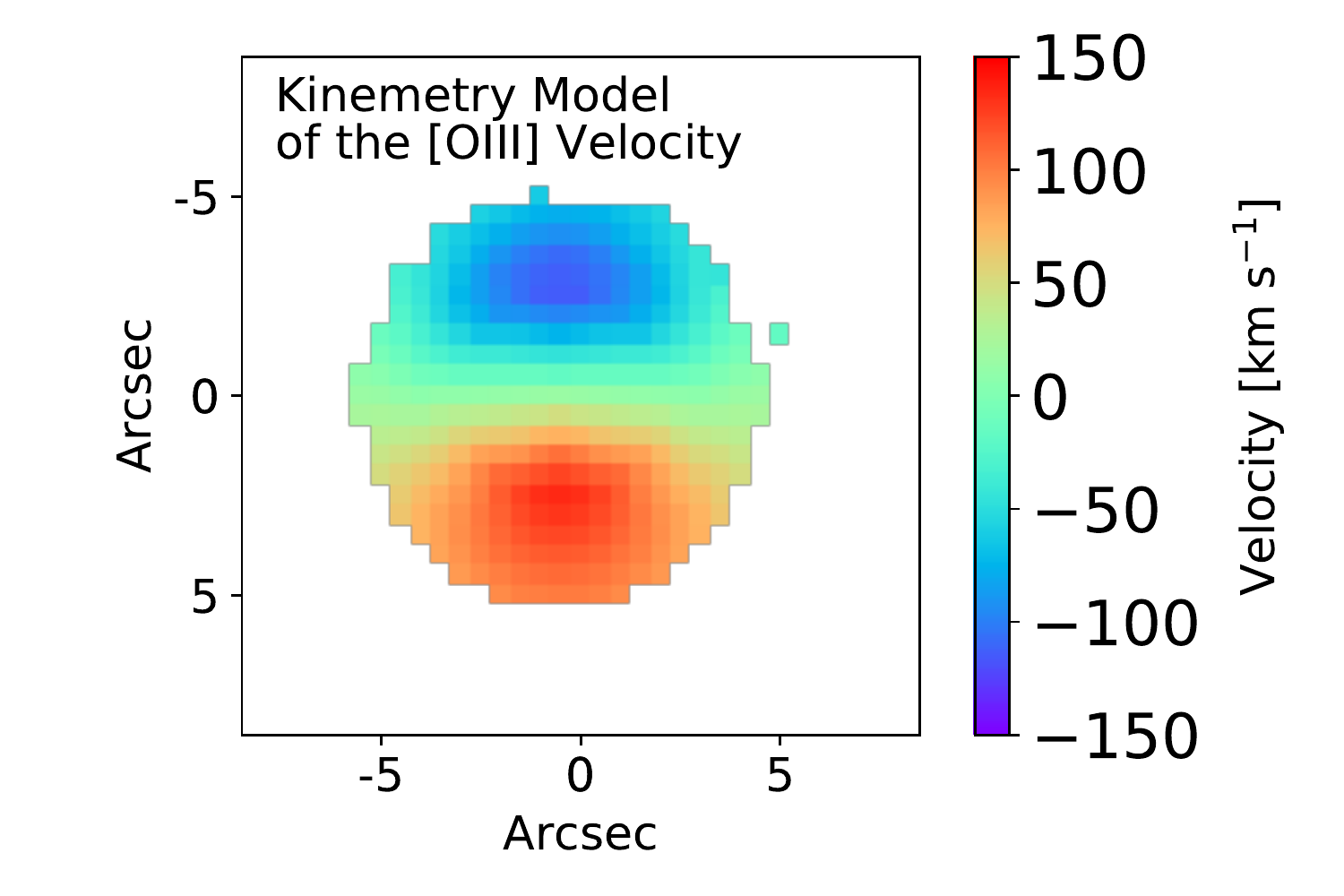}
 \hspace{-0.34in}
 \includegraphics[scale=0.45]{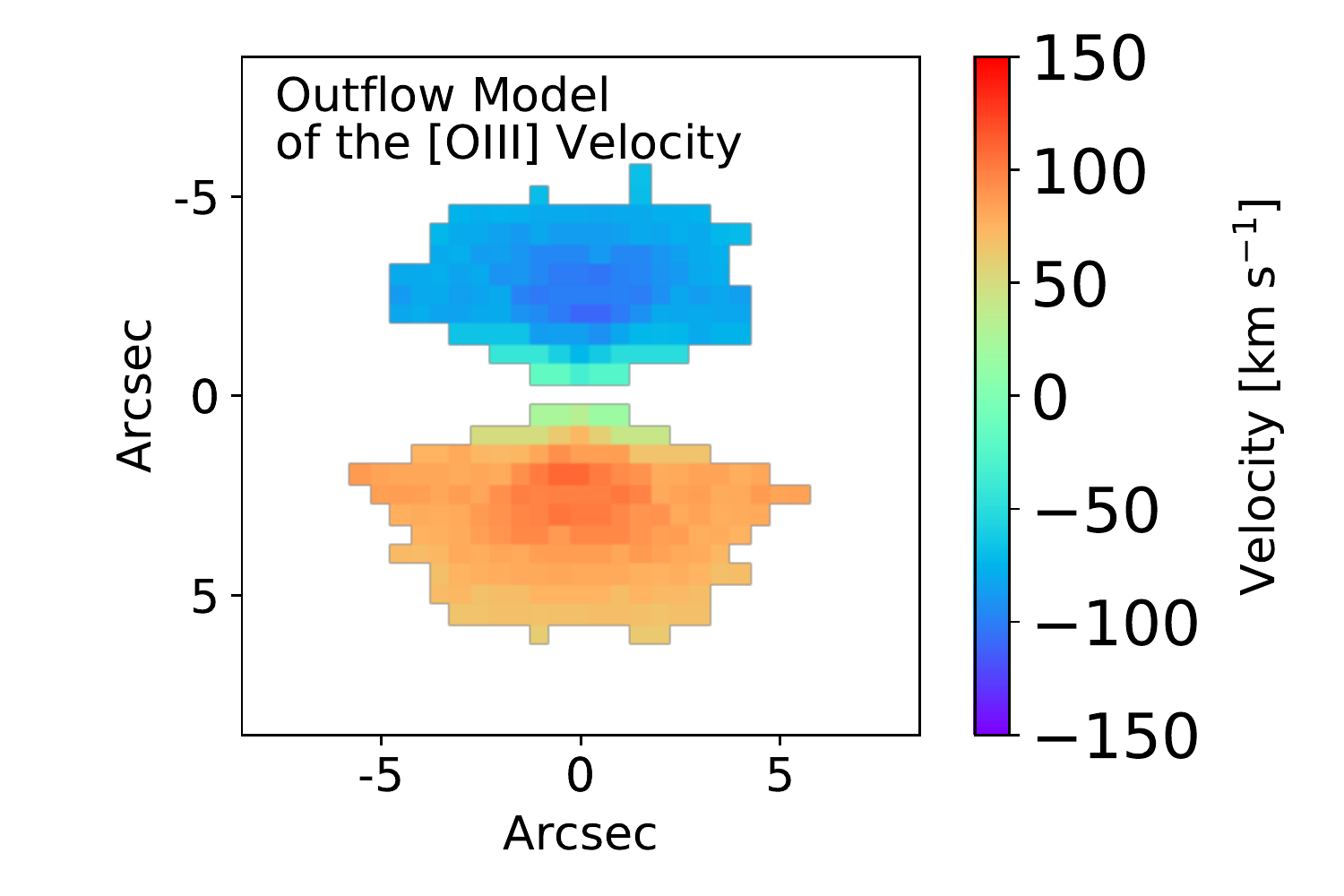}
  \hspace{-0.4in}

 \caption{IFS maps from MaNGA and models for J0749+4514. We plot the stellar velocity (top left) and the [OIII] gas velocity (top right). We show the \texttt{kinemetry} models (\citealt{kinemetry}) for the stellar velocity (bottom left). The [OIII] velocity map is anti-aligned to the stellar velocity and can be described as a counter-rotating disk (bottom middle). We also model a biconical outflow (\citealt{Muller-Sanchez2016}) along the plane of the galaxy (bottom right). }
 \label{MaNGA}

\end{figure*}
However, what may appear to be disk rotation in these figure*s could also mimic a symmetric slower velocity bicone in these spatially-resolved velocity maps. We further investigate the three galaxies that demonstrate disk-like rotation and find that individual component fits of J0009$-$0036 and J0803+3926 have large velocity dispersions ($\sigma >$ 500 km s$^{-1}$). These large dispersions occur at large spatial scales. This indicates that they are outflow-dominated at these spatial positions. J1152+1903 has a few spatial positions with large velocity dispersions of its individual components, however they are overall better characterized as narrower ($\sigma <$ 500 km s$^{-1}$). All three galaxies are double-peaked at all spatial positions. This is distinct from spatially compact outflows, where the outflow is confined to small spatial positions, and the double-peaked profile only appears at the center of the galaxy. We conclude that these three galaxies, while their spatially-resolved velocity maps mimic disk rotation, are dominated by outflows on all spatial scales due to the large velocity dispersions of their individual components (J0009$-$0036 and J0803+3926) and double-peaked profiles at all spatial positions (J0009$-$0036, J0803+3926, and J1152+1903). 

Although the galaxies are dominated by outflows on all spatial scales, we now examine whether there could be smaller contributions from disk rotation. If this is the case, then the analytic models should take this into account by including the parameters for a rotating disk structure in the model in addition to those for a biconical outflow.

To investigate this situation, we make a spatially-resolved velocity map of the narrower component of the double-peaked profile for all 18 galaxies. We choose to track the narrower component because broader components are most often associated with outflow-dominated kinematics. We use the same criteria as above, where we identify a galaxy as rotation-dominated if its spatially-resolved velocity map is symmetric about zero velocity. We find that the velocity offset of the narrow component is not consistent with disk rotation for any of the 18 galaxies.

After our analysis of the kinematics of the 18 galaxies in this sample, we find that they are well described by outflow kinematics on all spatial scales with minimal small contributions from disk rotation. This is consistent with our findings of the biases of the selection of this double-peaked sample. We have selected a sample of energetic and large AGN outflows. They are distinct from the population of outflows in local Seyfert galaxies, for example the galaxy in \citet{Fischer2017} was selected for the biconical morphology of the [OIII]$\lambda$5007 emission in imaging (not for double-peaked narrow lines). These types of outflows occur on smaller scales and the gas kinematics can be described by illuminated disk rotation on larger scales. This is consistent with the biases discussed in Section \ref{comparesize}. The outflows in this work are larger and more energetic than samples of lower-luminosity AGNs, which often only have small-scale outflows (e.g., \citealt{Fischer2017}).

While we have determined that the large-scale kinematics of the 18 galaxies in this sample are best described as outflow-dominated, we investigate a single galaxy in more depth to determine the role of rotating structure in the kinematics of the galaxy. We examine the galaxy J0749+4514 more closely in a case study of an `Outflow Composite' galaxy from \citet{Nevin2016} that is most likely dominated by outflow components on large scales with some rotation on small scales. This galaxy is the only galaxy from our original sample of 71 galaxies that has been observed by the SDSS-IV Mapping Nearby Galaxies at Apache Point (MaNGA) IFS survey (\citealt{Gunn2006,Bundy2015,Drory2015,Law2015,Yan2016b,Yan2016,Law2016,Abolfathi2017,Blanton2017}). So while this is not one of the 18 galaxies selected for the biconical modeling in this work, it offers a unique opportunity to determine how disk rotation shapes the kinematics, and directly investigate how outflow modeling with longslit data compares to outflow modeling with IFS data.

\begin{deluxetable*}{ccccccc}

\tabletypesize{\scriptsize}
\tablewidth{0pt}
\tablecolumns{6}
\tablecaption{J0749+4514 Outflow Model Parameters}
\tablehead{
\colhead{Model} & 
\colhead{$i$} & 
\colhead{$\mathrm{PA}_{\mathrm{bicone}}$} &
\colhead{$r_{t}$} &
\colhead{$\theta_{1,\mathrm{half}}$} &
\colhead{$\theta_{2,\mathrm{half}}$} &
\colhead{V$_{\mathrm{max}}$} \\
&[$^{\circ}$]  &[$^{\circ}$E of N] &[kpc] &[$^{\circ}$] &[$^{\circ}$] & [km s$^{-1}$]   }

\startdata

IFS & 11 $\pm$ 4 & -2 $\pm$ 3 & 3.2 $\pm$ 0.2 & 39 $\pm$ 6 & 66 $\pm$ 4 & 188 $\pm$ 14 \\

Longslit & $ 11 \substack{+ 26 \\- 7 }$ & $ 60 \substack{+ 110 \\- 50 }$ & $ 1 \substack{+ 4 \\- 1 }$ & $ 44 \substack{+ 39 \\- 33 }$ & -- & $ 300 \substack{+ 630 \\- 180 }$ 

\enddata

\tablecomments{Best-fit parameters for the IFS and longslit model of the outflow in J0749+4514 with $1\sigma$ error bars. Column 1: data used in model. Column 2: outflow inclination. Column 3: position angle of the bicone axis on the sky. Column 4: turnover radius in kpc. Column 5: inner half opening angle. Column 6: outer half opening angle if applicable. Column 7: maximum velocity. The best-fit longslit model is the symmetric bicone, so it has no outer opening angle. The IFS model for the outflow has a wall of finite thickness that can be described with both an inner and an outer opening angle.} 
\label{J0749outflow} 
\end{deluxetable*}

\begin{figure*}
\centering

\includegraphics[scale=1]{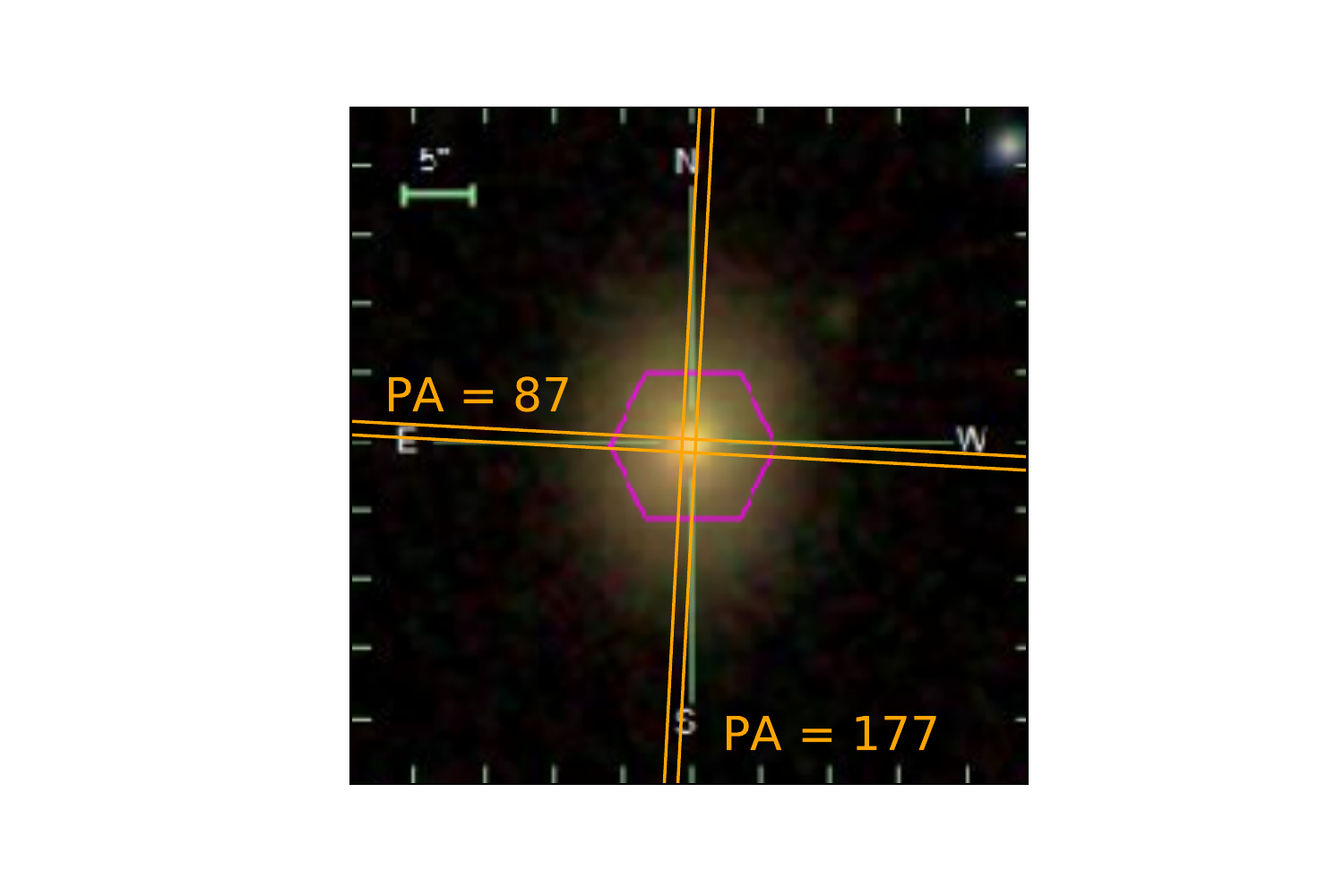}
\includegraphics[scale=0.17]{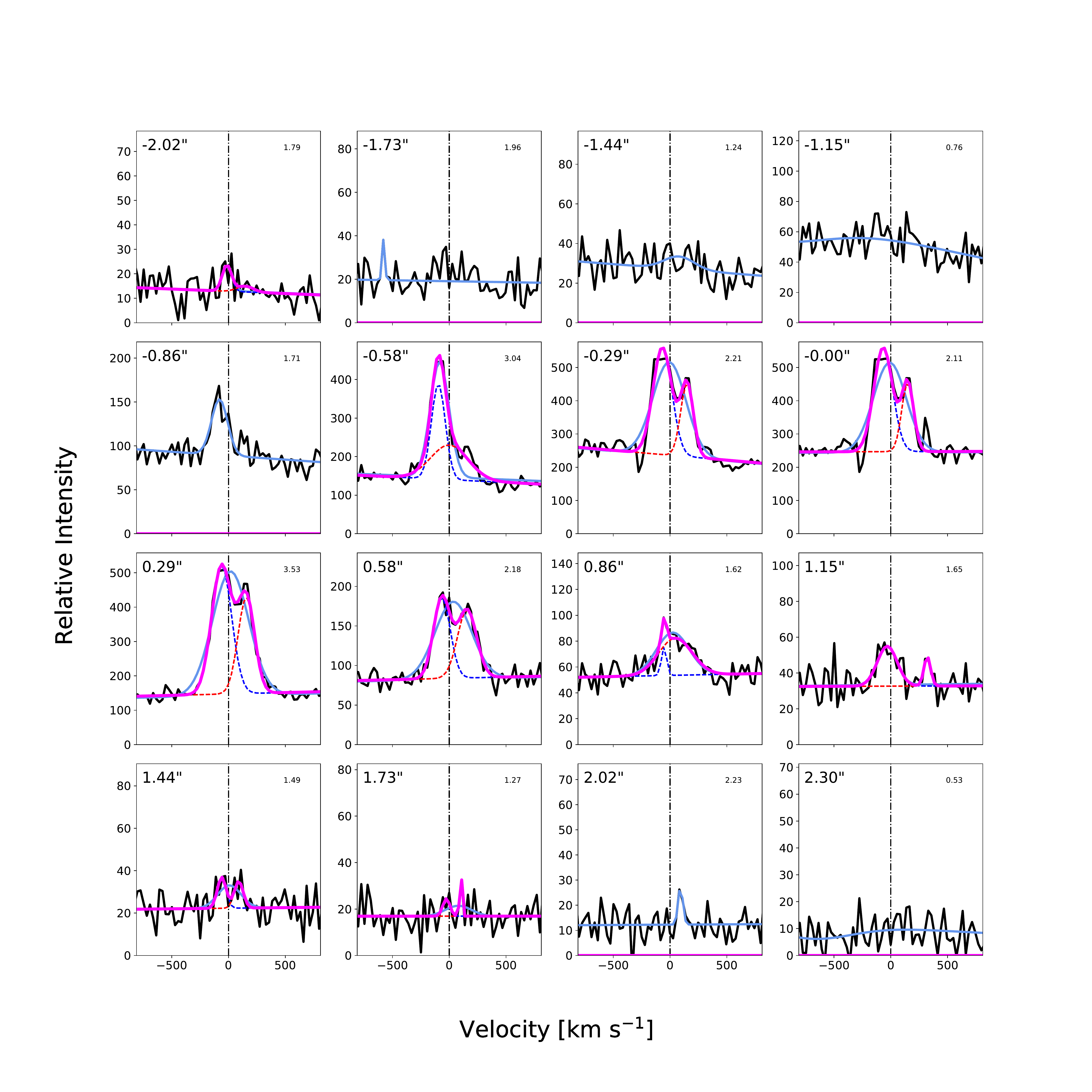}
\hspace{-0.75cm}
\includegraphics[scale=0.17]{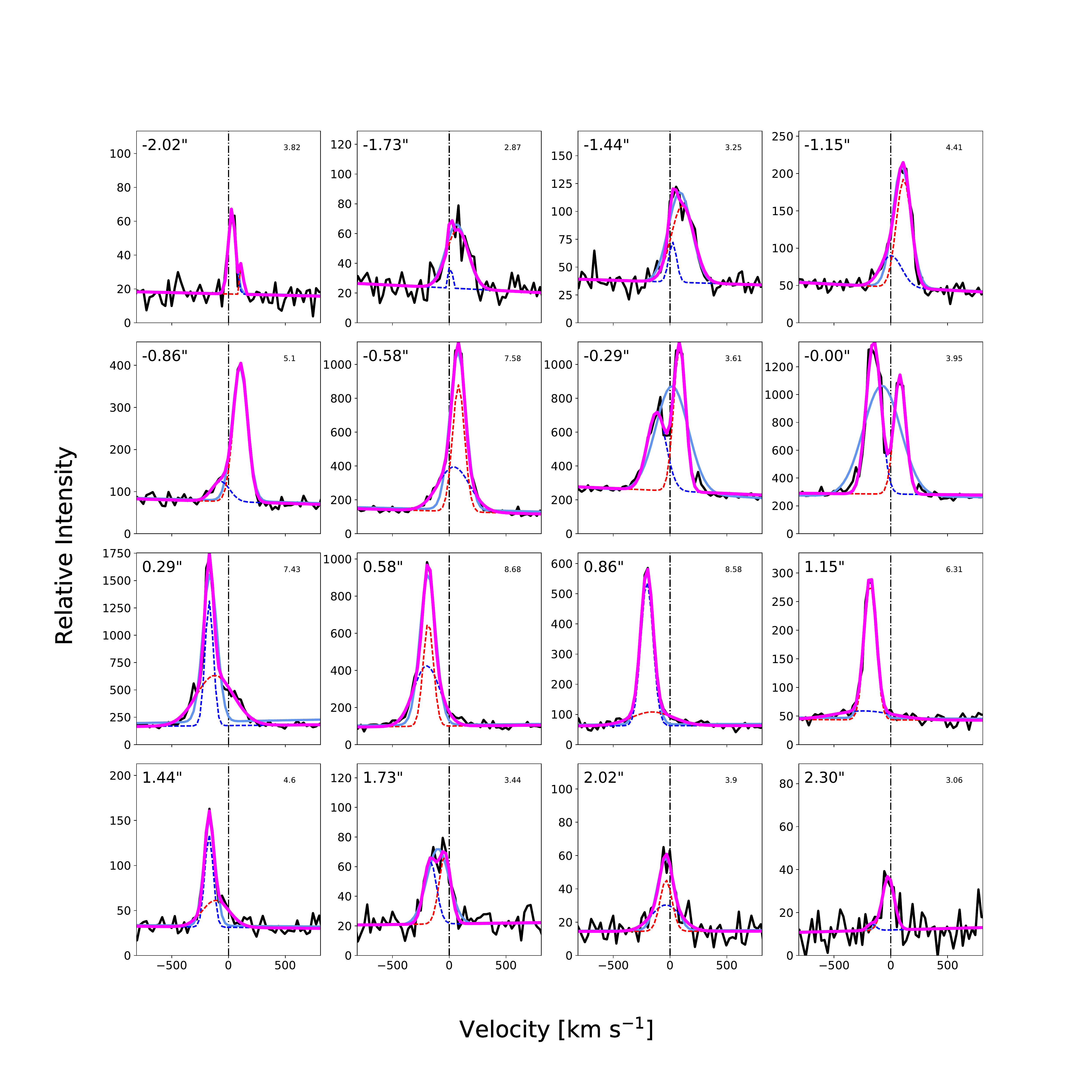}
\hspace{-0.75cm}

\caption{The longslit PAs for J0749+4514 overlaid on the SDSS $gri$ galaxy image and the MaNGA field of view (top). Longslit PA 87 (left bottom) and PA 177 (right bottom) observations. We label the spatial position in the upper left corner of each spatial panel, where positive directions are to the east (PA 87) and north (PA 177). The spatial positions of the longslit positions shown here are confined to the inner $2\farcs3$ radius of the galaxy, inside the magenta MaNGA fiber. We fit two Gaussians, blue and red corresponding to the blueshifted and redshifted profile respectively, and light blue corresponding to the one Gaussian fit. A vertical dashed line denotes systemic velocity. }
\label{Wang008PA177}
\end{figure*}

In IFS, this galaxy has misaligned ionized gas with respect to the stellar disk (Figure \ref{MaNGA}). This is a kinematic signature of either a counter-rotating disk produced by a merger or an outflow (\citealt{Allen2015}, \citealt{Muller-Sanchez2011}). To fully determine the kinematic origin of the [OIII]$\lambda$5007 emission, we model it as a rotating disk, a counter-rotating disk, and an outflow using the IFS data. First, we use the modeling code \texttt{kinemetry} to model the higher-order moments of the LOS velocity distribution of the stellar velocity map as a rotating disk (\citealt{kinemetry}). At each radius, a small number of harmonic terms in a Fourier expansion are used to determine the best fitting ellipse for the stellar velocity map. While \texttt{kinemetry} finds a good fit to the stellar velocity map, this same model is a bad fit for the [OIII] velocity as indicated by a high velocity residual and a high $\chi^2_{\nu}$ value (listed below). Second, we model the [OIII] velocity map as a counter-rotating disk. A counter-rotating disk can be produced by a merger, where gas is funneled to to the center of the galaxy, but is misaligned with the stellar velocity (e.g., \citealt{Allen2015}). Third, we model the [OIII]$\lambda$5007 emission using a biconical outflow model and give the parameters for the best-fit model in Table \ref{J0749outflow}. This analytic model from \citet{Muller-Sanchez2016} is a six parameter model that uses an inner and outer half opening angle, a position angle, a turnover radius, a maximum velocity, and an inclination to produce a biconical outflow model. 

We compare the $\chi^2_{\nu}$ values from the rotating disk, counter-rotating disk, and outflow model and find values of 1691, 7, and 32, respectively. We find that the velocity residuals ($<|$V$_{\mathrm{obs}}$-V$_{\mathrm{mod}}|>$) are 126, 10, and 18 km s$^{-1}$, respectively. The numerical best fit is the counter-rotating disk, but both the counter-rotating disk and the outflow are good fits to the data and also have low velocity residuals.
 
While the $\chi^2_{\nu}$ values indicate that both a counter-rotating disk and an outflow describe the data well, an outflow origin for the kinematics in J0749+4514 is a more likely explanation. First, J0749+4514 has BPT ratios from MaNGA consistent with an AGN origin for the ionized gas on all spatial scales. Second, this AGN is classified as an `Outflow Composite' in \citet{Nevin2016} and as discussed below, has double or triple peaked lines at many spatial positions, especially near the center, where one of these components is broader. As discussed for the full sample, double-peaked profiles at nearly all spatial positions indicates an outflow origin for the gas kinematics. Third, J0749+4514 is an isolated galaxy in SDSS imaging so it is unlikely that a merger is producing a counter-rotating disk. Fourth, outflows are more ubiquitous than counter-rotating disks (e.g., \citealt{Allen2015}). An outflow origin is a better explanation but does not fully explain all of the kinematics in the IFS data.

While the IFS data offers insight into the kinematics of this galaxy, we can also use the longslit data to investigate the kinematics and verify that any conclusions from the longslit observations are consistent with those from the IFS data. We present the MMT longslit data for this galaxy in Figure \ref{Wang008PA177}. PA 87 is the kinematic minor axis of the galaxy (as seen in stellar kinematics); it has double peaks at many spatial positions but is also spatially compact. PA 177 is aligned with the photometric major axis of the galaxy, which is also the kinematic major axis. PA 177 has double or triple peaks at all spatial positions as well, but one component dominates. The dominating component at the spatial extremes of PA 177 is anti-aligned with the stellar velocity maps from MaNGA and cannot be described as stellar rotation. It is redshifted to the south and blueshifted to the north, which is consistent with the [OIII]$\lambda$5007 maps from MaNGA. This dominating component is narrower ($\sigma <$ 500 km s$^{-1}$) at the spatial extremes of PA 177 and could be either a counter-rotating disk or an outflow, which is consistent with the interpretation from the IFS data for the [OIII]$\lambda$5007 maps. 

 This galaxy was classified as Outflow Composite due to the presence of three kinematic components in the longslit maps. For instance, at row $0\farcs288$ in PA 177 (Figure \ref{Wang008PA177}), there are three kinematic components. First, there is a dominating narrower component that is anti-aligned with respect to the stellar rotation as seen in the IFS maps. This component is the same component that dominates at large spatial positions. Second, a compact lower flux narrow component tracks the stellar rotation. Third, a broader lower flux component is centered around zero velocity.
 
There could be multiple kinematic explanations for these three components. First, outflows can have many kinematic components so all three could be attributed to an AGN outflow. Second, the low flux narrow compact component that is aligned with the stellar velocity map could be tracking the rotating disk while the other two components correspond to the walls of an outflow. Third, the dominating high flux narrow component could track a counter-rotating disk, while the low flux narrow compact component is the stellar disk and the broader component is a very small scale outflow. The presence of multiple interacting kinematic components in the longslit and IFS data may cause the observed ambiguity in the modeling using both the IFS and longslit data. This also underscores why we did not include this galaxy in the 18 outflows modeled in this work due to the presence of multiple kinematic processes.

We are conservative in selecting which galaxies we model as outflows, which is why we excluded this galaxy and others like it from the final sample for this work. This galaxy was excluded due to the spatially compact nature of PA 87 and the difficulty of tracking individual components for PA 177. But while we did not include J0749+4514 in the 18 galaxies selected for this work, we model it here as a biconical outflow to compare our longslit model to the IFS model for the outflow. As discussed above, due to the isolation of the galaxy, the ubiquity of AGN outflows, and the BPT ionization origin for the emission lines, the large-scale dominating component is most likely an outflow. We find consistent results from each model (within 2$\sigma$) and an outflow with parameters given in Table \ref{J0749outflow}.

The longslit outflow model has large uncertainties associated with the best-fit parameters. This is an indication of the success of our selection criteria in this paper to select the 18 galaxies to further model and correctly exclude those galaxies that will produce larger uncertainties on the best-fit parameters. Instead of just modeling the two walls of a biconical outflow, it is possible that we are incorrectly attributing small-scale rotational components to the walls of an outflow as discussed above. However, it is impractical to include a rotating component in our model in addition to the biconical outflow model because we do not have enough data points to satisfy the $n > 2k$ selection criterion for a bicone and a rotating disk. If we attempted to introduce more parameters, the model would be under constrained.

The outflow models done with longslit and IFS data agree within 1$\sigma$ errors for all but one parameter (PA$_{\mathrm{bicone}}$). We have shown in our sensitivity analysis in Section \ref{MCMC} that PA$_{\mathrm{bicone}}$ is not well determined. In this case, one of our observed PAs is aligned with the axis of the bicone from the IFS model, but the other does little to constrain the motion of the outflow. Therefore, we are unable to constrain this parameter in the longslit model. This is reflected in the large uncertainty associated with it. The values for PA$_{\mathrm{bicone}}$ from each model do agree to 2$\sigma$ uncertainty. Overall, the models agree but the longslit model has larger uncertainties associated with it, justifying our conservative selection criteria which allows us to use longslit data for modeling purposes only when it can better constrain the bicone geometry.
\clearpage

\section{18 Biconical Models}
\label{APb}
Here we show the best fit bicone models for all 18 galaxies.

\begin{figure*}
\centering
\hspace{-0.5in}
\includegraphics[scale=0.6]{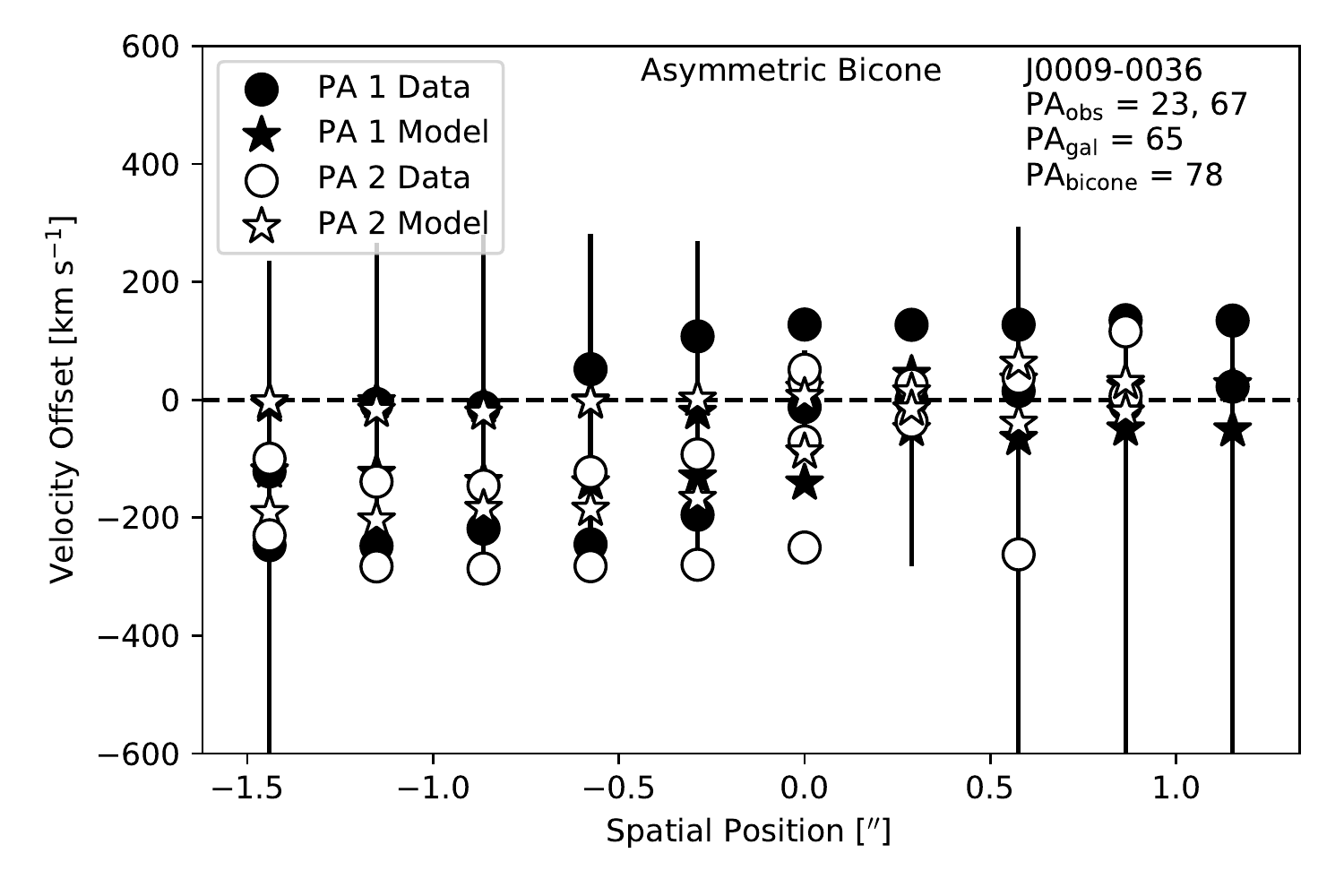}
\includegraphics[scale=0.6]{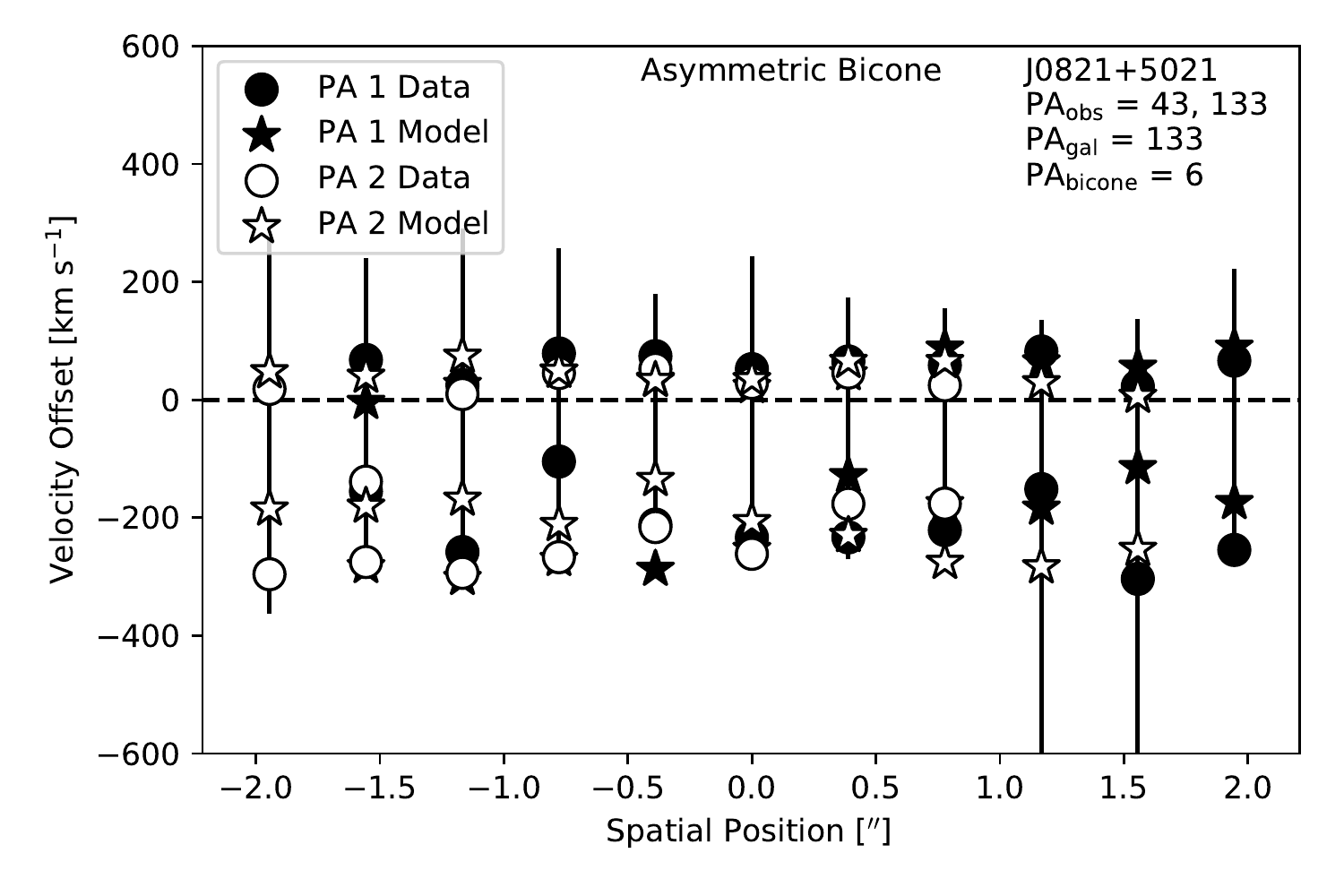}

\hspace{-0.5in}

\hspace{-0.5in}
\includegraphics[scale=0.6]{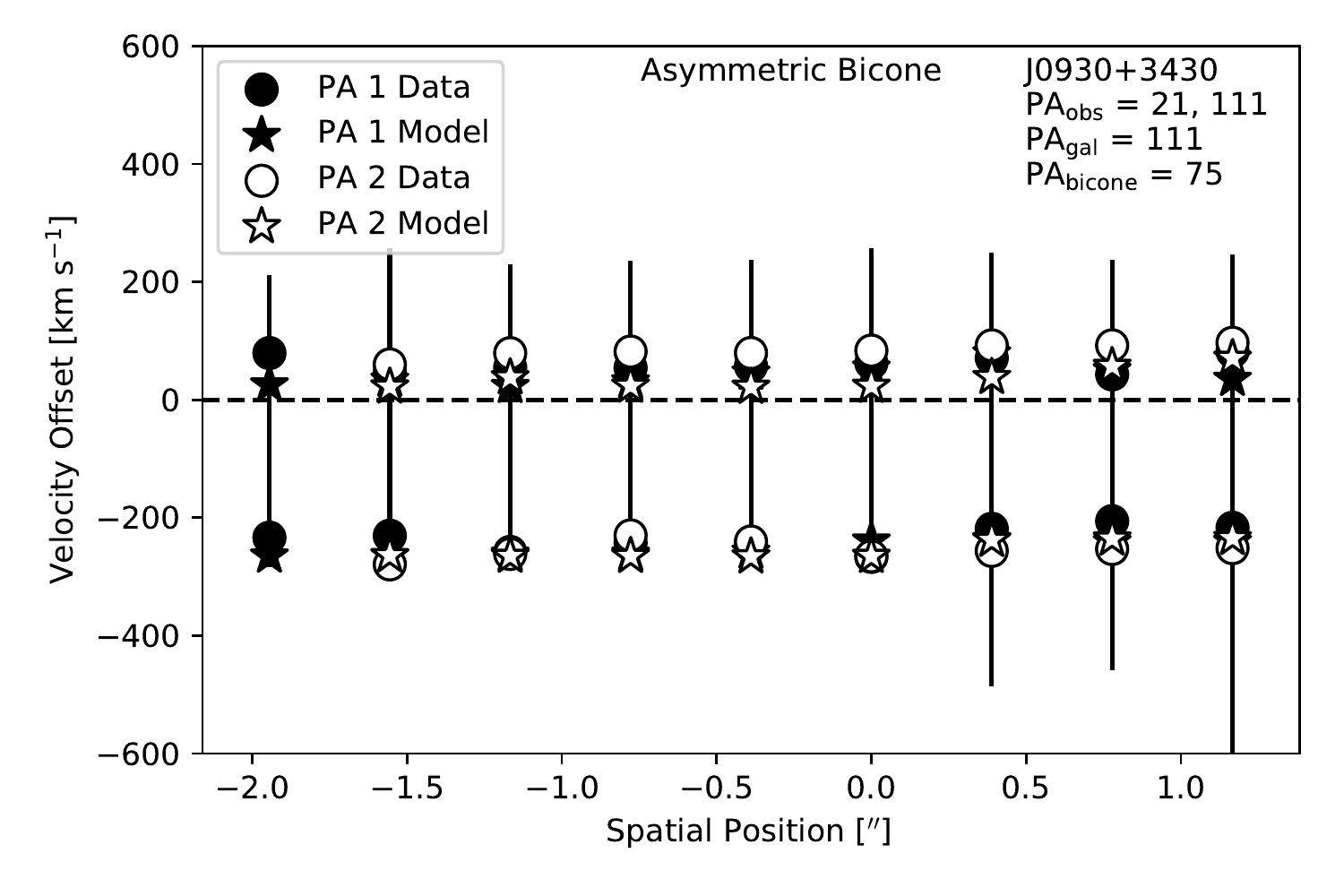}
\includegraphics[scale=0.6]{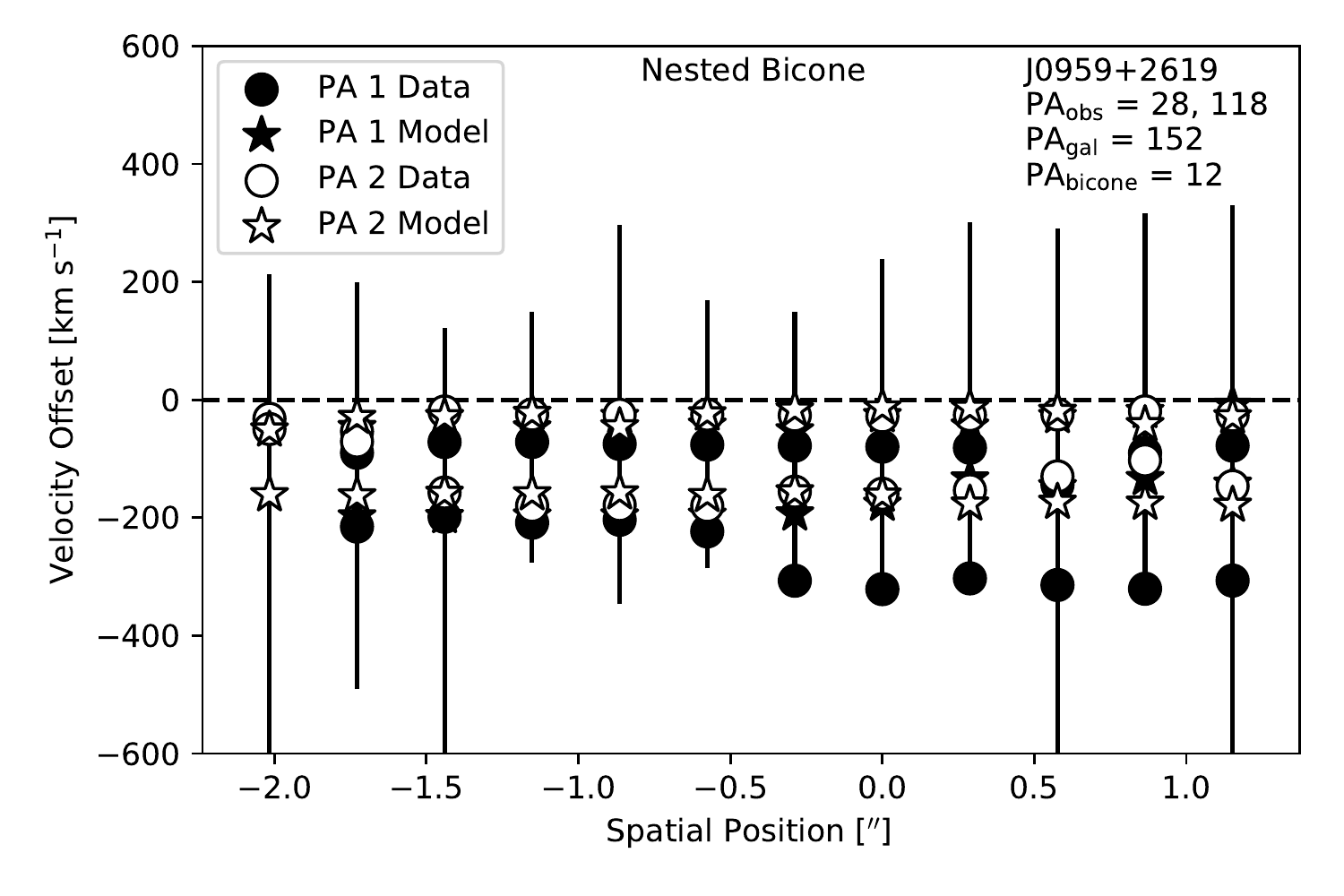}
\hspace{-0.5in}

\hspace{-0.5in}
\includegraphics[scale=0.6]{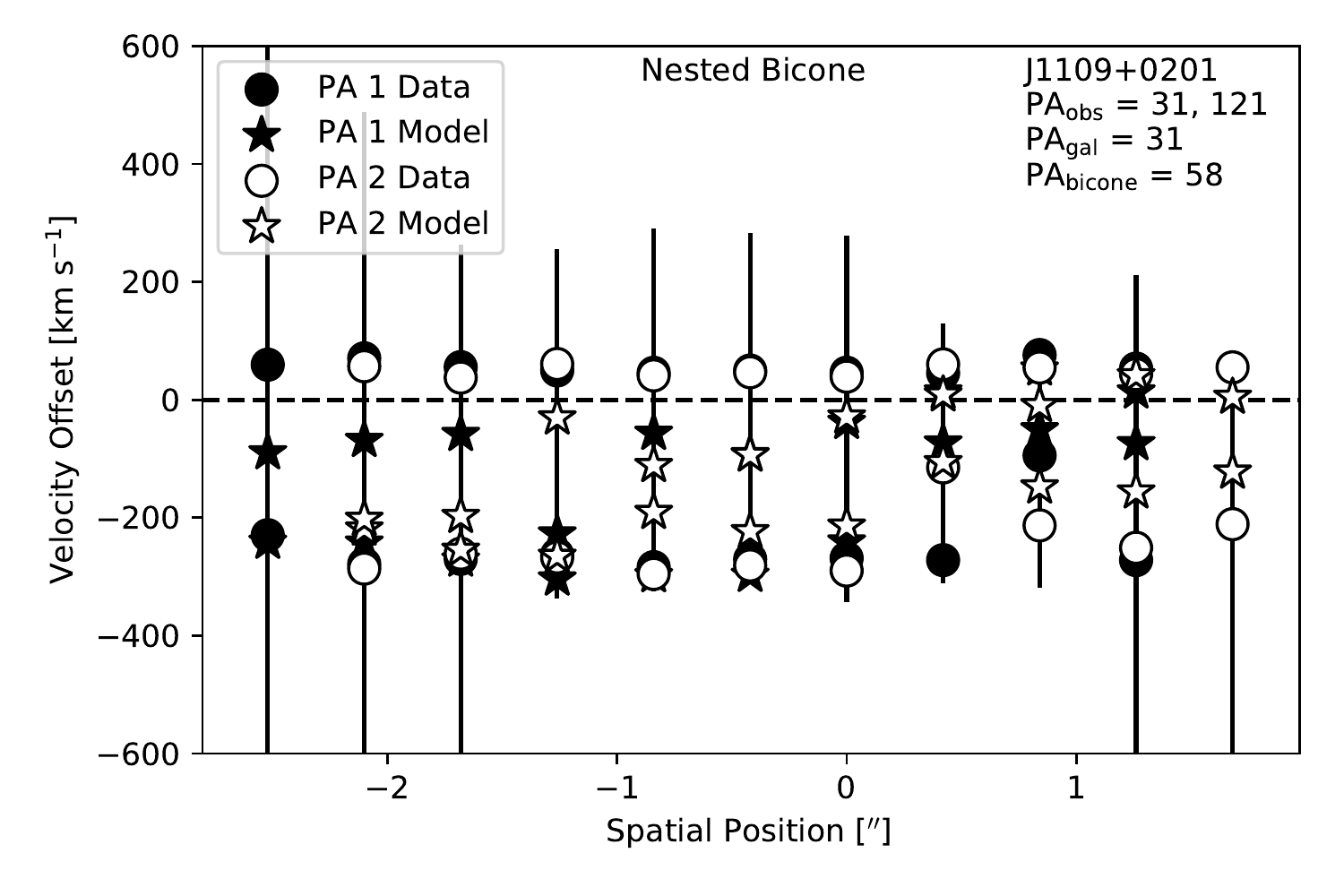}
\includegraphics[scale=0.6]{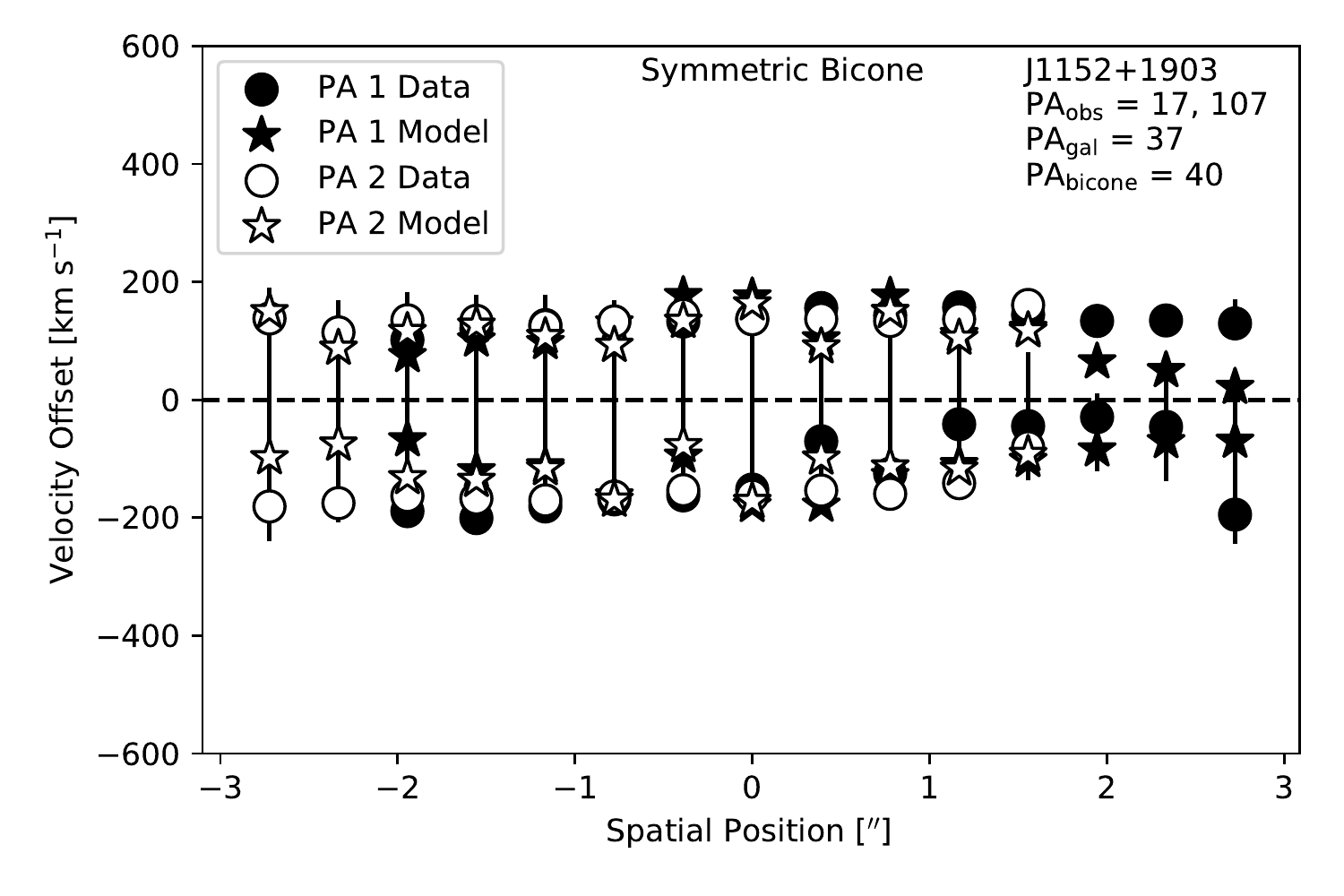}
\hspace{-0.5in}

\end{figure*}

\begin{figure*}

\hspace{-0.5in}
\includegraphics[scale=0.6]{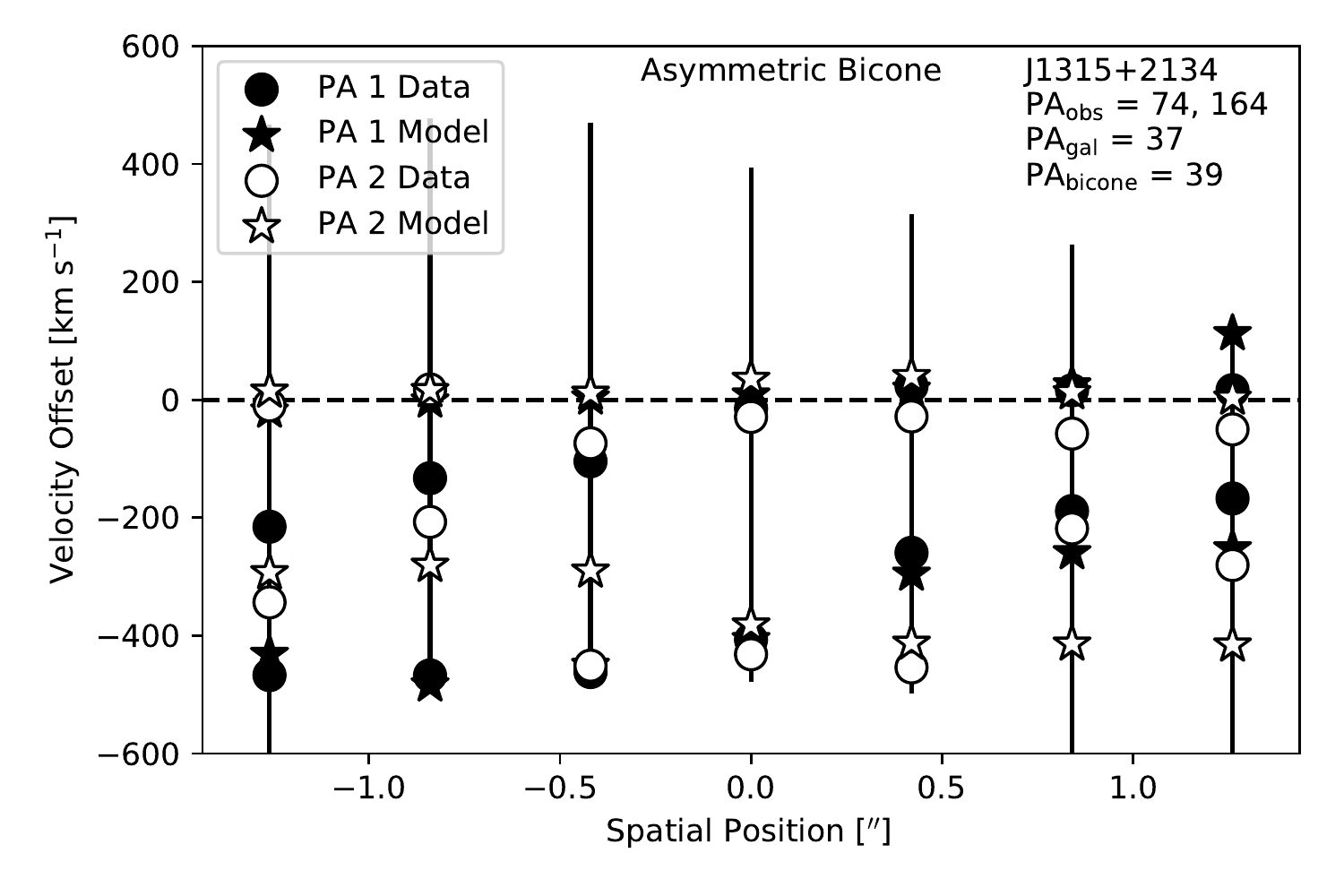}
\includegraphics[scale=0.6]{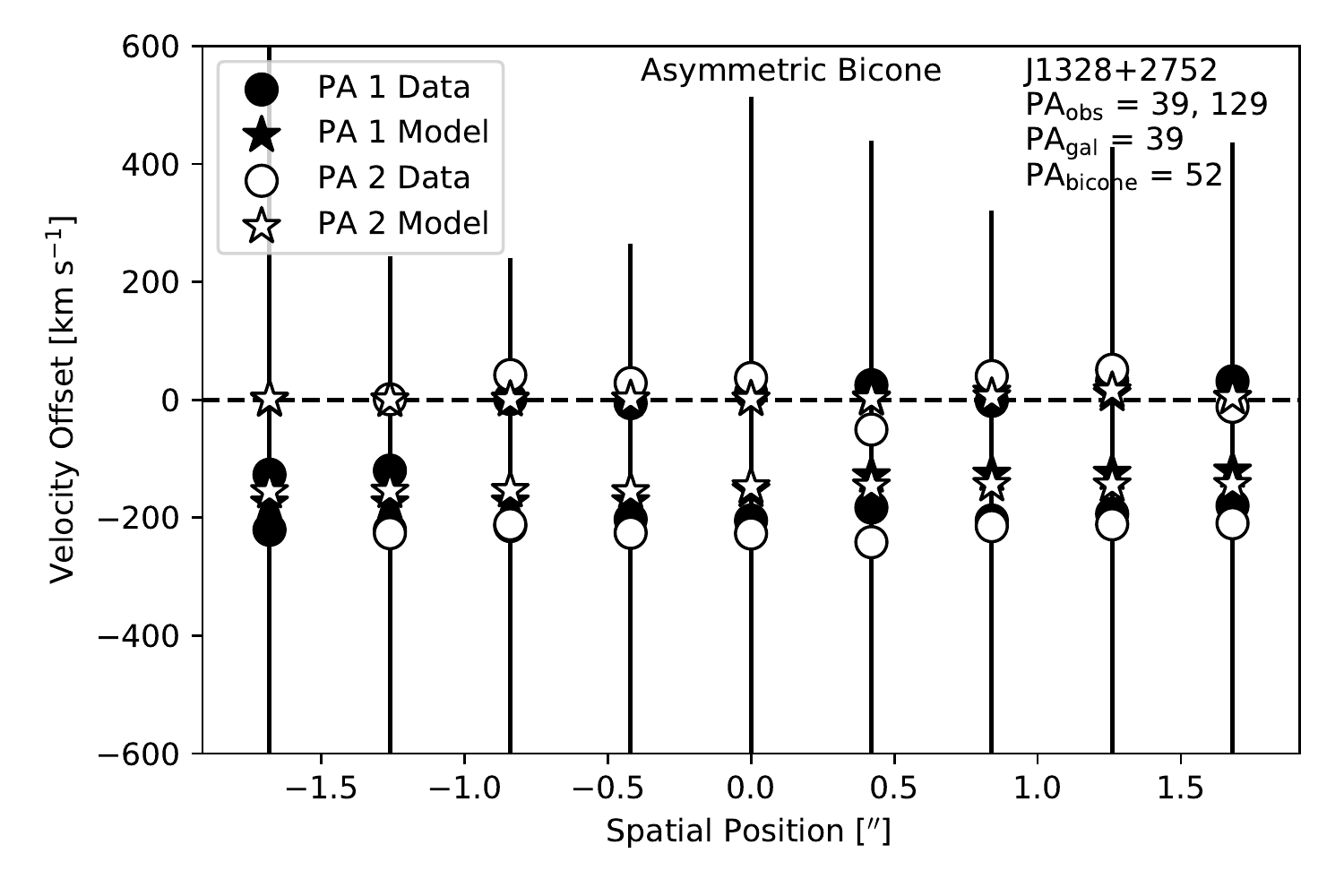}

\hspace{-0.5in}

\hspace{-0.5in}
\includegraphics[scale=0.6]{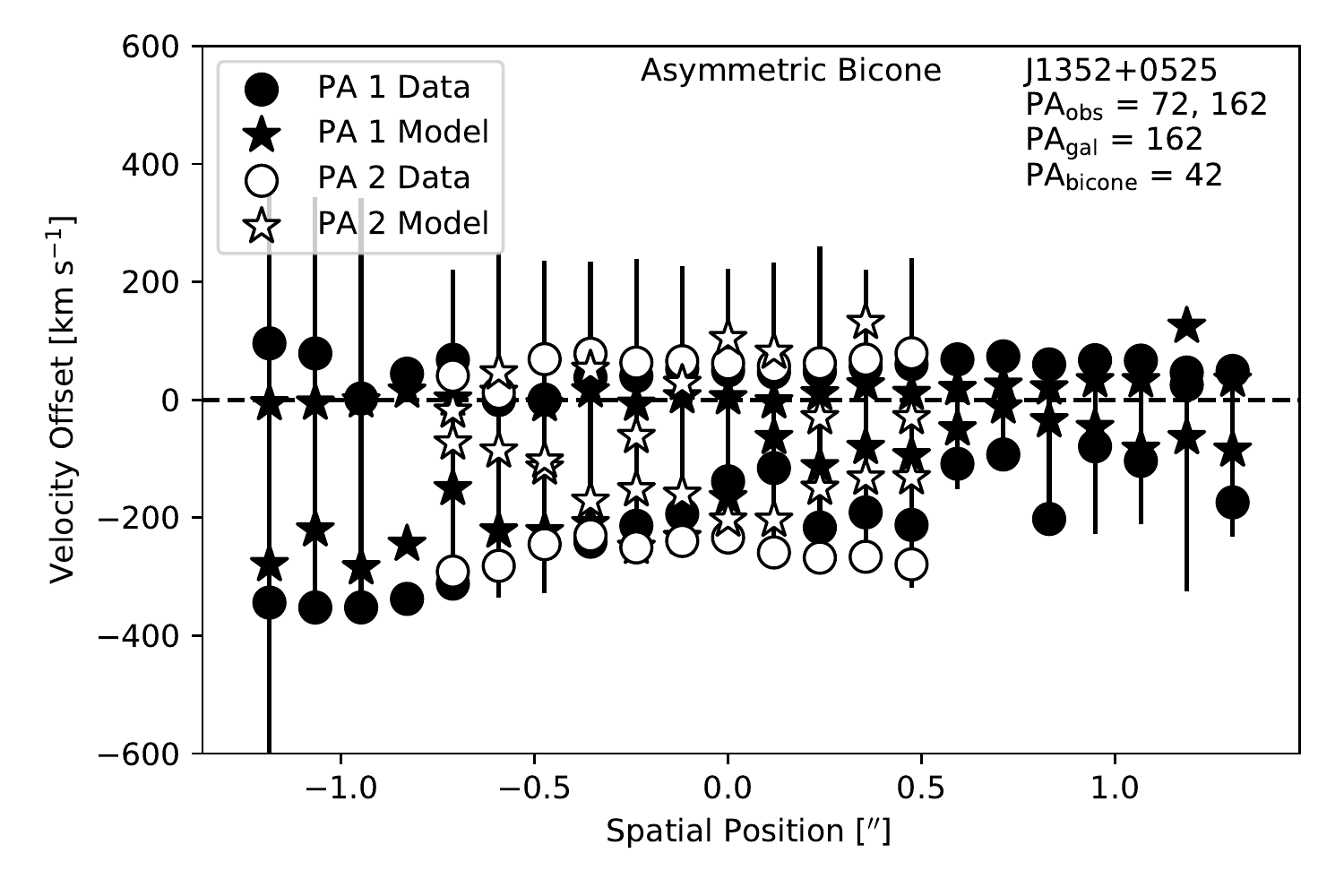}
\includegraphics[scale=0.6]{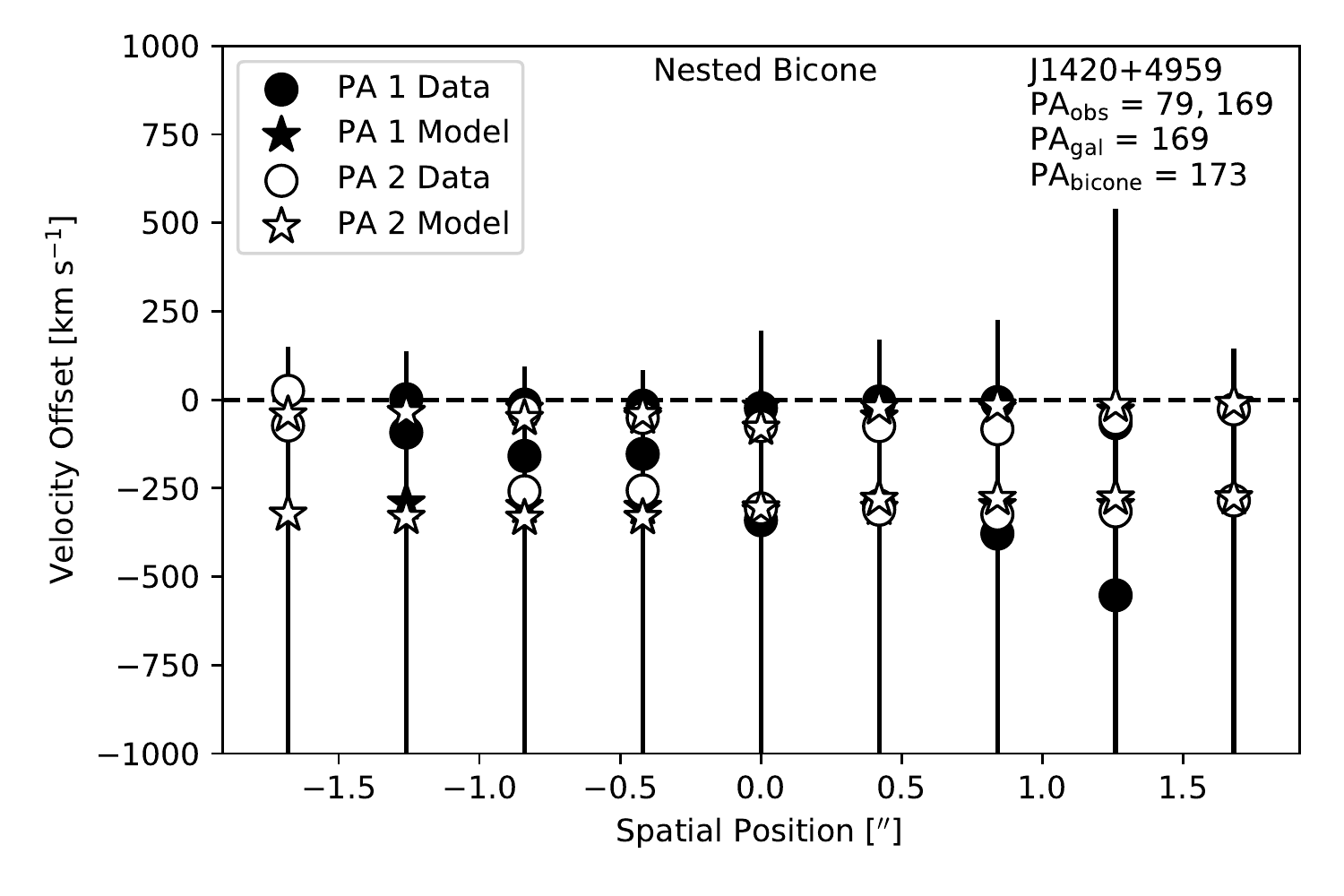}
\hspace{-0.5in}

\hspace{-0.5in}
\includegraphics[scale=0.6]{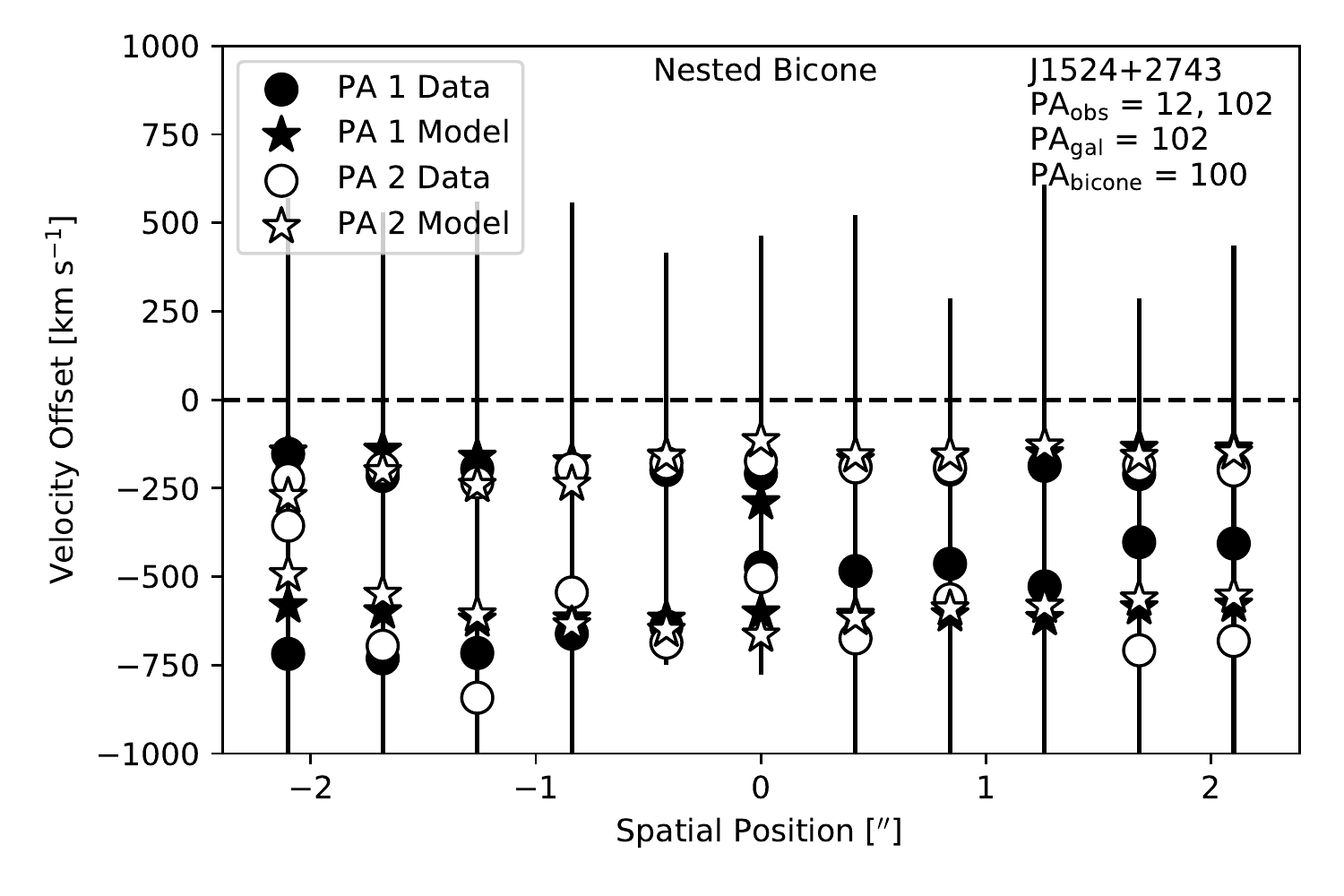}
\includegraphics[scale=0.6]{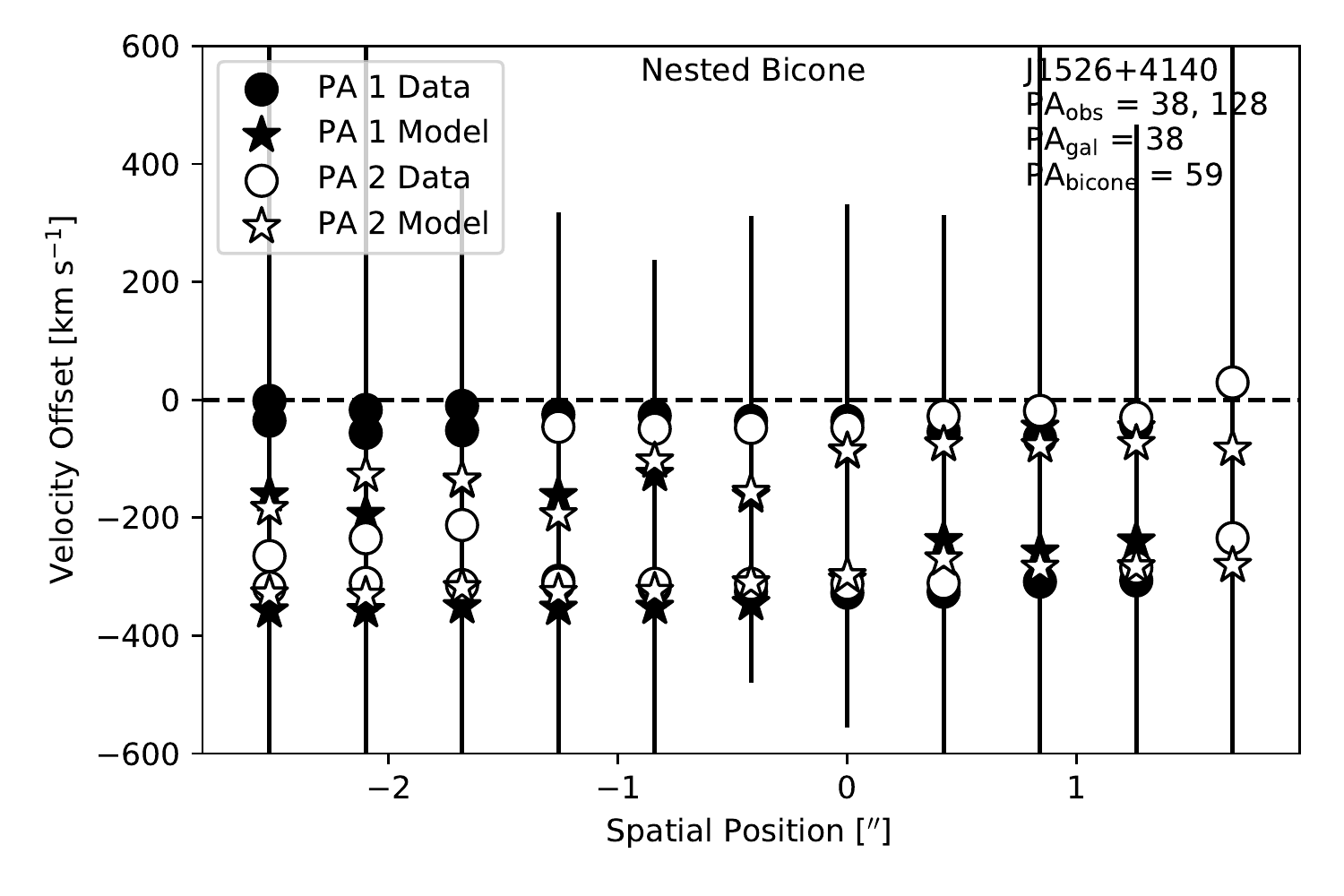}
\hspace{-0.5in}

\end{figure*}

\begin{figure*}
\hspace{-0.5in}
\includegraphics[scale=0.6]{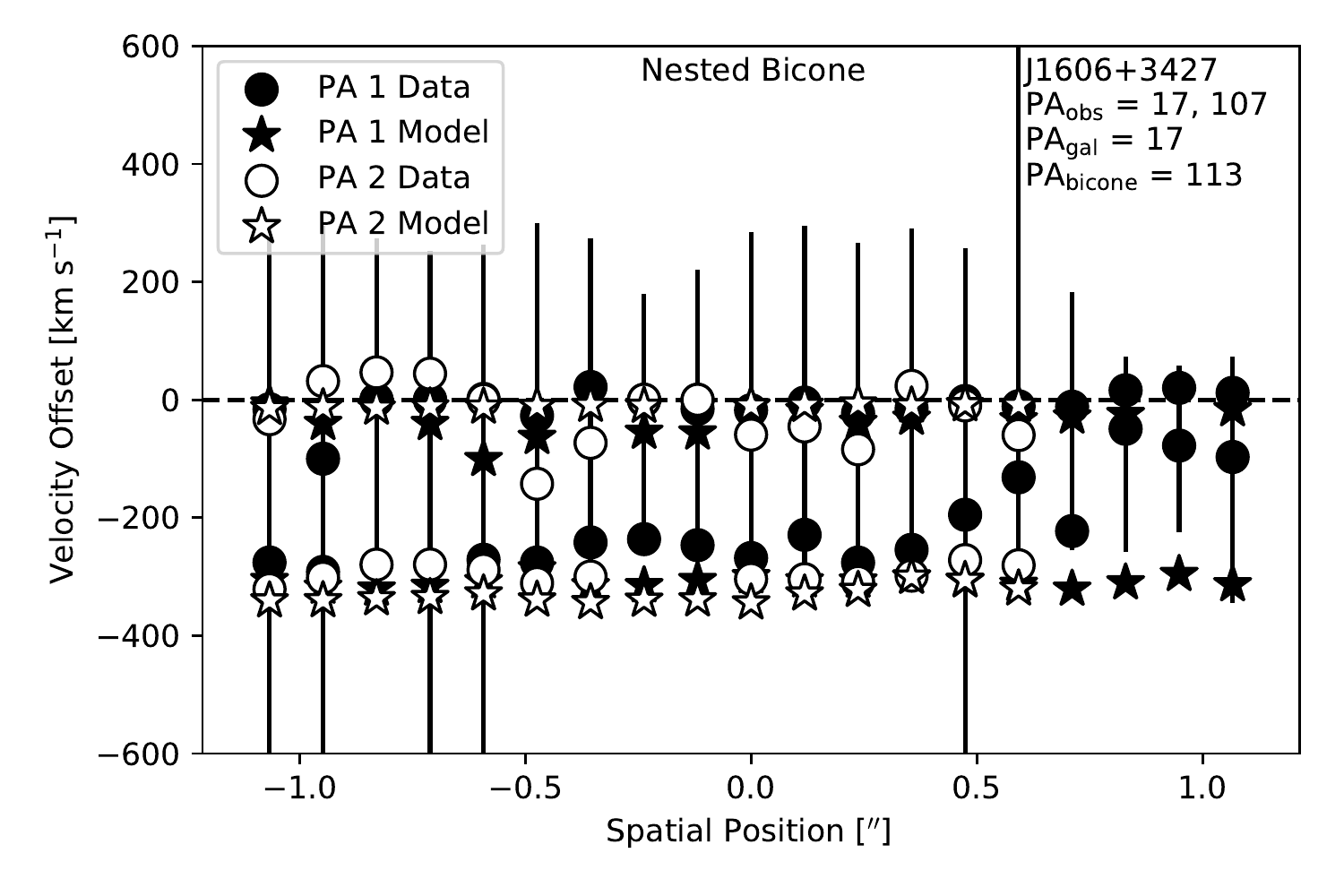}
\includegraphics[scale=0.6]{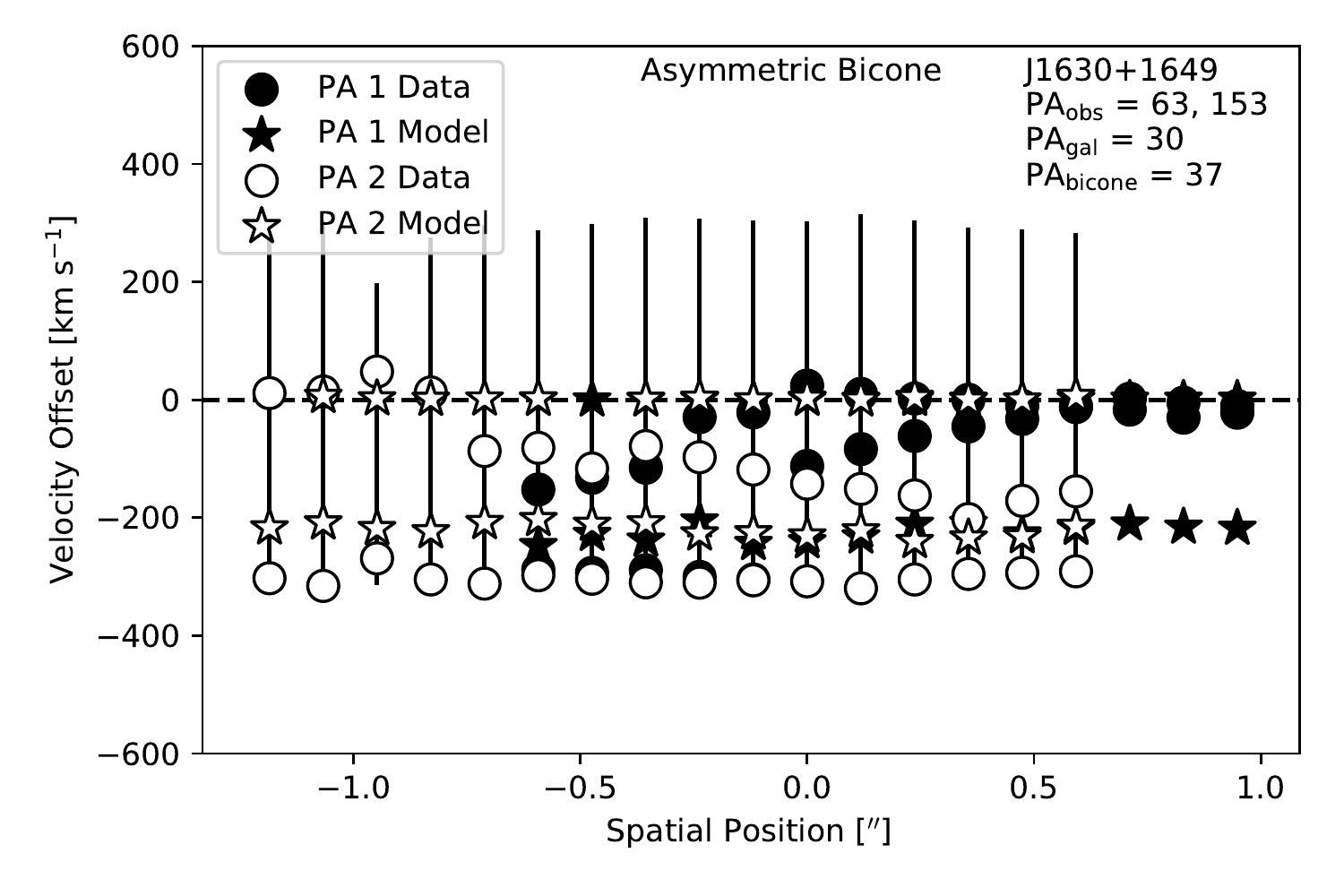}
\hspace{-0.5in}

\hspace{-0.5in}
\includegraphics[scale=0.6]{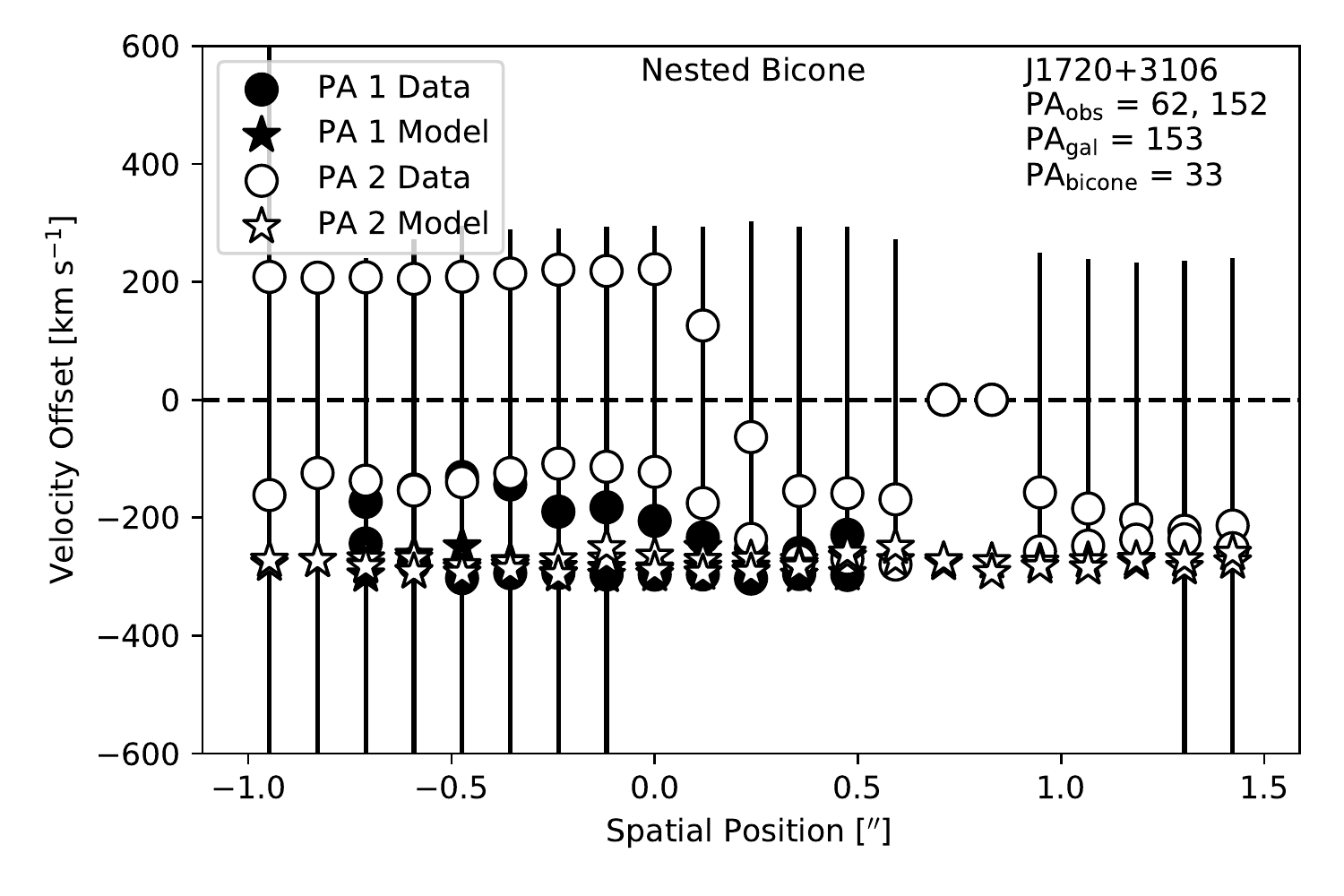}

\hspace{-0.5in}

\caption{As Figure \ref{anywallobs}, but for the remaining 15 galaxies not shown there.}
\end{figure*}

\end{document}